\documentclass[12pt]{article}

\usepackage{mdwlist}
\usepackage{enumerate}
\usepackage{graphicx}
\usepackage{float}
\usepackage{cite}
\usepackage{amsfonts}
\usepackage{amssymb}
\usepackage{amsmath}
\usepackage{relsize}
\usepackage[compatibility=false]{caption}
\usepackage{subcaption}
\usepackage{xcolor}

\usepackage{rotating}
\usepackage{booktabs}

\def\gtwid{\mathrel{\raise.3ex\hbox{$>$\kern-.75em\lower1ex\hbox{$\sim$}}}}
\def\ltwid{\mathrel{\raise.3ex\hbox{$<$\kern-.75em\lower1ex\hbox{$\sim$}}}}
\def\square{\kern1pt\vbox{\hrule height 1.2pt\hbox{\vrule width 1.2pt\hskip 3pt
   \vbox{\vskip 6pt}\hskip 3pt\vrule width 0.6pt}\hrule height 0.6pt}\kern1pt}

\begin{document}

\begin{titlepage}

\begin{flushright}
UFIFT-QG-26-09 \\
CCTP-2026-19 \\
ITCP/2026/19 
\end{flushright}

\vskip 0.5cm

\begin{center}
{\bf Dimensionally Regulated Graviton Self-Energy from Gravitons on de Sitter}
\end{center}

\vskip 0.25cm

\begin{center}
B. Yesilyurt$^{1*}$, N. C. Tsamis$^{2\star}$ and R. P. Woodard$^{1\dagger}$
\end{center}

\begin{center}
\it{$^{1}$ Department of Physics, University of Florida,\\
Gainesville, FL 32611, UNITED STATES}
\end{center}

\begin{center}
\it{$^{2}$ Institute of Theoretical Physics \& Computational Physics, \\
Department of Physics, University of Crete, \\
GR-700 13 Heraklion, GREECE}
\end{center}

\vspace{0.25cm}

\begin{center}
ABSTRACT
\end{center}
The graviton self-energy $-i [\mbox{}^{\mu\nu} \Sigma^{\rho\sigma}](x;x')$
is the 1PI 2-point function of quantum gravity. It can be used to 
quantum-correct the linearized Einstein equation in order to infer quantum 
corrections to gravitational radiation and to the force of gravity. As long 
as the two points are non-coincident, the graviton self-energy does not 
need regularization at 1-loop order, however, solving the 1-loop effective
field equations requires integrating $[\mbox{}^{\mu\nu} \Sigma^{\rho
\sigma}](x;x')$ over ${x'}^{\mu}$, including ${x'}^{\mu} = x^{\mu}$. To
facilitate this we employ dimensional regularization and BPHZ renormalization
to compute the 1-loop contribution to the graviton self-energy, in pure
quantum gravity, on de Sitter background, in the simplest gauge. We then solve 
the effective field equations to derive 1-loop corrections to the plane wave 
mode functions of gravitational radiation, and to the Newtonian potential. 
Although it seems possible to renormalize in such a way that the mode function 
shows no secular growth, the Newtonian potential grows like the third power 
of the number of e-foldings.

\begin{flushleft}
\end{flushleft}
\begin{flushleft}
$^{*}$ b.yesilyurt@ufl.edu \\
$^{\star}$ tsamis@physics.uoc.gr \\
$^{\dagger}$ woodard@phys.ufl.edu
\end{flushleft}

\end{titlepage}

\section{Introduction}

The symmetries of cosmology are homogeneity and isotropy. We can
take the background geometry to be,
\begin{equation}
ds^2 = -dt^2 + a^2(t) d\vec{x} \cdot d\vec{x} = a^2 [-d\eta^2
+ d\vec{x} \cdot d\vec{x}] \;\; \Longrightarrow \;\; H(t) 
\equiv \tfrac{\dot{a}}{a} \;\; , \;\; \epsilon(t) \equiv 
-\tfrac{\dot{H}}{H^2} \; . \label{geometry}
\end{equation}
Here $a(t)$ is the scale factor, $H(t)$ is the Hubble parameter,
and $\epsilon(t)$ is the first slow roll parameter. 

Accelerated expansion ($H > 0$ with $0 \leq \epsilon < 1$) is known
as inflation. Inflation alters the energy-time uncertainty relation
so that massless, long wave length virtual particles can persist 
indefinitely \cite{Woodard:2014jba}. Particles which are both massless
and not conformally invariant, such as massless, minimally coupled 
scalars and gravitons, experience explosive particle production 
during inflation. For the limiting case of de Sitter ($\epsilon = 0$,
$a(t) = e^{H t}$) the occupation number of gravitons with a single 
polarization and wave vector $\vec{k}$ is,
\begin{equation}
N(t,k) = \Bigl[ \tfrac{H a(t)}{2 k}\Bigr]^2 \; . \label{occnum}
\end{equation}
This is what produces the tensor power spectrum \cite{Starobinsky:1979ty,
Starobinsky:1982ee}, and it is natural to expect significant loop
corrections as well. Because $N(t,k)$ is negligible for the ultraviolet 
($k \gg H a(t)$) and large for the infrared ($k \ll H a(t)$), we can 
reliably study these corrections using quantum general relativity 
as a low energy effective field theory in the sense of Donoghue 
\cite{Donoghue:1993eb,Donoghue:1994dn,Donoghue:2017ovt}.

The exponential time dependence evident in expression (\ref{occnum})
causes loop corrections that would be constant in flat space background
to become time dependent in de Sitter background. One way to study this 
is by using the graviton 1PI (one-particle-irreducible) 2-point function 
$-i [\mbox{}^{\mu\nu} \Sigma^{\rho\sigma}](x;x')$ to quantum-correct the 
linearized Einstein equation,
\begin{equation}
\mathcal{L}^{\mu\nu\rho\sigma} \kappa h_{\rho\sigma}(x) - \int \!\!
d^4x' [\mbox{}^{\mu\nu} \Sigma^{\rho\sigma}](x;x') \kappa h_{\rho\sigma}(x')
= 8 \pi G T^{\mu\nu}(x) \; . \label{linEin}
\end{equation}
Here $\mathcal{L}^{\mu\nu\rho\sigma}$ is the Lichnerowicz operator, the 
graviton field $h_{\mu\nu}$ is defined in terms of the full metric
$g_{\mu\nu} \equiv a^2 [\eta_{\mu\nu} + \kappa h_{\mu\nu}]$, with $\kappa^2
\equiv 16 \pi G$, and $T^{\mu\nu}$ is the variation of the matter action
with respect to $h_{\mu\nu}$. By setting the stress tensor to zero one
can use equation (\ref{linEin}) to work out corrections to the mode function
$u(t,k)$ of plane wave gravitational radiation,
\begin{equation}
T^{\mu\nu} = 0 \qquad \Longrightarrow \qquad h_{\mu\nu}(x) = u(t,k) 
e^{i \vec{k} \cdot \vec{x}} \epsilon_{\mu\nu}(\vec{k},\lambda) \; . 
\label{radiation}
\end{equation}
Choosing the stress tensor to be that of a static, point mass allows one
to work out corrections to the Newtonian potential $\Psi(t,r)$,
\begin{equation}
T^{\mu\nu} = M a \delta^3(\vec{x}) \delta^{\mu}_{~0} \delta^{\nu}_{~0} \qquad 
\Longrightarrow \qquad \kappa h_{00}(x) = - 2 \Psi(t,r) \; . \label{Newtonian}
\end{equation}

Quantum gravity on de Sitter background is {\it hard}. Only a handful of
of graviton loop corrections to 1PI 2-point functions have been computed.
These include:
\begin{itemize*}
\item{The self-mass $-i M^2(x;x')$ of a massless, minimally coupled 
scalar \cite{Kahya:2007bc,Glavan:2021adm};}
\item{The self-mass of a massless, conformally coupled scalar 
\cite{Boran:2014xpa,Boran:2017fsx,Glavan:2020ccz};}
\item{The self-energy $-i [\mbox{}_{i} \Sigma_{j}](x;x')$ of a massless 
\cite{Miao:2005am}, or massive \cite{Miao:2012bj}, Dirac fermion; and}
\item{The vacuum polarization $i [\mbox{}^{\mu} \Pi^{\nu}](x;x')$ 
\cite{Leonard:2013xsa,Glavan:2016bvp}.}
\end{itemize*}
For each of these results, which were dimensionally regulated 
\cite{Bollini:1972ui,tHooft:1972tcz} and fully renormalized in the sense 
of BPHZ (Bogoliubov and Parasiuk \cite{Bogoliubov:1957gp}, Hepp 
\cite{Hepp:1966eg}, and Zimmermann \cite{Zimmermann:1968mu,
Zimmermann:1969jj}, the loop involved only a single graviton propagator
and the 1PI 2-point function had at most two free indices. In contrast,
the graviton self-energy $-i [\mbox{}^{\mu\nu} \Sigma^{\rho\sigma}](x;x')$
involves two graviton propagators and has four free indices. It is so 
complicated that the original calculation \cite{Tsamis:1996qk} was 
unregulated, in $D=4$ spacetime dimensions, and away from coincidence
($x^{\mu} \neq {x'}^{\mu}$). That suffices to check the Ward identity,
but it is not enough for the effective field equation (\ref{linEin}),
which involves an integral over all ${x'}^{\mu}$, including coincidence.
The purpose of this paper is to perform a dimensionally regulated and
fully renormalized computation. We will also use the self-energy in 
equation (\ref{linEin}) to check estimates for the leading secular 
dependence of $u(t,k)$ and $\Psi(t,r)$ that were obtained by applying a 
set of plausible rules to extend the noncoincident self-energy to a guess 
for the fully renormalized result \cite{Tan:2021lza,Tan:2022xpn},
\begin{eqnarray}
u(t,k) &\!\!\! \longrightarrow \!\!\!& \tfrac{H}{\sqrt{2 k^3}} \Bigl\{1 + 
\tfrac{\kappa^2 H^2}{3 \pi^2} \!\times\! \ln^2[a(t)] + \dots \Bigr\} 
\; , \label{guessu} \qquad \\ 
\Psi(t,r) &\!\!\! \longrightarrow \!\!\!& \tfrac{G M}{a(t) r} \Bigl\{1
- \tfrac{\kappa^2 H^2}{2 \pi^2} \!\times\! \ln^3[a(t)] + \dots \Bigr\} 
\; . \label{guessPsi} \qquad
\end{eqnarray}

In Section 2 we review the Feynman rules. The primitive, dimensionally
regulated result for $-i [\mbox{}^{\mu\nu} \Sigma^{\rho\sigma}](x;x')$
is derived in Section 3. BPHZ renormalization is accomplished in Section 
4, and the effective field equation (\ref{linEin}) is solved for $u(t,k)$
and $\Psi(t,r)$ in Section 5. Our conclusions comprise Section 6. Some
expansions for products of propagators are given in an Appendix. 

\section{Feynman Rules}

The purpose of this section is to present those of the Feynman rules 
whcih are needed to compute $-i [\mbox{}^{\mu\nu} \Sigma^{\rho\sigma}](x;x')$
at 1-loop order on de Sitter background. We begin by expressing the invariant
Lagrangian in terms of the graviton field and the conformally rescaled metric
$\widetilde{g}_{\mu\nu} \equiv a^{-2} \times g_{\mu\nu}$. Then the gauge
fixing functional is presented, along with explicit expressions for the ghost
and graviton propagators.

The invariant Lagrangian of general relativity in $D$ spacetime dimensions
with cosmological constant $\Lambda$ is,
\begin{equation}
\mathcal{L}_{\rm inv} = \tfrac{[R - (D-2) \Lambda] \sqrt{-g}}{16 \pi G}
\qquad \kappa^2 \equiv 16 \pi G \quad , \quad \Lambda \equiv (D-1) H^2 \; .
\label{Linv1}
\end{equation}
By judicious partial integrations one can express this as \cite{Tsamis:1992xa},
\begin{eqnarray}
\lefteqn{\mathcal{L}_{\rm inv} \longrightarrow a^{D-2} \sqrt{-\widetilde{g}}
\, \widetilde{g}^{\alpha\beta} \widetilde{g}^{\gamma\delta} \widetilde{g}^{
\rho\sigma} \Bigl\{ \tfrac12 h_{\alpha \rho , \gamma} h_{\delta \sigma , \beta}
- \tfrac12 h_{\alpha\beta , \rho} h_{\sigma\gamma , \delta} + \tfrac14 
h_{\alpha\beta , \rho} h_{\gamma\delta , \sigma} } \nonumber \\
& & \hspace{2.5cm} - \tfrac14 h_{\alpha \rho , \gamma} h_{\beta\sigma ,\delta} 
\Bigr\} + (\tfrac{D}2 - 1) a^{D-1} H \sqrt{-\widetilde{g}} \,
\widetilde{g}^{\gamma\delta} \widetilde{g}^{\rho\sigma} h_{\rho \sigma ,
\gamma} h_{\delta 0} \; . \qquad \label{Linv2}
\end{eqnarray}

A linearization instability precludes adding a de Sitter invariant gauge
fixing function \cite{Tsamis:1992xa,Miao:2009hb}. We instead fix the gauge
by breaking spatial special conformal transformations, while preserving
homogeneity, isotropy and dilatation invariance \cite{Tsamis:1992xa,
Woodard:2004ut},
\begin{equation}
\mathcal{L}_{\rm GF} = -\tfrac12 a^{D-2} \eta^{\mu\nu} F_{\mu} F_{\nu} \quad ,
\quad F_{\mu} = \eta^{\rho\sigma} \Bigl[ h_{\mu \rho , \sigma} - \tfrac12
h_{\rho\sigma , \mu} + {\scriptstyle (D-2)} a H h_{\mu \rho} 
\delta^0_{~\sigma} \Bigr] . \label{gauge}
\end{equation}
As usual, the ghost Lagrangian $\mathcal{L}_{gh} = -a^{D-2} \overline{c}^{\mu}
\delta F_{\mu}$ is formed by making a linearized coordinate transformation
by the ghost field $x^{\mu} \rightarrow x^{\mu} + \kappa c^{\mu}$,
\begin{eqnarray}
\lefteqn{ \mathcal{L}_{\rm gh} = -a^{D-2} \overline{c}^{\mu ,\nu} \Bigl[
2\widetilde{g}_{\rho (\mu} \partial_{\nu)} + \widetilde{g}_{\mu\nu ,\rho}
+ 2 a H \widetilde{g}_{\mu\nu} \delta^0_{~\rho} \Bigr] c^{\rho} } \nonumber \\
& & \hspace{3.5cm} + (a^{D-2} \overline{c}^{\mu} )_{,\mu} \eta^{\rho\sigma}
\Bigl[ \widetilde{g}_{\nu \rho} \partial_{\sigma} + \tfrac12 
\widetilde{g}_{\rho\sigma ,\nu} + a H \widetilde{g}_{\rho\sigma} 
\delta^0_{~\nu} \Bigr] c^{\nu} \; . \qquad \label{Lghost}
\end{eqnarray}

In this gauge the graviton and ghost propagators can be expressed using
three scalar propagators times tensor factors which are spacetime constants,
\begin{eqnarray}
i[\mbox{}_{\mu\nu} \Delta_{\rho\sigma}](x;x') &\!\!\! = \!\!\!& \sum_{I=A,B,C}
[\mbox{}_{\mu\nu} T^I_{\rho\sigma}] \times i \Delta_I(x;x') \; , \qquad
\label{gravprop} \\
i[\mbox{}_{\mu} \Delta_{\rho}](x;x') &\!\!\! = \!\!\!& 
\overline{\eta}_{\mu\rho} \times i\Delta_A(x;x') - \delta^0_{~\mu} 
\delta^0_{~\rho} \times  i\Delta_B(x;x') \; . \qquad \label{ghostprop}
\end{eqnarray}
An overline indicates the suppression of a tensor's temporal indices,
$\overline{\eta}_{\mu\nu} \equiv \eta_{\mu\nu} + \delta^0_{~\mu} 
\delta^0_{~\nu}$. The various tensor factors in (\ref{gravprop}) are,
\begin{eqnarray}
[\mbox{}_{\mu\nu} T^A_{\rho\sigma}] &\!\!\! \equiv \!\!\!& 2 
\overline{\eta}_{\mu (\rho} \overline{\eta}_{\nu) \sigma} - \tfrac{2}{D-3} 
\overline{\eta}_{\mu\nu} \overline{\eta}_{\rho\sigma} \quad , \quad
[\mbox{}_{\mu\nu} T^B_{\rho\sigma}] \equiv -4 \delta^0_{~(\mu} 
\overline{\eta}_{\nu) (\rho} \delta^0_{~\sigma)} \; , \qquad \label{TAB} \\
{[}\mbox{}_{\mu\nu} T^C_{\rho\sigma}{]} &\!\!\! = \!\!\!& \tfrac{2}{(D-2) (D-3)}
[{\scriptstyle (D-3)} \delta^0_{~\mu} \delta^0_{~\nu} + \overline{\eta}_{\mu\nu}
] [{\scriptstyle (D-3)} \delta^0_{~\rho} \delta^0_{~\sigma} + 
\overline{\eta}_{\rho\sigma}] \; . \label{TC}
\end{eqnarray}
Here and henceforth, parenthesized indices are symmetrized.

The propagators $i \Delta_{I}(x;x')$ correspond to minimally coupled scalars
with masses $M_A^2 = 0$, $M_B^2 = (D-2) H^2$ and $M_C^2 = 2(D-3) H^2$,
\begin{equation}
[ \mathcal{D} - M_I^2 a^D ] i\Delta_{I}(x;x') = i\delta^D(x \!-\! x') 
\qquad , \qquad \mathcal{D} \equiv \partial^{\alpha} a^{D-2} \partial_{\alpha}
\; . \label{propeqn}
\end{equation}
The three propagators depend mostly on the de Sitter invariant length
function $y(x;x') \equiv a a' H^2 \Delta x^2$,
\begin{eqnarray}
i \Delta_A &\!\!\! = \!\!\!& \tfrac{H^{D-2}}{(4\pi)^{\frac{D}2}} \Bigl\{
\tfrac{\Gamma(\frac{D}2)}{\frac{D}2 - 1} ( \tfrac{4}{y})^{\frac{D}2-1}
+ \tfrac{\Gamma(\frac{D}2 + 1)}{\frac{D}2 - 2} (\tfrac{4}{y})^{\frac{D}2-2} 
\Bigr\} + k \Bigl[\ln(a a') - \pi \cot(\tfrac{D \pi}{2}) \Bigr] \nonumber \\
& & \hspace{1cm} + \tfrac{H^{D-2}}{(4\pi)^{\frac{D}2}} \sum_{n=1}^{\infty} 
\Bigl[\tfrac{\Gamma(n + D - 1)}{n \Gamma(n + \frac{D}2)} (\tfrac{y}4 )^n - 
\tfrac{\Gamma(n + \frac{D}2 + 1)}{(n - \frac{D}2 + 2) \Gamma(n + 2)} 
(\tfrac{y}4 )^{n - \frac{D}2 +2} \Bigr] , \qquad \label{Adef} \\
i \Delta_B &\!\!\! = \!\!\!& \tfrac{H^{D-2}}{(4\pi)^{\frac{D}2}} \Bigl\{
\tfrac{\Gamma(\frac{D}2)}{\frac{D}2 - 1} ( \tfrac{4}{y})^{\frac{D}2-1}
+ \sum_{n=0}^{\infty} \Bigl[ \tfrac{\Gamma(n + \frac{D}2)}{\Gamma(n +2)} 
( \tfrac{y}4 )^{n - \frac{D}2 +2} - \tfrac{\Gamma(n + D - 2)}{
\Gamma(n + \frac{D}2)} (\tfrac{y}4 )^n \Bigr] \Bigr\} , \label{Bdef} \qquad \\
i \Delta_C &\!\!\! = \!\!\!& \tfrac{H^{D-2}}{(4\pi)^{\frac{D}2}} \Bigl\{ 
\tfrac{\Gamma(\frac{D}2)}{\frac{D}2 - 1} ( \tfrac{4}{y})^{\frac{D}2-1}
- \sum_{n=0}^{\infty} \Bigl[ \tfrac{(n - \frac{D}2 + 3) \Gamma(n + 
\frac{D}2 - 1)}{\Gamma(n + 2)} (\tfrac{y}4)^{n - \frac{D}2 +2} \nonumber \\
& & \hspace{6.5cm} - \tfrac{(n+1) \Gamma(n + D - 3)}{\Gamma(n + 
\frac{D}2)} (\tfrac{y}4 )^n \Bigr] \Bigr\} . \qquad \label{Cdef}
\end{eqnarray}
The constant $k$ is,
\begin{equation}
k \equiv \tfrac{H^{D-2}}{(4\pi)^{\frac{D}2}} \tfrac{\Gamma(D-1)}{
\Gamma(\frac{D}2)} \; . \label{kdef}
\end{equation}
Note that the infinite series corrections in expressions 
(\ref{Adef}-\ref{Cdef}) cancel in $D=4$ spacetime dimensions. This means
that the $D=4$ limits of the propagators involve only one or two terms,
\begin{eqnarray}
\lim_{D=4} i\Delta_A(x;x') &\!\!\! = \!\!\!& \tfrac1{4 \pi^2} 
\tfrac{1}{a a' \Delta x^2} - \tfrac{H^2}{8 \pi^2} \ln(\tfrac14 H^2
\Delta x^2) \; , \qquad \label{DA4} \\
\lim_{D=4} i\Delta_B(x;x') &\!\!\! = \!\!\!& \tfrac1{4 \pi^2} 
\tfrac{1}{a a' \Delta x^2} \; , \qquad \label{DB4} \\
\lim_{D=4} i\Delta_C(x;x') &\!\!\! = \!\!\!& \tfrac1{4 \pi^2} 
\tfrac{1}{a a' \Delta x^2} \; . \qquad \label{DC4}
\end{eqnarray}
Because the infinite series corrections in (\ref{Adef}-\ref{Cdef})
vanish more and more rapidly at coincidence with increasing $n$, only a 
few terms ever need to be retained.

The graviton propagator (\ref{gravprop}) can be expressed in a useful way,
\begin{eqnarray}
\lefteqn{ i[\mbox{}_{\mu\nu} \Delta_{\rho\sigma}](x;x') = 
[2 \eta_{\mu (\rho} \eta_{\sigma)\nu} - \tfrac{2}{D-2} \eta_{\mu\nu}
\eta_{\rho\sigma}] i \Delta_C(x;x') } \nonumber \\
& & \hspace{0cm} + [\mbox{}_{\mu\nu} T^A_{\rho\sigma}] [i\Delta_A(x;x')
\!-\! i\Delta_C(x;x')] + [\mbox{}_{\mu\nu} T^B_{\rho\sigma}] 
[i\Delta_B(x;x') \!-\! i\Delta_C(x;x')] \; . \qquad \label{simpprop}
\end{eqnarray}
One advantage of this form is that the Poincar\'e invariant tensor factor
on the first line is the same as for the flat space limit. Another
advantage is that only this term is quadratically divergent; the $A$-$C$
and $B$-$C$ propagator differences on the second line are only 
logarithmically divergent. 

Some special cases occur with significant frequency. Two of these 
consist of diagonal and cross-diagonal contractions,
\begin{eqnarray}
\tfrac18 \!\times\! i[\mbox{}^{\rho}_{~\rho} \Delta^{\sigma}_{~\sigma}] -
\tfrac14 \!\times\! i[\mbox{}^{\rho\sigma} \Delta_{\rho\sigma}] =
-\tfrac14 D (D\!-\!1) i\Delta_A - \tfrac12 (D\!-\!1) i\Delta_B - 
\tfrac12 i\Delta_C \; , \label{case1} \\
-\tfrac12 i [\mbox{}^{\rho}_{~\rho} \Delta^{\mu\nu}] + i [\mbox{}^{\mu\rho} 
\Delta^{\nu}_{~\rho}] = \overline{\eta}^{\mu\nu} 
[D i\Delta_A \!+\! i\Delta_B] - \delta^{\mu}_{~0} \delta^{\nu}_{~0} 
[(D\!-\!1) i\Delta_B \!+\! 2 i\Delta_C] \; . \label{case2}
\end{eqnarray}
It is also interesting when one of the indices is temporal,
\begin{eqnarray}
i[\mbox{}^{\rho\nu} \Delta_{\rho 0}] &\!\!\! = \!\!\!& (D\!-\!1) 
\delta^{\nu}_{~0} i\Delta_B + 2 (\tfrac{D-3}{D-2}) \delta^{\nu}_{~0} 
i\Delta_C \; , \qquad \label{case3} \\
i[\mbox{}^{\mu\nu} \Delta^{\rho}_{~0}] &\!\!\! = \!\!\!& 2 \delta^{(\mu}_{~~0}
\overline{\eta}^{\nu) \rho} i\Delta_B - [2 (\tfrac{D-3}{D-2}) \delta^{\mu}_{~0}
\delta^{\nu}_{~0} \!+\! \tfrac{2}{D-2} \overline{\eta}^{\mu\nu}] 
\delta^{\rho}_{~0} i\Delta_C \; . \qquad \label{case4}
\end{eqnarray}
There is also the case when two indices are temporal,
\begin{equation}
i[\mbox{}^{\mu}_{~0} \Delta^{\nu}_{~0}] = -\overline{\eta}^{\mu\nu} 
i\Delta_B + 2 (\tfrac{D-3}{D-2}) \delta^{\mu}_{~0} \delta^{\nu}_{~0} 
i\Delta_C \; . \label{case5}
\end{equation}

\section{Primitive Results}

This section is the heart of the paper. In it we compute the dimensionally
regulated, primitive result for the 1-loop correction to $-i [\mbox{}^{\mu\nu}
\Sigma^{\rho\sigma}](x;x')$. We begin with a discussion of the tensor and
functional basis in terms of which the answer is expressed. The actual
computation begins with the simplest diagram, which involves a single graviton 
4-point vertex, then the ghost loop is presented and, finally, the graviton 
loop between a pair of 3-point vertices. 

\subsection{The Tensor \& Functional Basis}

The graviton self-energy $-i [\mbox{}^{\mu\nu} \Sigma^{\rho\sigma}](x;x')$ 
is a second rank, bi-tensor density which is symmetric under the interchange
of $\mu$ with $\nu$, and also under the interchange of $\rho$ with $\sigma$. 
It can be expressed as a superposition of 21 tensor differential operators 
$[\mbox{}^{\mu\nu} D_i^{\rho\sigma}]$ acting on three functions $f_{j}(x -x')$ 
of the difference between the two coordinates,
\begin{equation}
-i [\mbox{}^{\mu\nu} \Sigma^{\rho\sigma}](x;x') = \sum_{i=1}^{21} \sum_{j=1}^3
C_{ij}(a,a',\partial) \times [\mbox{}^{\mu\nu} D^{\rho\sigma}_{i}] \times 
f_{j}(x \!-\! x') \; . \label{genform}
\end{equation}
The 4-index tensor differential operators $[\mbox{}^{\mu\nu} 
D^{\rho\sigma}_{i}]$ are listed in Table~\ref{Tbasis}.
\begin{table}[H]
\setlength{\tabcolsep}{8pt}
\def\arraystretch{1.5}
\centering
\begin{tabular}{|@{\hskip 1mm }c@{\hskip 1mm }||c||c|c||c|c|}
\hline
$i$ & $[\mbox{}^{\mu\nu} D^{\rho\sigma}_i]$ & $i$ & 
$[\mbox{}^{\mu\nu} D^{\rho\sigma}_i]$
& $i$ & $[\mbox{}^{\mu\nu} D^{\rho\sigma}_i]$ \\
\hline\hline
1 & $\eta^{\mu\nu} \eta^{\rho\sigma}$ & 8 & 
$\partial^{\mu} \partial^{\nu} \eta^{\rho\sigma}$ & 15 & 
$\delta^{(\mu}_{~~0} \partial^{\nu)} 
\delta^{\rho}_{~0} \delta^{\sigma}_{~0}$ \\
\hline
2 & $\eta^{\mu (\rho} \eta^{\sigma) \nu}$ & 9 & 
$\delta^{(\mu}_{~~0} \eta^{\nu) (\rho}
\delta^{\sigma)}_{~~0}$ & 16 & 
$\delta^{\mu}_{~0} \delta^{\nu}_{~0} \partial^{\rho}
\partial^{\sigma}$ \\
\hline
3 & $\eta^{\mu\nu} \delta^{\rho}_{~0} \delta^{\sigma}_{~0}$ & 10 & 
$\delta^{(\mu}_{~~0} \eta^{\nu) (\rho} \partial^{\sigma)}$ & 17 & 
$\partial^{\mu} \partial^{\nu} \delta^{\rho}_{~0}
\delta^{\sigma}_{~0}$ \\
\hline
4 & $\delta^{\mu}_{~0} \delta^{\nu}_{~0} \eta^{\rho\sigma}$ & 11 & 
$\partial^{(\mu} \eta^{\nu) (\rho} \delta^{\sigma)}_{~~0}$ & 18 & 
$\delta^{(\mu}_{~~0} \partial^{\nu)}
\delta^{(\rho}_{~~0} \partial^{\sigma)}$ \\
\hline
5 & $\eta^{\mu\nu} \delta^{(\rho}_{~~0} \partial^{\sigma)}$ & 12 & 
$\partial^{(\mu} \eta^{\nu)(\rho} \partial^{\sigma)}$ & 19 & 
$\delta^{(\mu}_{~~0} \partial^{\nu)}
\partial^{\rho} \partial^{\sigma}$ \\
\hline
6 & $\delta^{(\mu}_{~~0} \partial^{\nu)} \eta^{\rho\sigma}$ & 13 & 
$\delta^{\mu}_{~0} \delta^{\nu}_{~0} \delta^{\rho}_{~0} 
\delta^{\sigma}_{~0}$ & 20 & 
$\partial^{\mu} \partial^{\nu} 
\delta^{(\rho}_{~~0} \partial^{\sigma)}$ \\
\hline
7 & $\eta^{\mu\nu} \partial^{\rho} \partial^{\sigma}$ & 14 & 
$\delta^{\mu}_{~0} \delta^{\nu}_{~0} \delta^{(\rho}_{~~0} 
\partial^{\sigma)}$ & 21 & 
$\partial^{\mu} \partial^{\nu} \partial^{\rho} \partial^{\sigma}$ \\
\hline
\end{tabular}
\caption{\footnotesize The 21 basis tensors.}
\label{Tbasis}
\end{table} 
\noindent Note that we only need unprimed derivatives because one can 
reflect primed derivatives into unprimed one when acting on a function 
of the difference of $x^{\mu}$ and ${x'}^{\mu}$,
\begin{equation}
\partial'_{\mu} f_{j}(x \!-\! x') = -\partial_{\mu} f_{j}(x \!-\! x')
\; . \label{reflect}
\end{equation}

The first of the three functions is the $D$-dimensional delta function,
\begin{equation}
f_1(x \!-\! x') \equiv \delta^D(x \!-\! x') \; . \label{f1}
\end{equation}
All divergences reside in the coefficients $C_{i1}(a,a',\partial)$ acting
on this term, although these coefficients can be contain finite contributions 
as well. The two remaining functions are integrable in the effective field 
equation (\ref{linEin}), and depend on the conformal coordinate interval,
\begin{equation}
\Delta x^2 \equiv \Bigl\Vert \vec{x} \!-\! \vec{x}' \Bigr\Vert^2 -
\Bigl( \vert \eta \!-\! \eta'\vert \!-\! i \varepsilon\Bigr)^2 \; .
\label{Deltaxsq}
\end{equation}
These two functions are,
\begin{equation}
f_2(x \!-\! x') \equiv \ln^2(\tfrac14 H^2 \Delta x^2) \qquad , \qquad
f_3(x \!-\! x') \equiv \ln(\tfrac14 H^2 \Delta x^2) \; . \label{f2f3}
\end{equation}
Important differential relations exist between the functions,
\begin{eqnarray}
(\Delta \eta \partial^2 + 2 \partial_0) f_2 = -4 \partial_0 f_3 \qquad 
& , & \qquad \Delta \eta \partial^4 f_2 = -2 \partial_0 \partial^2 f_3
\; , \qquad \label{REL1} \\
(\Delta \eta \partial^2 + 2 \partial_0) f_3 = 0 \qquad & , & \qquad
\partial^4 f_3 = 16 \pi^2 i f_1 \; . \qquad \label{REL2}
\end{eqnarray} 
The coefficients $C_{i2}(a,a',\partial)$ and $C_{i3}(a,a',\partial)$
are finite and are evaluated for $D=4$. 

The factors within the various $C_{ij}(a,a'\partial)$ have any derivatives
standing to the right, so that they can be reflected (\ref{reflect}). As a 
1PI 2-point function, the graviton self-energy 
must also obey Reflection Invariance,
\begin{equation}
-i [\mbox{}^{\mu\nu} \Sigma^{\rho\sigma}](x;x') = -i [\mbox{}^{\rho\sigma}
\Sigma^{\mu\nu}](x';x) \; , \label{Reflect}
\end{equation}
This implies relations between seven pairs of the coefficient functions,
\begin{equation}
C_{(i+1)j}(a,a',\partial) = s_i \times C_{ij}(a',a,-\partial) \quad 
{\rm for} \quad i \in \{3,5,7,10,14,16,19\} \; . \label{Crelations}
\end{equation}
The factors $s_i$ are $\pm 1$ according to the rules,
\begin{equation}
s_i = \begin{cases}
+1 & i \in \{3,7,16\} \\ -1 & i \in \{5,10,14,19\} \end{cases} \; .
\label{pmfactors}
\end{equation}

\subsection{The 4-Point Contribution}

\begin{figure}[ht]
\centering
\vskip -1cm
\includegraphics[width=8cm]{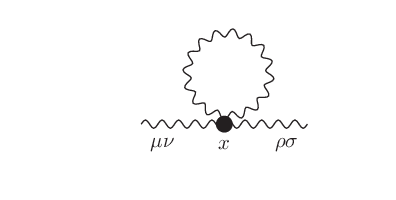}
\vskip -1cm
\caption{\footnotesize Contribution to the graviton self-energy from a single 
4-point vertex (\ref{Sigma4pt}). Wavy lines stand for gravitons.}
\label{graviton4pt}
\vskip -0.5cm
\end{figure}

The simplest 1-loop contribution to the self-energy comes from a
single 4-point vertex,
\begin{equation}
-i [\mbox{}^{\mu\nu} \Sigma^{\rho\sigma}_{\rm 4pt}](x;x') =
\Bigl\langle \Omega \Bigl\vert \Bigl[\tfrac{i \delta^2 S}{\delta h_{\mu\nu}(x)
\delta h_{\rho\sigma}(x')}\Bigr]_{h^2} \Bigr\vert \Omega \Bigr\rangle 
\label{Sigma4pt}
\end{equation}
The subscript ``$h^2$'' indicates that two graviton fields are retained 
after the variation. The diagramatic representation of expression (\ref{Sigma4pt})
is depicted in Figure~\ref{graviton4pt}. 

The 130 quadratic contributions to the second variation in expression 
(\ref{Sigma4pt}) are listed in Table 4 of \cite{Tsamis:1996qm}, however,
this formidable result simplifies considerably when the expectation 
value is taken. To understand this, consider the first and second 
variations of the first term of the invariant Lagrangian  
(\ref{Linv2}),
\begin{equation}
\mathcal{L}_1 \equiv \tfrac12 a^{D-2} \sqrt{-\widetilde{g}} \, 
\widetilde{g}^{\alpha\beta} \widetilde{g}^{\gamma\delta} 
\widetilde{g}^{\kappa\lambda} h_{\alpha \kappa , \gamma} 
h_{\delta \lambda , \beta} \; . \label{L1}
\end{equation}
The first variation results in all the fields being evaluated at 
$x^{\mu}$,
\begin{eqnarray}
\lefteqn{\tfrac{\delta S_1}{\delta h_{\mu\nu}(x)} = -\partial_{\beta} 
\Bigl\{a^{D-2} \sqrt{-\widetilde{g}} \, \widetilde{g}^{\alpha\beta} 
\widetilde{g}^{\gamma (\mu} \widetilde{g}^{\nu) \kappa} 
h_{\alpha \kappa , \gamma} \Bigr\} } \nonumber \\
& & \hspace{4.5cm} + \tfrac12 a^{D-2} \tfrac{\partial}{
\partial h_{\mu\nu}} \Bigl\{ \sqrt{-\widetilde{g}} \, 
\widetilde{g}^{\alpha\beta} \widetilde{g}^{\gamma\delta} 
\widetilde{g}^{\kappa\lambda} \Bigr\} h_{\alpha \kappa , \gamma}
h_{\delta \lambda , \beta} \; . \qquad \label{varS1}
\end{eqnarray}
Note the distinction between the variational derivative on the left
and the ordinary derivative on the second line of the right hand side.
Taking the second variation does not change the fact that all the 
fields are evaluated at the same point $x^{\mu}$,
\begin{eqnarray}
\lefteqn{\tfrac{\delta^2 S_1}{\delta h_{\mu\nu}(x) \delta h_{\rho\sigma}(x')} 
= -\partial_{\beta} \Bigl\{a^{D-2} \sqrt{-\widetilde{g}} \, 
\widetilde{g}^{\beta (\rho} \widetilde{g}^{\sigma) (\mu}  
\widetilde{g}^{\nu) \kappa} \partial_{\kappa} \delta^D(x \!-\! x')
\Bigr\} } \nonumber \\
& & \hspace{0cm} -\partial_{\beta} \Bigl\{ a^{D-2} \tfrac{\partial}{
\partial h_{\rho\sigma}} \Bigl[ \sqrt{-\widetilde{g}} \, 
\widetilde{g}^{\alpha\beta} \widetilde{g}^{\gamma (\mu} 
\widetilde{g}^{\nu ) \kappa} \Bigr] h_{\alpha \gamma , \kappa} 
\delta^D(x \!-\! x') \Bigr\} \nonumber \\
& & \hspace{1cm} + a^{D-2} \tfrac{\partial}{\partial h_{\mu\nu}} \Bigl\{ 
\sqrt{-\widetilde{g}} \, \widetilde{g}^{\alpha\beta} 
\widetilde{g}^{\gamma (\rho} \widetilde{g}^{\sigma) \kappa} \Bigr\} 
h_{\alpha \gamma , \kappa} \partial_{\beta} \delta^D(x \!-\! x') \nonumber \\
& & \hspace{2cm} + \tfrac12 a^{D-2} \tfrac{\partial^2}{\partial h_{\mu\nu}
\partial h_{\rho\sigma}} \Bigl\{ \sqrt{-\widetilde{g}} \, 
\widetilde{g}^{\alpha\beta} \widetilde{g}^{\gamma\delta} 
\widetilde{g}^{\kappa\lambda} \Bigr\} h_{\alpha\gamma , \kappa}
h_{\lambda \delta , \beta} \delta^D(x \!-\! x') \; . \qquad
\label{var2S1}
\end{eqnarray}
The next step is to expand each term in (\ref{var2S1}) to quadratic
order in the graviton field, and then take the expectation value,
which produces coincident, sometimes differentiated, propagators. 

The coincidence limits of the scalar propagators (\ref{Adef}-\ref{Cdef}) 
are,
\begin{eqnarray}
i\Delta_A(x;x) &\!\!\! = \!\!\!& k [2\ln(a) - \pi {\rm cot}(\tfrac{D \pi}{2})]
= -\tfrac{k \pi {\rm cot}(\frac{D \pi}{2})}{a^{D-4}} + O(D\!-\!4) \; , 
\label{coincA} \qquad \\
i\Delta_B(x;x) &\!\!\! = \!\!\!& -\tfrac{k}{D-2} = -\tfrac{H^2}{16 \pi^2} +
O(D\!-\!4) \; , \label{coincB} \qquad \\
i\Delta_C(x;x) &\!\!\! = \!\!\!& -\tfrac{k}{(D-3) (D-2)} = \tfrac{H^2}{16 \pi^2} 
+ O(D\!-\!4) \; . \label{coincC} \qquad 
\end{eqnarray}
The coincidence limits of singly differentiated propagators are,
\begin{eqnarray}
\lim_{x' \rightarrow x} \partial_{\mu} i\Delta_A(x;x') &\!\!\! = \!\!\!&
k H a \delta^0_{~\mu} = \tfrac{H^3}{8\pi^2} a \delta^0_{~\mu} + O(D\!-\!4) \; , 
\label{coincA'} \qquad \\
\lim_{x' \rightarrow x} \partial_{\mu} i\Delta_B(x;x') &\!\!\! = \!\!\!& 0 =
\lim_{x' \rightarrow x} \partial_{\mu} i\Delta_C(x;x') \; . \qquad
\label{coincBC'}
\end{eqnarray}
And the coincidence limits of mixed second derivatives are,
\begin{eqnarray}
\lim_{x' \rightarrow x} \partial_{\mu} \partial'_{\nu} i\Delta_A(x;x') 
&\!\!\! = \!\!\!& -(\tfrac{D-1}{D}) k H^2 g_{\mu\nu} = - \tfrac{3 H^4}{32\pi^2} 
g_{\mu\nu} + O(D\!-\!4) \; , \label{coincA''} \qquad \\
\lim_{x' \rightarrow x} \partial_{\mu} \partial'_{\nu} i\Delta_B(x;x') 
&\!\!\! = \!\!\!& \tfrac{k H^2}{D} g_{\mu\nu} = \tfrac{H^4}{32\pi^2} 
g_{\mu\nu} + O(D\!-\!4) \; , \label{coincB''} \qquad \\
\lim_{x' \rightarrow x} \partial_{\mu} \partial'_{\nu} i\Delta_C(x;x') 
&\!\!\! = \!\!\!& -\tfrac{2 k H^2}{(D-2)D} g_{\mu\nu} = -\tfrac{H^4}{32\pi^2} 
g_{\mu\nu} + O(D\!-\!4) \; . \label{coincC''}
\end{eqnarray}
Three facts emerge concerning these coincidence limits 
(\ref{coincA}-\ref{coincC''}):
\begin{enumerate*}
\item{Only the undifferentiated $A$-type propagator (\ref{coincA}) 
is divergent;}
\item{The only nonzero, singly differentiated propagator is $\partial_{\mu}
i\Delta_A$ in expression (\ref{coincA'}), and only the $\mu = 0$ component
is nonzero; and}
\item{Second derivatives are are proportional to the Minkowski metric,
\begin{equation}
\lim_{x' \rightarrow x} \partial_{\mu} \partial'_{\nu} i\Delta_I(x;x')
= \tfrac1{D} \eta_{\mu\nu} \times \lim_{x' \rightarrow x} \partial \!\cdot\!
\partial' i\Delta_I(x;x') \; . \label{coninc''} 
\end{equation}}
\end{enumerate*}

We compute the divergent contributions by hand. The various 4-point 
divergences are proportional to the constant $\mathcal{K}_4$,
\begin{equation}
\mathcal{K}_4 \equiv \tfrac{H^{D-4}}{(4\pi)^{\frac{D}2}} \tfrac{\Gamma(D-1)}{
\Gamma(\frac{D}2)} \, \pi {\rm cot}(\tfrac{D \pi}{2}) = \tfrac1{4\pi^2}
\tfrac{H^{D-4}}{D-4} + {\rm Finite} \; . \label{K4def}
\end{equation}
These divergences come from the undifferentiated $A$-type propagator
(\ref{coincA}),
\begin{equation}
i\Delta_A(x;x) = -\tfrac{\mathcal{K}_4 H^2}{a^{D-4}} + O(D\!-\! 4) \; .
\label{4ptdiv}
\end{equation}
Only the expectation value of the first term in (\ref{var2S1}) can produce 
an undifferentiated $A$-type propagator. There is such a contribution from 
each of the first four terms of the invariant Lagrangian (\ref{Linv2}), and
their sum takes the form $\partial_{\alpha} \{ a^{D-2} 
T^{\alpha\beta\mu\nu\rho\sigma} \partial_{\beta}i \delta^D(x -x') \}$ with,
\begin{eqnarray}
T^{\alpha\beta\mu\nu\rho\sigma} &\!\!\! = \!\!\!& \sqrt{-\widetilde{g}} \, 
\Bigl\{-\widetilde{g}^{\alpha (\rho} \widetilde{g}^{\sigma) (\mu} 
\widetilde{g}^{\nu) \beta} + \tfrac12 \widetilde{g}^{\mu\nu} 
\widetilde{g}^{\alpha (\rho} \widetilde{g}^{\sigma) \beta} + \tfrac12 
\widetilde{g}^{\rho\sigma} \widetilde{g}^{\alpha (\mu} \widetilde{g}^{\nu) \beta}
\nonumber \\
& & \hspace{5.3cm} - \tfrac12 \widetilde{g}^{\alpha\beta} \widetilde{g}^{\rho\sigma}
\widetilde{g}^{\mu\nu} + \tfrac12 \widetilde{g}^{\alpha\beta} 
\widetilde{g}^{\rho (\mu} \widetilde{g}^{\nu) \sigma} \Bigr\} . \qquad
\label{1stfour}
\end{eqnarray}
The variation of the final term in (\ref{Linv2}) also contributes,
\begin{eqnarray}
\lefteqn{-(\tfrac{D}2 \!-\! 1) \partial_{\alpha} \Bigl\{ a^{D-1} H 
\sqrt{-\widetilde{g}} \, \widetilde{g}^{\mu\nu} \widetilde{g}^{\alpha (\rho} 
\delta^{\sigma)}_{0} i \delta^D(x \!-\! x') \Bigr\} } \nonumber \\
& & \hspace{4cm} + (\tfrac{D}2 \!-\! 1) a^{D-1} H \sqrt{-\widetilde{g}} \, 
\widetilde{g}^{\rho\sigma} \widetilde{g}^{\alpha (\mu} \delta^{\nu)}_{0}
\partial_{\alpha} i \delta^D(x \!-\! x') \; . \qquad \label{lasttwo}
\end{eqnarray}

\begin{table}[H]
\setlength{\tabcolsep}{8pt}
\def\arraystretch{1.5}
\centering
\begin{tabular}{|@{\hskip 1mm }c@{\hskip 1mm }||c|}
\hline
$i$ & Coefficient\ of\ $\kappa^2 [\mbox{}^{\mu\nu} D_i^{\rho\sigma}] \,
i\delta^D(x \!-\! x')$ \\
\hline\hline
$1$ & $-\mathcal{K}_4 a a' H^2 (\partial_0^2 \!-\! 2 a a' H^2) - 
\tfrac{a a' H^2}{32 \pi^2} (\partial^2 \!+\! 2 \partial_0^2 \!-\! a a' H^2)$ \\
\hline
$2$ & $\mathcal{K}_4 a a' H^2 (\partial_0^2 \!-\! 2 a a' H^2) +
\tfrac{a a' H^2}{32 \pi^2} (\partial^2 \!+\! 2 \partial_0^2 \!+\! 22 a a' H^2)$ \\
\hline
$3$ & $-\mathcal{K}_4 a^2 H^2 (\partial^2 \!-\! a H \partial_0 \!+\! 3 a^2 H^2)
- \tfrac{a^2 H^2}{16 \pi^2} (\partial^2 \!-\! 4 a H \partial_0 \!+\! 4 a^2 H^2)$ \\
\hline
$5$ & $2 \mathcal{K}_4 a^2 H^2 (\partial_0 \!+\! 2 a H) + \tfrac{a^2 H^2}{8 \pi^2} 
(\partial_0 \!+\! 2 a H)$ \\
\hline
$7$ & $\tfrac{a^2 H^2}{32 \pi^2}$ \\
\hline
$9$ & $2 \mathcal{K}_4 a a' H^2 (\partial^2 \!+\! 2 a a' H^2) + \tfrac{a a' H^2}{16 \pi^2}
(2 \partial^2 \!-\! 29 a a' H^2)$ \\
\hline
$11$ & $-2 a^2 \mathcal{K}_4 a^2 H^2(\partial_0 \!+\! 2 a H) - \tfrac{a^2 H^2}{16 \pi^2}
(\partial_0 \!+\! 5 a' H)$ \\
\hline
$12$ & $-\tfrac{a a' H^2}{16 \pi^2}$ \\
\hline
$13$ & $-14 \mathcal{K}_4 (a a')^2 H^4 - \tfrac{27 (a a')^2 H^4}{8 \pi^2}$ \\
\hline
$14$ & $+ \tfrac{5 a^3 H^3}{8 \pi^2}$ \\
\hline
$16$ & $\mathcal{K}_4 a^2 H^2 + \tfrac{a^2 H^2}{16 \pi^2}$ \\
\hline
$18$ & $-2 \mathcal{K}_4 a a' H^2 - \tfrac{a a' H^2}{8 \pi^2}$ \\
\hline
$19$ & $0$ \\
\hline
$21$ & $0$ \\
\hline
\end{tabular}
\caption{\footnotesize Primitive 4-point contributions to 
$-i [\mbox{}^{\mu\nu} \Sigma^{\rho\sigma}](x;x')$.}
\label{Primitive4pt}
\end{table}

The basic strategy for extracting the divergent part of the 4-point
contribution is to first expand expressions (\ref{1stfour}) and (\ref{lasttwo})
to second order in the graviton field, and then take the expectation values.
The expansion involves the conformally transformed volume element and the
inverse,
\begin{eqnarray}
\sqrt{-\widetilde{g}} &\!\!\! = \!\!\!& 1 + \tfrac{\kappa}{2} h + \kappa^2 
(\tfrac18 h^2 \!-\! \tfrac14 h^{\gamma\delta} h_{\gamma\delta}) + \dots 
\; , \label{sqrtg} \\
\widetilde{g}^{\gamma\delta} &\!\!\! = \!\!\!& \eta^{\gamma\delta} - \kappa
h^{\gamma\delta} + \kappa^2 h^{\gamma\epsilon} h^{\delta}_{~\epsilon} +
\dots \label{inverse}
\end{eqnarray}
The expectation value of the $\kappa^2$ part of (\ref{sqrtg}) is,
\begin{equation}
\langle \tfrac18 h^2 \!-\! \tfrac14 h^{\gamma\delta} h_{\gamma\delta} \rangle
= -\tfrac{D (D-1)}{4} i\Delta_A + {\rm Finite} \; . \label{2trace}
\end{equation}
The total contribution from terms of this form is,
\begin{eqnarray}
\lefteqn{ \tfrac{\kappa^2 H^2 D (D-1) \mathcal{K}_4}{8} \Bigl\{
\eta^{\mu\nu} \partial^{(\rho} a^2 \partial^{\sigma)} \!-\! 2 \eta^{\mu)(\rho}
\partial^{\sigma)} a^2 \partial^{(\nu} \!+\! \eta^{\rho\sigma} \partial^{(\mu}
a^2 \partial^{\nu)} } \nonumber \\
& & \hspace{1.8cm} - (D\!-\!2) \eta^{\mu\nu} \delta^{(\rho}_{0} \partial^{\sigma)}
a^3 H \!+\! (D\!-\!2) a^3 H \eta^{\rho\sigma} \delta^{(\mu}_0 \partial^{\nu)}
\Bigr\} i \delta^D(x \!-\! x') \; . \qquad \label{2Trcont}
\end{eqnarray}
The order $\kappa^2$ part of (\ref{inverse}) is always associated with the
product of an order $\kappa$ term from (\ref{sqrtg}) and one from 
(\ref{inverse}),
\begin{equation}
\langle h^{\gamma\epsilon} h^{\delta}_{~\epsilon} \!-\! \tfrac12 h 
h^{\gamma\delta} \rangle = D i\Delta_A \overline{\eta}^{\gamma\delta} + 
{\rm Finite} \; . \label{1trace}
\end{equation}
The total contribution from terms of this form is,
\begin{eqnarray}
\lefteqn{ \tfrac{\kappa^2 H^2 D \mathcal{K}_4}{2} \Bigl\{2 \eta^{\mu)(\rho}
\overline{\partial}^{\sigma)} a^2 \partial^{(\nu} \!+\! 2 
\overline{\eta}^{\mu)(\rho} \partial^{\sigma)} a^2 \partial^{(\nu} \!+\!
2 \eta^{\mu) (\rho} \partial^{\sigma)} a^2 \overline{\partial}^{(\nu} 
-\overline{\eta}^{\mu\nu} \partial^{(\rho} a^2 \partial^{\sigma)} }
\nonumber \\
& & \hspace{0.5cm} - \eta^{\mu\nu} [ \overline{\partial}^{(\rho} a^2 
\partial^{\sigma)} \!+\! \partial^{(\rho} a^2 \overline{\partial}^{\sigma)}]
\!-\! \overline{\eta}^{\rho\sigma} \partial^{(\mu} a^2 \partial^{\nu)} 
\!-\! \eta^{\rho\sigma} [ \overline{\partial}^{(\mu} a^2 \partial^{\nu)} 
\!+\! \partial^{(\mu} a^2 \overline{\partial}^{\nu)}] \nonumber \\
& & \hspace{1.8cm} + (D\!-\!2) \overline{\eta}^{\mu\nu} \delta^{(\rho}_{0}
\partial^{\sigma)} a^3 H \!-\! (D\!-\!2) a^3 H \overline{\eta}^{\rho\sigma}
\delta^{(\mu}_{0} \partial^{\nu)} \Bigr\} i\delta^D(x \!-\! x') \; . \qquad
\label{1Trcont}
\end{eqnarray}
That leaves products of two order $\kappa$ terms from (\ref{inverse}). The
total contribution from terms of this form is,
\begin{eqnarray}
\lefteqn{\tfrac{\kappa^2 H^2 \mathcal{K}_4}{D-3} \Bigl\{- 
\overline{\eta}^{\mu\nu} [\partial^{(\rho} a^2 \overline{\partial}^{\sigma)} 
\!-\! \overline{\partial}^{(\rho} a^2 \partial^{\sigma)} \!+\! 
\overline{\eta}^{\rho\sigma} [\partial^{(\mu} a^2 \overline{\partial}^{\nu)} 
\!-\! \overline{\partial}^{(\mu} a^2 \partial^{\nu)} ] + 4 a^3 H 
\delta^{(\rho}_{0} \overline{\eta}^{\sigma) (\mu} \overline{\partial}^{\nu)} } 
\nonumber \\
& & \hspace{1cm} + 2 a^3 H \overline{\eta}^{\mu\nu} \delta^{(\rho}_{0} 
\overline{\partial}^{\sigma)} \!\!+\! 4 a^3 H \delta^{(\mu}_{0} 
\overline{\eta}^{\nu)(\rho} \overline{\partial}^{\sigma)} \!\!+\! 2 a^3 H 
\overline{\eta}^{\rho\sigma} \delta^{(\mu}_{0} \partial^{\nu)} \Bigr\} 
i \delta^D(x \!-\! x') \qquad \nonumber \\
& & \hspace{-0.7cm} + \tfrac{\kappa^2 H^2 (D-1) \mathcal{K}_4}{D-3} \Bigl\{
\overline{\eta}^{\mu\nu} \partial^{(\rho} a^2 \overline{\partial}^{\sigma)}
\!+\! \tfrac12 \eta^{\mu\nu} \overline{\partial}^{(\rho} a^2 
\overline{\partial}^{\sigma)} \!+\! \overline{\eta}^{\rho\sigma} 
\overline{\partial}^{(\mu} a^2 \partial^{\nu)} \!+\! \tfrac12 \eta^{\rho\sigma}
\overline{\partial}^{(\mu} a^2 \partial^{\nu)} \qquad \nonumber \\
& & \hspace{2cm} -\overline{\eta}^{\mu) (\rho} \partial^{\sigma)} a^2 
\overline{\partial}^{(\nu} \!-\! \overline{\eta}^{\mu) (\rho} 
\overline{\partial}^{\sigma)} a^2 \partial^{(\nu} \!-\! \eta^{\mu) (\rho}
\overline{\partial}^{\sigma)} a^2 \overline{\partial}^{(\nu} \qquad 
\nonumber \\
& & \hspace{3.5cm} - 2 a^3 H \delta^{(\rho}_0 \overline{\eta}^{\sigma) (\mu}
\overline{\partial}^{\nu)} \!-\! 2 a^3 H \delta^{(\mu}_{0} 
\overline{\eta}^{\nu)(\rho} \overline{\partial}^{\sigma)} \Bigr\} 
i\delta^D(x \!-\! x') . \qquad \label{0Trcont}
\end{eqnarray}

Table~\ref{Primitive4pt} gives all the 4-point contributions to $-i
[\mbox{}^{\mu\nu} \Sigma^{\rho\sigma}](x;x')$. This includes the divergent 
contributions (\ref{2Trcont}), (\ref{1Trcont}) and (\ref{0Trcont}) from 
undifferentiated $A$-type propagators, as well as finite contributions 
from the $B$-type and $C$-type propagators, and from differentiated 
propagators. Note that we have represented functions of $D$ times 
$\mathcal{K}_4$ as pure numbers times $\mathcal{K}_4$ plus finite 
contributions,
\begin{equation}
f(D) \!\times\! \mathcal{K}_4 \longrightarrow f(4) \!\times\! \mathcal{K}_4 + 
\tfrac{f'(4)}{4 \pi^2} \; . \label{extractDiv}
\end{equation}
Only those coefficients not determined by reflection are given; the results 
for $i=4$, 6, 8, 10, 15, 17 and 20 follow from  relations 
(\ref{Crelations}-\ref{pmfactors}).

\subsection{The Ghost Contribution}

\vskip .5cm
\begin{figure}[ht]
\centering
\includegraphics[width=6cm]{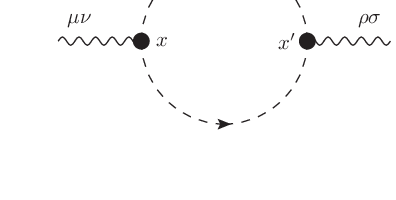}
\vskip -1cm
\caption{\footnotesize Contribution to the graviton self-energy from the product
of two $\overline{c} c h$ vertices (\ref{Sigmagh}). Wavy lines stand for gravitons
and dashed lines stand for ghosts.}
\label{ghost3pt}
\end{figure}

A more complicated 1-loop contribution to the self-energy comes from the
ghost loop depicted in Figure~\ref{ghost3pt}. The analytic expression is,
\begin{equation}
-i [\mbox{}^{\mu\nu} \Sigma^{\rho\sigma}_{\rm gh}](x;x') =
\Bigl\langle \Omega \Bigl\vert \tfrac{i \delta S_{\rm gh}}{\delta h_{\mu\nu}(x)}
\times \tfrac{i \delta S_{\rm gh}}{\delta h_{\rho\sigma}(x')}
\Bigr\vert \Omega \Bigr\rangle \label{Sigmagh}
\end{equation}
Varying the ghost action (\ref{Lghost}) gives,
\begin{eqnarray}
\lefteqn{ \tfrac{i \delta S_{\rm gh}}{\delta h_{\mu\nu}} = -i\kappa a^{D-2}
[ \overline{c}^{(\mu}_{~~,\alpha} + \overline{c}_{\alpha}^{~,(\mu}] 
c^{\nu) ,\alpha} + i\kappa \partial_{\alpha} [a^{D-2} \overline{c}^{(\mu , \nu)} 
c^{\alpha}] - 2 i \kappa H a^{D-1} \overline{c}^{(\mu ,\nu)} c^0 } \nonumber \\
& & \hspace{0cm} + i \kappa (a^{D-2} \overline{c}^{\alpha})_{,\alpha} 
c^{(\mu , \nu)} \!+\! i \kappa \eta^{\mu\nu} \Bigl\{-\tfrac{i}{2} 
\partial_{\alpha} [ (a^{D-2} \overline{c}^{\beta})_{,\beta} c^{\alpha}] \!+\! 
a H (a^{D-2} \overline{c}^{\alpha} )_{,\alpha} c^0 \Bigr\} . 
\label{ghostvertex} \qquad
\end{eqnarray}
The various interactions are listed in Table 2 of \cite{Tsamis:1996qk}
for $D=4$, and for arbitrary $D$ in Table 2 of \cite{Tsamis:2005je}.

It is useful to reduce the contribution from the product of two vertices. 
Taking account of the Fermi minus sign, and suppressing the cumbersome
symmetrizations, the product of the 1st vertex at $x^{\mu}$ and the 4th 
vertex at ${x'}^{\mu}$ gives,
\begin{eqnarray}
\lefteqn{ -i\kappa a^{D-2} \!\cdot\! \partial^{\alpha} \overline{c}^{\mu}(x) 
\!\cdot\! \partial_{\alpha} c^{\nu}(x) \times -2i \kappa H {a'}^{D-1} \!\cdot\! 
{\partial'}^{\rho} \overline{c}^{\sigma}(x') \!\cdot\! c^0(x') } \nonumber \\
& & \hspace{2cm} \longrightarrow +2 \kappa^2 a^{D-2} {a'}^{D-1} 
\partial^{\alpha} i [\mbox{}^{\mu} \Delta^0](x;x') \times \partial_{\alpha}
{\partial'}^{\rho} i[\mbox{}^{\sigma} \Delta^{\nu}](x';x) \; . \qquad
\label{ghost1}
\end{eqnarray}
Substituting for the ghost-anti-ghost propagator (\ref{ghostprop}), and
restoring the symmetrizations over $\mu\nu$ and $\rho\sigma$, gives,
\begin{eqnarray}
\lefteqn{-2 \kappa^2 H a^{D-2} {a'}^{D-1} \partial^{\alpha} 
i\Delta_{B}(x;x') } \nonumber \\
& & \hspace{1.7cm} \times \Bigl\{\delta^{(\mu}_{~~0} \overline{\eta}^{\nu) (\rho} 
{\partial'}^{\sigma)} \partial_{\alpha} i\Delta_{A}(x';x) - \delta^{\mu}_{~0} 
\delta^{\nu}_{~0} \delta^{(\rho}_{~~0} {\partial'}^{\sigma)} \partial_{\alpha} 
i\Delta_{B}(x';x) \Bigr\} , \qquad \label{ghost2} \\
& & = -2 \kappa^2 H a^{D-2} {a'}^{D-1} \partial^{\alpha} i\Delta_{B}
\nonumber \\
& & \hspace{1.7cm} \times \Bigl\{\delta^{(\mu}_{~~0} \eta^{\nu) (\rho} 
{\partial'}^{\sigma)} \partial_{\alpha} i\Delta_{A} + \delta^{\mu}_{~0} 
\delta^{\nu}_{~0} \delta^{(\rho}_{~~0} {\partial'}^{\sigma)} \partial_{\alpha} 
[i\Delta_A - i\Delta_{B}] \Bigr\} . \qquad \label{ghost3}
\end{eqnarray}

\begin{table}[H]
\setlength{\tabcolsep}{8pt}
\def\arraystretch{1.5}
\centering
\begin{tabular}{|@{\hskip 1mm }c@{\hskip 1mm }||c|}
\hline
$i$ & Coefficient\ of\ $\kappa^2 [\mbox{}^{\mu\nu} D_i^{\rho\sigma}] \,
i\delta^D(x \!-\! x')$ \\
\hline\hline
$1$ & $\mathcal{K}_3 [\tfrac{7}{12} \partial^4 \!+\! 3 a a' H^2 \partial^2
\!+\! a a' H^2 \partial_0^2 \!-\! 2 {a'}^4 H^4] - \tfrac{a a' H^2}{2^7 \pi^2} 
[3\partial_0^2 \!+\! {a'}^2 H^2]$ \\
\hline
$2$ & $\mathcal{K}_3 [\tfrac13 \partial^4 \!-\! \tfrac43 a a' H^2 \partial^2
\!-\! \tfrac{16}{3} a a' H^2 \partial_0^2 \!+\! 16 {a'}^4 H^4] \!+\! 
\tfrac{a a' H^2}{2^6 \pi^2} [2 \partial_0^2 \!+\! 13 a a' H^2]$ \\
\hline
$3$ & $\mathcal{K}_3 a^2 H^2 [\tfrac43 \partial^2 \!-\! 8 a H \partial_0 
\!-\! 16 a^2 H^2] - \tfrac{a^2 H^2}{2^7 \pi^2} [\partial^2 \!-\! 12 a H 
\partial_0 \!+\! 10 a^2 H^2]$ \\
\hline
$5$ & $\mathcal{K}_3 a H [\tfrac23 \partial^2 \!-\! \tfrac23 a H \partial_0
\!+\! 4 a^2 H^2] + \tfrac{a^2}{2^6 \pi^2} [\partial_0 \!+\! 3 a H]$ \\
\hline
$7$ & $-\mathcal{K}_3 [\tfrac12 \partial^2 \!+\! \tfrac13 a H \partial_0
\!+\! \tfrac73 a^2 H^2]$ \\
\hline
$9$ & $\mathcal{K}_3 a a' H^2 [\tfrac{16}{3} \partial^2 \!+\! 20
a a' H^2] - \tfrac{a a'}{16 \pi^2} (\partial^2 \!-\! 3 a a' H^2)$ \\
\hline
$11$ & $\mathcal{K}_3 a H [\tfrac43 \partial^2 \!+\! 4 a^2 H^2] + 
\tfrac{a^3 H^3}{2^4 \pi^2}$ \\
\hline
$12$ & $-\mathcal{K}_3 [\tfrac13 \partial^2 \!-\! 4 a a' H^2]$ \\
\hline
$13$ & $\tfrac{28}{3} \mathcal{K}_3 (a a')^2 H^4 + \tfrac{13 (a a')^2 
H^4}{2^5 \pi^2}$ \\
\hline
$14$ & $\tfrac{28}{3} \mathcal{K}_3 a^3 H^3 - \tfrac{5 a^3 H^3}{2^5 \pi^2}$ \\
\hline
$16$ & $-\tfrac{14}{3} \mathcal{K}_3 a^2 H^2 + \tfrac{a^2 H^2}{2^6 \pi^2}$ \\
\hline
$18$ & $4 \mathcal{K}_3 a a' H^2 - \tfrac{a a' H^2}{2^5 \pi^2}$ \\
\hline
$19$ & $\tfrac{2}{3} \mathcal{K}_3 a H$ \\
\hline
$21$ & $\tfrac{2}{3} \mathcal{K}_3$ \\
\hline
\end{tabular}
\caption{\footnotesize Primitive local ghost loop contributions to $-i [\mbox{}^{\mu\nu}
\Sigma^{\rho\sigma}](x;x')$.}
\label{PrimitiveGhostDiv}
\end{table} 

Expression (\ref{ghost3}) implies that we need expansions for two products
of differentiated propagators,
\begin{equation}
(a a')^{D-2} \partial^{\alpha} {\partial'}^{\sigma} i\Delta_A \!\times\!
\partial_{\alpha} i\Delta_B \quad {\rm and} \quad (a a')^{D-2} 
\partial^{\alpha} {\partial'}^{\sigma} [i\Delta_A \!-\! i\Delta_B] \!\times\!
\partial_{\alpha} i\Delta_B \; . \label{neededproducts}
\end{equation}
We can read these expansions off from Tables~\ref{TableEDiv} and \ref{TableEFin}
of the Appendix (which is section 7). Note first that we need only a handful of
terms from the expansion (\ref{propexp}) of each propagator,
\begin{eqnarray}
i\Delta_A &\!\!\! = \!\!\!& i\Delta_{\rm cf} + A_0 + {\rm not\ needed} \; , \\
i\Delta_B &\!\!\! = \!\!\!& i\Delta_{\rm cf} + B_0 + {\rm not\ needed} \; , \\
i\Delta_A - i\Delta_B &\!\!\! = \!\!\!& A_0 + {\rm not\ needed} \; ,
\end{eqnarray}
where $i\Delta_{\rm cf}$, $A_0$ and $B_0$ are defined by expressions 
(\ref{conformal}), (\ref{A0}) and (\ref{B0}), respectively. From 
Tables~\ref{TableEDiv} and \ref{TableEFin} we read,
\begin{eqnarray}
\lefteqn{ (a a')^{D-2} \partial^{\alpha} {\partial'}^{\sigma} [i\Delta_A
\!-\! i\Delta_B] \!\times\! \partial_{\alpha} i\Delta_B } \nonumber \\
& & \hspace{0cm} = (a a')^{D-2} \partial^{\alpha} {\partial'}^{\sigma} 
[A_0 \!+\! \dots] \!\times\! \partial_{\alpha} [i\Delta_{\rm cf} \!+\! \dots] 
\; , \qquad \\
& & \hspace{0cm} \longrightarrow a a' H^2 \mathcal{K}_3 \Bigl\{-
\tfrac{D (D-2) (D^2 - 2 D - 2)}{8 (D-1)} \partial^{\sigma} +
\tfrac{D (D-2)^2}{8} a H \delta^{\sigma}_{0} \Bigr\} i \delta^D(x \!-\! x')
\qquad \nonumber \\
& & \hspace{2cm} + \tfrac{a a' H^2 [\partial^{\sigma} - a H \delta^{\sigma}_0]
\partial^4}{2^{10} \cdot \pi^4} \ln^2(\tfrac14 H^2 \Delta x^2) \qquad 
\nonumber \\
& & \hspace{4cm} - \tfrac{a a' H^2 [(5 \partial^{\sigma} - 12 a' H 
\delta^{\sigma}_0) \partial^2 + 6 a H \partial_0 \partial^{\sigma}] 
\partial^2}{2^9 \cdot 3 \cdot \pi^4} \ln(\tfrac14 H^2 \Delta x^2) \; , \qquad 
\label{EGghost}
\end{eqnarray} 
where the constant $\mathcal{K}_3$ is,
\begin{equation}
\mathcal{K}_3 \equiv \tfrac{H^{D-4}}{(4\pi)^{\frac{D}2}} 
\tfrac{\Gamma(\frac{D}2 - 1)}{2 (D-3) (D-4)} = \tfrac1{32 \pi^2}
\tfrac{H^{D-4}}{D-4} + {\rm Finite} \; . \label{K3def}
\end{equation}
Putting everything together for the second part of expression (\ref{ghost3}), 
and consulting Table~\ref{Tbasis} of basis tensors gives,
\begin{eqnarray}
\lefteqn{ -2 \kappa^2 a' H \!\times\! \delta^{\mu}_{~0} \delta^{\nu}_{~0} 
\delta^{(\rho}_{~~0} (a a')^{D-2} {\partial'}^{\sigma)} \partial^{\alpha} 
[i\Delta_A - i\Delta_{B}] \!\times\! \partial_{\alpha} i\Delta_{B} } 
\nonumber \\
& & \hspace{-0.5cm} \longrightarrow [-4 \mathcal{K}_3 \!+\! 
\tfrac{5}{2^5 \cdot \pi^2}] (a a')^2 H^4 \!\times\! \kappa^2 
[\mbox{}^{\mu\nu} D_{13}^{\rho\sigma}] i \delta^D(x \!-\! x') + 
[4 \mathcal{K}_3 \!+\! \tfrac{17}{2^5 \cdot 3 \cdot \pi^2} ] a {a'}^2 H^3
\nonumber \\
& & \hspace{1cm} \times \kappa^2 [\mbox{}^{\mu\nu} D_{14}^{\rho\sigma}] 
i \delta^D(x \!-\! x') + \tfrac{(a a')^2 H^4 \partial^4}{2^9 \cdot \pi^4}
\!\times\! \kappa^2 [\mbox{}^{\mu\nu} D_{13}^{\rho\sigma}] \ln^2(\tfrac14 
H^2 \Delta x^2) \nonumber \\
& & \hspace{0cm} - \tfrac{a {a'}^2 H^3 \partial^4}{2^9 \cdot \pi^4} \!\times\! 
\kappa^2 [\mbox{}^{\mu\nu} D_{14}^{\rho\sigma}] \ln^2(\tfrac14 H^2 \Delta x^2) 
- \tfrac{a {a'}^3 H^3 \partial^2}{2^6 \cdot \pi^4} \!\times\! \kappa^2 
[\mbox{}^{\mu\nu} D_{13}^{\rho\sigma}] \ln(\tfrac14 H^2 \Delta x^2) 
\nonumber \\
& & \hspace{4cm} + \tfrac{a {a'}^2 H^3 [5 \partial^4 + 6 a H \partial_0 
\partial^2]}{2^8 \cdot 3 \cdot \pi^4} \!\times\! \kappa^2 [\mbox{}^{\mu\nu} 
D_{14}^{\rho\sigma}] \ln(\tfrac14 H^2 \Delta x^2) \; . \qquad 
\label{EGghost2}
\end{eqnarray}
Table~\ref{PrimitiveGhostDiv} gives the local contributions from the ghost 
loop and Table~\ref{PrimitiveGhostFin} gives the finite, nonlocal 
contributions.

\subsection{The 3-Point Contribution}

\begin{figure}[ht]
\centering
\includegraphics[width=6cm]{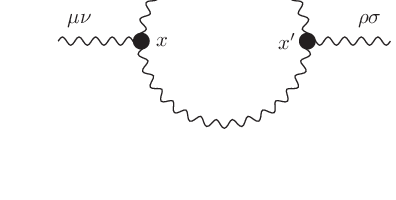}
\vskip -1cm
\caption{\footnotesize Contribution to the graviton self-energy from the product
of two 3-point vertices (\ref{Sigma3pt}). Wavy lines stand for gravitons.}
\label{graviton3pt}
\end{figure}

The most complicated 1-loop contribution to the self-energy comes from the
product of two 3-point vertices,
\begin{equation}
-i [\mbox{}^{\mu\nu} \Sigma^{\rho\sigma}_{\rm 3pt}](x;x') =
\Bigl\langle \Omega \Bigl\vert \Bigl[\tfrac{i \delta S}{\delta h_{\mu\nu}(x)}
\Bigr]_{h^2} \times \Bigl[\tfrac{i \delta S}{\delta h_{\rho\sigma}(x')}\Bigr]_{h^2} 
\Bigr\vert \Omega \Bigr\rangle \label{Sigma3pt}
\end{equation}
As before, the subscript ``$h^2$'' indicates that two graviton fields are retained 
after the variation. The diagramatic representation of expression (\ref{Sigma3pt})
is depicted in Figure~\ref{graviton3pt}. 

The distinguished 3-point vertex occurring in expression (\ref{Sigma3pt}) is,
\begin{eqnarray}
\lefteqn{ \tfrac{\delta S}{\delta h_{\mu\nu}} = \partial_{\alpha} \Bigl\{
a^{D-2} \sqrt{-\widetilde{g}} \Bigl[-\widetilde{g}^{\alpha\gamma}
\widetilde{g}^{\delta (\mu} \widetilde{g}^{\nu) \beta} \!+\! \tfrac12
\widetilde{g}^{\alpha \gamma} \widetilde{g}^{\beta\delta} \widetilde{g}^{\mu\nu}
\!+\! \tfrac12 \widetilde{g}^{\alpha (\mu} \widetilde{g}^{\nu)\beta} 
\widetilde{g}^{\gamma\delta}
} \nonumber \\
& & \hspace{0cm} -\tfrac12 \widetilde{g}^{\alpha\beta} \widetilde{g}^{\gamma\delta}
\widetilde{g}^{\mu\nu} \!+\! \tfrac12 \widetilde{g}^{\alpha\beta} \widetilde{g}^{
\gamma (\mu} \widetilde{g}^{\nu) \delta} \Bigr] h_{\gamma\delta ,\beta} \Bigr\} + 
a^{D-2} \tfrac{\partial}{\partial h_{\mu\nu}} \Bigl\{ \sqrt{-\widetilde{g}} \, 
\widetilde{g}^{\alpha\beta} \widetilde{g}^{\gamma\delta} \widetilde{g}^{\kappa\lambda}
\Bigr\} \qquad \nonumber \\
& & \hspace{1cm} \times \Bigl[ \tfrac12 h_{\alpha\gamma , \kappa} h_{\lambda 
\delta , \beta} \!-\! \tfrac12 h_{\alpha\beta , \kappa} h_{\lambda \gamma , \delta}
\!+\! \tfrac14 h_{\alpha\beta \kappa} h_{\gamma\delta , \lambda} \!-\! \tfrac14
h_{\alpha\gamma , \kappa} h_{\beta\delta , \lambda} \Bigr] \qquad \nonumber \\
& & \hspace{-0.5cm} - (\tfrac{D}2 \!-\! 1) \partial_{\alpha} \Bigl\{ a^{D-1} H
\sqrt{-\widetilde{g}} \, \widetilde{g}^{\mu\nu} \widetilde{g}^{\alpha\beta}
h_{\beta 0} \Bigr\} + (\tfrac{D}2 \!-\! 1) a^{D-1} H \sqrt{-\widetilde{g}} \,
\widetilde{g}^{\alpha \gamma} \widetilde{g}^{\delta (\mu} \delta^{\nu)}_{0}
h_{\gamma\delta ,\alpha} \nonumber \\
& & \hspace{4cm} + (\tfrac{D}2 \!-\! 1) a^{D-1} H \tfrac{\partial}{\partial h_{\mu\nu}}
\Bigl\{ \sqrt{-\widetilde{g}} \, \widetilde{g}^{\alpha\beta} 
\widetilde{g}^{\gamma\delta} \Bigr\} h_{\alpha\beta , \gamma} h_{\delta 0} \; .
\qquad \label{3ptVertex}
\end{eqnarray}
Expanding this to second order in the graviton field naively produces 45
terms, but some of these are the same so that only 42 are independent. They can
be found for $D=4$ in Table~3 of \cite{Tsamis:1996qm}, and Table~1 of 
\cite{Tsamis:1996qk}.\footnote{These tables actually actually lists 43 terms, 
but two of these 43 are the same.}

\begin{table}
\setlength{\tabcolsep}{8pt}
\def\arraystretch{1.5}
\centering
\begin{tabular}{|@{\hskip 1mm }c@{\hskip 1mm }||c|}
\hline
$i$ & Coefficient\ of\ $\kappa^2 [\mbox{}^{\mu\nu} D_i^{\rho\sigma}] \,
i\delta^D(x \!-\! x')$ \\
\hline\hline
$$ & $-\tfrac{1}{30} \mathcal{K}_3 [23 \partial^4 \!+\! 130 a a' H^2 \partial^2
\!-\! 1010 a a' H^2 \partial_0^2 \!+\! 1940 (a a')^2 H^4]$ \\
$1$ & $-\tfrac{a a' H^2 [53 \partial^2 + 80 \partial_0^2 + 261 a a' H^2]}{2^8 
\cdot 3 \cdot \pi^2}$ \\
\hline
$$ & $-\tfrac{1}{30} \mathcal{K}_3 [61 \partial^4 \!+\! 340 a a' H^2 \partial^2
\!+\! 200 a a' H^2 \partial_0^2 \!+\! 280 (a a')^2 H^4]$ \\
$2$ & $+ \tfrac{a a' H^2 [50 \partial^2 + 88 \partial_0^2 + 287 a a' H^2]}{2^7 
\cdot 3 \cdot \pi^2}$ \\
\hline
$3$ & $\mathcal{K}_3 a^2 H^2 [62 \partial^2 \!+\! 300 a^2 H^2] \!-\!
\tfrac{a^2 H^2 [23 \partial^2 + 32 \partial_0^2 + 105 a H \partial_0
+ 241 a^2 H^2]}{2^7 \cdot 3 \cdot \pi^2}$ \\
\hline
$5$ & $-\tfrac23 \mathcal{K}_3 a H [29 \partial^2 \!+\! 83 a H \partial_0
\!+\! 146 a^2 H^2] \!+\! \tfrac{a^2 H^2 [38 \partial_0 - 25 a H]}{2^7 
\cdot 3 \cdot \pi^2}$ \\
\hline
$7$ & $\tfrac{1}{30} \mathcal{K}_3 [23 \partial^2 \!+\! 290 a H \partial_0
\!-\! 130 a^2 H^2] \!+\! \tfrac{29 a^2 H^2}{2^7 \cdot 3 \cdot \pi^2}$ \\
\hline
$9$ & $\tfrac{4}{3} \mathcal{K}_3 a a' H^2 [10 \partial^2 \!+\! 39 a a' H^2]
\!+\! \tfrac{a a' H^2 [71 \partial^2 + 136 \partial_0^2 + 198 a a' H^2]}{2^6 
\cdot 3 \cdot \pi^2}$ \\
\hline
$11$ & $\tfrac{4}{3} \mathcal{K}_3 a H [2 \partial^2 \!-\! 2 a H \partial_0 \!-\! 
76 a^2 H^2] \!-\! \tfrac{a^2 H^2 [29 \partial_0 - 24 a H]}{2^6 \cdot 3 \cdot \pi^2}$ \\
\hline
$12$ & $\tfrac{1}{15} \mathcal{K}_3 [61 \partial^2 \!+\! 380 a a' H^2] \!-\!
\tfrac{13 a a' H^2}{2^6 \cdot \pi^2}$ \\
\hline
$13$ & $-\tfrac{2812}{3} \mathcal{K}_3 (a a')^2 H^4 \!+\! \tfrac{a a' H^2 
[28 \partial^2 + 129 a a' H^2]}{2^5 \cdot 3 \cdot \pi^2}$ \\
\hline
$14$ & $300 \mathcal{K}_3 a^3 H^3 \!-\! \tfrac{a^2 H^2 [76 \partial_0 + 203 a H]}{
2^6 \cdot 3 \cdot \pi^2}$ \\
\hline
$16$ & $-\tfrac{134}{3} a^2 H^2 \!+\! \tfrac{a^2 H^2}{2^4 \cdot \pi^2}$ \\
\hline
$18$ & $\tfrac{500}{3} \mathcal{K}_3 a a' H^2 \!-\! \tfrac{19 a a' H^2}{2^4 
\cdot 3 \cdot \pi^4}$ \\
\hline
$19$ & $-\tfrac{50}{3} \mathcal{K}_3 a H$ \\
\hline
$21$ & $-\tfrac{14}{5} \mathcal{K}_3$ \\
\hline
\end{tabular}
\caption{\footnotesize Primitive local 3-point contributions to 
$-i [\mbox{}^{\mu\nu} \Sigma^{\rho\sigma}](x;x')$.}
\label{Primitive3ptDiv}
\end{table} 

\begin{sidewaystable}[p]
\setlength{\tabcolsep}{8pt}
\def\arraystretch{1.5}
\centering
\begin{tabular}{|@{\hskip 1mm }c@{\hskip 1mm }||c|c|}
\hline
$i$ & Coefficient\ of\ $\kappa^2 [\mbox{}^{\mu\nu} D_i^{\rho\sigma}] 
\ln^2(\tfrac14 H^2 \Delta x^2)$ & Coefficient\ of\ $\kappa^2 [\mbox{}^{\mu\nu}
D_i^{\rho\sigma}] \ln(\tfrac14 H^2 \Delta x^2)$ \\
\hline\hline
$1$ & $-\tfrac{[\partial^8 + 4 (a - a') H \partial_0 \partial^6
+ 40 a a' H^2 \partial^6]}{2^{13} \cdot 3 \cdot \pi^4}$ & $\tfrac{- 3 a ^2 a'^2 H^4 
\partial^2 \partial_0^2 + \frac{3}{2} a a'^2 H^3 \partial^2\partial_0^3 - \frac{3}{2} 
a^2 a' H^3 \partial^2 \partial_0^3+(-\frac{3}{4} a ^2 H^2 + 5 a a' H^2 - 
\frac{3}{4}H^2 a'^2) \partial^4 \partial_0^2}{2^{9} \cdot 3 \cdot \pi^4}$ \\
$$ & $+\tfrac{[(6 a^2 - 16 a a' + 6 {a'}^2) H^2 \partial_0^2 \partial^4 
- 12 (a a')^2 H^4 \partial^4]}{2^{13} \cdot 3 \cdot \pi^4}$ & $+\tfrac{(\frac{3}{2} a a'^2 
H^3 - \frac{3}{2} a ^2  a') H^3 \partial^4 \partial_0 + (\frac{1}{2} a H - \frac{1}{2} a'H)
\partial^6 \partial_0 - 6 a ^2 a'^2 H^4 \partial_0^4 + (\frac{3}{2} a ^2 a'^2 H^4 + 
8 a a' H^2 \partial^2 + \frac{7}{8} \partial^4)\partial^4 }{2^{9} \cdot 3 \cdot \pi^4}$ \\
\hline
$2$ & $-\tfrac{[\partial^8 + 4 (a - a') H \partial_0 \partial^6
- 6 (a -a')^2 H^2 \partial^6]}{2^{11} \cdot 3 \cdot \pi^4}$ 
$$ & $\tfrac{(16 a ^2 \partial^2 H^3 a'- 16 a  \partial^2 H^3 a'^2 -36 a ^2 a'^2 H^4 
\partial_0 -12 a a'  H^2 \partial^2\partial_0 ) \partial^2\partial_0}{2^{9} 
\cdot 3 \cdot \pi^4} $ \\ & $+\tfrac{[24 a a' H^2 \partial_0^2 \partial^4 - 24 a a' (a - a') H^3 
\partial_0 \partial^4 - 60 (a a')^2 H^4 \partial^4]}{2^{11} \cdot 3 
\cdot \pi^4}$ & $+\tfrac{(+ 14 a ^2a'^2 H^4-2a'H \partial^2\partial_0 + 2 a H
\partial^2\partial_0 +8 a  a' H^2 \partial^2 - 3 a'^2H^2 \partial^2 -3 a ^2  H^2
\partial^2 +\frac{\partial^4}{2})\partial^4}{2^{9} 
\cdot 3 \cdot \pi^4}$ \\
\hline
$3$ & $\tfrac{[a (3 a - 13 a') H^2 \partial^6 + 3 a a' (3 a + a') H^3
\partial_0 \partial^4 - 3 (a a')^2 H^4 \partial^4]}{2^{11} \cdot 3 \cdot \pi^4}$ & 
$\tfrac{(-\frac{13}{2} a ^2  a'H^3 \partial^2 - 6 a  a'^2 H^3\partial^2 + 6 a ^2 a'^2 
H^4  \partial_0 ) \partial^2 \partial_0+ (\frac{9}{2} a ^2 a'^2 H^4 + \frac{13}{2} 
a a' H^2\partial^2 -\frac{3}{2} a ^2  H^2\partial^2)\partial^4 }{2^{9} \cdot 3 \cdot \pi^4}$ \\
\hline
$5$ & $\tfrac{[(4 a - 6 a') H \partial^6 - (3 a^2 - 4 a a' + 3 {a'}^2) H^2
\partial_0 \partial^4 + 6 a^2 a' H^3 \partial^4]}{2^{11} \cdot 3 \cdot \pi^4}$ & 
$\tfrac{(+3 a ^2 a'^2 H^4  + 3 a ^2 a' H^3 \partial_0 + 3 a a'^2 H^3 \partial_0 - 
2 a a' H^2 \partial^2 + \frac{3}{2} a'^2 H^2 \partial^2 +\frac{3}{2} a^2 H^2
\partial^2)\partial^2 \partial_0}{2^{9} \cdot 3 \cdot \pi^4}$ \\  
$$ & $$ & $+\tfrac{12 a ^2 H^4 a'^2 \partial_0^3 - (8 a^2 a' H^3 + \frac{15}{2} a a'^2 
H^3 +3  H a'\partial^2 - 2 a H \partial^2) \partial^4}{2^{9} \cdot 3 \cdot \pi^4}$ \\
\hline
$7$ & $\tfrac{[3 \partial^6 - (6 a - 8 a') H \partial_0 \partial^4
+ 12 {a'}^2 H^2 \partial^4]}{2^{12} \cdot 3 \cdot \pi^4}$ & $\tfrac{(6 a^2  a' H^3 
- 6 a  a'^2H^3 - 3 a a' H^2 \partial_0 - 2 a'H \partial^2 + \frac{3}{2} a H \partial^2)
\partial^2 \partial_0  + 12 a ^2 a'^2 H^4\partial_0^2}{2^{9} \cdot 3 \cdot \pi^4}$ \\
$$ & $$ & $+\tfrac{(3 a ^2  a'^2 H^4-3 a a'H^2 \partial^2-\frac{3}{2} {a'}^2 H^2 
\partial^2-\frac{3}{4} \partial^4)\partial^2 }{2^{9} \cdot 3 \cdot \pi^4}$ \\
\hline
$9$ & $-\tfrac{[(3 a^2 + 4 a a' + 3 {a'}^2) H^2 \partial^6 - 6 a a' (a - a')
H^3 \partial_0 \partial^4 + 54 (a a')^2 H^4 \partial^4}{2^{10} \cdot 3 \cdot \pi^4}$ & 
$\tfrac{(-16 a ^2 a' \partial_0+16 a a'^2 \partial_0-2 a ^2 a'^2H +24 a^3  a' H
+24 a a'^3 H) H^3\partial^4 (6 a a'+3 a'^2+3 a ^2)H^2 \partial^6 }{2^{9} 
\cdot 3 \cdot \pi^4}$ \\
\hline
$11$ & $\tfrac{[(a + a') H \partial^6 - 2 (7 a - 3 a') a' H^2 \partial_0
\partial^4 + 2 a a' (a - a') H^3 \partial^4]}{2^{10} \cdot 3 \cdot \pi^4}$ & 
$\tfrac{(42 a ^2  a'^2 H^4 +10 aa' H^2 \partial^2-6  a'^2 H^2 \partial^2)
\partial^2 \partial_0 + (-9 a^2 a'H^3 + 33 a a'^2H^3+  a' H \partial^2 +a H 
\partial^2)\partial^4}{2^{9} \cdot 3 \cdot \pi^4}$ \\
\hline
$12$ & $\tfrac{[\partial^6 + 12 (a - a') H \partial_0 \partial^4 - 
6 (a-a')^2 H^2 \partial^4]}{2^{11} \cdot 3 \cdot \pi^4}$ & $\tfrac{(-18 a ^2 a' H^2
+ 18 aa'^2H^2 +6 a' \partial^2-6 a \partial^2)H\partial^2 \partial_0}{2^{10} 
\cdot 3 \cdot \pi^4}$\\ 
$$ & $$ & $\tfrac{(+24 a ^2 H^4 a'^2-18 a  a' H^2 \partial^2+3 a'^2 \partial^2 H^2 
+ 3 a ^2 H^2\partial^2 -\frac{1}{2} \partial^4) \partial^2 }{2^{10} \cdot 3 \cdot \pi^4}$ \\
\hline
$13$ & $-\tfrac{7 (a a')^2 H^4 \partial^4}{2^9 \cdot \pi^4}$ & $\tfrac{ (26 a ^2 {a'}^2 
+ 24 a ^3 a' +24 a {a'}^3)H^4\partial^4}{2^{9} 
\cdot 3 \cdot \pi^4}$ \\
\hline
$14$ & $\tfrac{ a a' (a - a') H^3 \partial^4}{2^{10} \cdot \pi^4}$ & $\tfrac{-12 a ^2 
{a'}^2  H^4 \partial^2\partial_0-16 a ^2   a'H^3\partial^4+ 6 a {a'}^2 H^3 
\partial^4 }{2^{9} \cdot 3 \cdot \pi^4}$ \\
\hline
$16$ & $\tfrac{3 a a' H^2 \partial^4}{2^{10} \cdot \pi^4}$ & 
$\tfrac{-18 a ^2 \partial^2 H^4 a'^2-9 a  \partial^4 H^2 a' }{2^{9} 
\cdot 3 \cdot \pi^4}$ \\
\hline
$18$ & $-\tfrac{(3 a^2 - 8 a a' + 3 {a'}^2) H^2 \partial^4}{2^{10} 
\cdot 3 \cdot \pi^4}$ & $\tfrac{ (6 a ^2 a'\partial_0-6 a  a'^2 \partial_0+12 a ^2 
a'^2 H)H^3 \partial^2 - 24 a ^2 H^4 a'^2 \partial_0^2 +( - 12 a a'+3 a'^2+3 a ^2)
H^2\partial^4 }{2^{9} 
\cdot 3 \cdot \pi^4}$ \\
\hline
$19$ & $\tfrac{[(2 a' - 3 a) H] \partial^4}{2^{10} \cdot 3 \cdot \pi^4}$ & 
$\tfrac{[(3 a - 2 a') H \partial^4 + 6 a a' H^2 \partial_0 \partial^2 + 6 a a' 
(2 a - a') H^3 \partial^2 - 24 (a a')^2 H^4 \partial_0]}{2^9 \cdot 3 \cdot \pi^4}$ \\
\hline
$21$ & $-\tfrac{\partial^4}{2^{10} \cdot 3 \cdot \pi^4}$ & 
$\tfrac{[\partial^4 + 12 a a' H^2 \partial^2 - 24 (a a')^2 H^4]
}{2^9 \cdot 3 \cdot \pi^4}$ \\
\hline
\end{tabular}
\caption{\footnotesize Primitive nonlocal ghost loop contributions to 
$-i [\mbox{}^{\mu\nu} \Sigma^{\rho\sigma}](x;x')$.}
\label{PrimitiveGhostFin}
\end{sidewaystable}

The procedure for reducing expression (\ref{Sigma3pt}) is much the same as
(\ref{Sigmagh}), except there are $42^2 = 1764$ graviton vertex pairs instead 
of the $10^2 = 100$ ghost vertex pairs. It is instructive to consider a single 
vertex pair,
\begin{eqnarray}
\lefteqn{\Bigl\langle \tfrac{i\kappa}{2} \partial_{\alpha} \Bigl\{ a^{D-2}
\eta^{\mu\nu} h^{\alpha\beta}(x) h_{,\beta}(x) \Bigr\} \times \tfrac{i (D-2)
\kappa}{2} \partial'_{\gamma} \Bigl\{ {a'}^{D-1} H h^{\rho\sigma}(x')
h^{\gamma}_{~0}(x') \Bigr\} \Bigr\rangle } \nonumber \\
& & \hspace{-0.5cm} = -(\tfrac{D-2}{4}) \kappa^2 H \eta^{\mu\nu} \partial_{\alpha}
\partial'_{\gamma} \Bigl\{ a^{D-2} {a'}^{D-1} \Bigl[ i[\mbox{}^{\alpha\beta}
\Delta^{\rho\sigma}](x;x') \!\times\! \partial_{\beta} i[\mbox{}^{\delta}_{~\delta} 
\Delta^{\gamma}_{~0}](x;x') \nonumber \\
& & \hspace{5.3cm} + i[\mbox{}^{\alpha\beta} \Delta^{\gamma}_{~0}](x;x') \!\times\!
\partial_{\beta} i [\mbox{}^{\delta}_{~\delta} \Delta^{\rho\sigma}](x;x') 
\Bigr] \Bigr\} \; . \qquad \label{3ptStep1}
\end{eqnarray}
The next step is to insert the expansion (\ref{gravprop}) of the graviton
propagators in terms if $i\Delta_A$, $i\Delta_B$ and $i\Delta_C$, and then
expand each of the scalar propagators (\ref{propexp}) to the order we require.
Because the terms inside the curly brackets of expression (\ref{3ptStep1}) 
involve just a single derivative, we see from Table~\ref{TableB} that only
$i\Delta_{cf}$ and $A_0$ are needed. The result is,
\begin{eqnarray}
\lefteqn{ \Bigl\{ \dots \Bigr\} \longrightarrow -\tfrac{4 \delta^{\gamma}_0 a'}{D-2} 
\Bigl\{2 \eta^{\alpha (\rho} \eta^{\sigma) \beta} \!-\! \tfrac{2 \eta^{\alpha\beta}
\eta^{\rho\sigma} }{D-2} \Bigr\} \!\times\! (a a')^{D-2} i\Delta_{\rm cf} \,
\partial_{\beta} i\Delta_{\rm cf} -\tfrac{4 \eta^{\rho\sigma}a'}{D-2}} \nonumber \\
& & \hspace{-0.5cm} \times \Bigl\{2 \delta^{(\alpha}_{0} \eta^{\beta) \gamma} \!-\! 
\tfrac{2 \eta^{\alpha\beta} \delta^{\gamma}_{0}}{D-2} \Bigr\} \!\times\! 
(a a')^{D-2} i\Delta_{\rm cf} \, \partial_{\beta} i\Delta_{\rm cf} - 
\tfrac{4 \delta^{\gamma}_0 a'}{D-2} \Bigl\{ 2 \overline{\eta}^{\alpha (\rho} 
\overline{\eta}^{\sigma) \beta} \!\!-\!\! \tfrac{2 \overline{\eta}^{\alpha\beta} 
\overline{\eta}^{\rho\sigma}}{D-3} \Bigr\} \qquad \nonumber \\
& & \hspace{0cm} \times (a a')^{D-2} A_0 \, \partial_{\beta} 
i\Delta_{\rm cf} \!-\! \tfrac{4 \overline{\eta}^{\rho\sigma} a'}{D-3} \Bigl\{2
\delta^{(\alpha}_0 \eta^{\beta)\gamma} \!\!-\!\! \tfrac{2 \eta^{\alpha\beta}
\delta^{\gamma}_0}{D-2} \Bigr\} \!\times\! (a a')^{D-2} i\Delta_{\rm cf}
\, \partial_{\beta} A_0 \; . \qquad \label{3ptStep2}
\end{eqnarray}
We then read off the relevant products from Table~\ref{TableB} and reduce the 
result to standard form. The contribution from the final term of expression
(\ref{3ptStep2}) is,
\begin{eqnarray}
\lefteqn{-(\tfrac{D-2}{4}) \kappa^2 H \eta^{\mu\nu} \partial_{\alpha}
\partial'_{\gamma} \!\times\! - \tfrac{4 \overline{\eta}^{\rho\sigma} a'}{D-3} 
\Bigl\{2 \delta^{(\alpha}_0 \eta^{\beta)\gamma} \!\!-\!\! \tfrac{2 
\eta^{\alpha\beta} \delta^{\gamma}_0}{D-2} \Bigr\} \!\times\! (a a')^{D-2} 
i\Delta_{\rm cf} \, \partial_{\beta} A_0 } \nonumber \\
& & \hspace{-0.5cm} \longrightarrow 2 \kappa^2 H \eta^{\mu\nu} 
\overline{\eta}^{\rho\sigma} \partial_{\alpha} \partial'_{\gamma} \Bigl\{
[\delta^{\alpha}_0 \eta^{\beta\gamma} \!+\! \delta^{\beta}_0 \eta^{\alpha\gamma}
\!-\! \eta^{\alpha\beta} \delta^{\gamma}_0 ] \tfrac{a {a'}^2 H^2 \partial_{\beta}
\partial^2}{2^7 \cdot \pi^4} \ln(\tfrac14 H^2 \Delta x^2) \Bigr\} , \qquad \\
& & \hspace{0.5cm} =-\tfrac{\kappa^2 H^3 a {a'}^2 [\partial^4 \partial_0 +
(a - 2 a') H \partial^4 + 4 a' H \partial^2 \partial_0^2 + 2 a a' H^2 \partial^2 
\partial_0]}{2^6 \cdot \pi^4} \nonumber \\
& & \hspace{5.8cm} \times \Bigl\{[\mbox{}^{\mu\nu} D_1^{\rho\sigma}] \!+\!
[\mbox{}^{\mu\nu} D_3^{\rho\sigma}] \Bigr\} \ln(\tfrac14 H^2 \Delta x^2) \; .
\qquad \label{3ptStep3}
\end{eqnarray}
Tables~\ref{Primitive3ptDiv}, \ref{Primitive3ptFin} and \ref{Primitive3ptFinSingleLog}
list the total local and nonlocal 3-point contributions from ghost and graviton loops.

\section{Renormalization}

The purpose of this section is to absorb the divergences of 
Tables~\ref{Primitive4pt} and \ref{Primitive3ptDiv} using BPHZ counterterms. 
The section begins with a discussion of the general principles that constrain 
these counterterms. The two invariant 1-loop counterterms are given next. 
Then discussion turns to the plethora of noninvariant counterterms. The
section ends with the fully renormalized result.

\subsection{General Principles}

Counterterms are gauge dependent. Because we are not using the background 
field gauge \cite{DeWitt:1967ub,Abbott:1981ke}, and because our gauge 
(\ref{gauge}) breaks de Sitter invariance \cite{Tsamis:1992xa,Woodard:2004ut}, 
we must consider counterterms in addition to the usual $R^2$ and $C^2$ 
contributions. Certain general principles constrain the BPHZ counterterms 
needed to renormalize the graviton self-energy at 1-loop order on de Sitter 
background:

\begin{table}[H]
\setlength{\tabcolsep}{8pt}
\def\arraystretch{1.5}
\centering
\begin{tabular}{|@{\hskip 1mm }c@{\hskip 1mm }||c|}
\hline
$i$ & Coefficient\ of\ $\kappa^2 [\mbox{}^{\mu\nu} D_i^{\rho\sigma}] 
\ln^2(\tfrac14 H^2 \Delta x^2)$ \\
\hline\hline
$$ & $\frac{10 \left(161 a a' H (a'-a )\partial_0 +8 \left(a ^2-25 a  a'+a'^2\right)
\partial_0^2 +2 a  H^2 a' \left(24 a ^2-37 a  a'+24 a'^2\right)\right) H^2
\partial^4}{2^{13} \cdot 3 \cdot 5 \cdot \pi^4}$ \\
$1$ & $\frac{-960 a ^2 a'^2 (a'H (3 \partial_0+2 a  H)+(\partial_0-3 a  H)\partial_0 )
H^4\partial^2 }{2^{13} \cdot 3 \cdot 5 \cdot \pi^4}$ \\
$$ & $\frac{+5  \left(30 (a -a')\partial_0 + (3 a'+68 a )a'H +3 a ^2 H\right) H
\partial^6 + 960 a ^2 H^5 a'^2 ((a -a')\partial_0 +2 a a' H )\partial_0^2 + 
46 \partial^8}{2^{13} \cdot 3 \cdot 5 \cdot \pi^4}$ \\
\hline
$$ & $\frac{-10 \left(131 a a'H (a -a')\partial_0 +5 \left(a ^2+10 a a' + {a'}^2\right)
\partial_0^2 +12 a a' H^2 \left(2 a ^2-17 a a'+2 a'^2\right)\right)H^2 
\partial^4}{2^{12} \cdot 3 \cdot 5 \cdot \pi^4}$ \\
$2$ & $\frac{+240 a a' \left(4 a a'H^2 (a'-a )\partial_0-H(a -2 a') (2 a -a') 
\partial_0^2 + (a'-a )\partial_0^3 +8 a ^2  a'^2H^3\right)H^3 \partial^2 }{
2^{12} \cdot 3 \cdot 5 \cdot \pi^4}$ \\
& $\frac{+5 H \left[8 (a - a') \partial_0 + H a' (57 a'-86 a )+ 57 a^2 H\right]
\partial^6 + 480 a ^2 H^4 a'^2 [2 H a' (2 \partial_0 + 3 a H) - (\partial_0 + 4 a H)
\partial_0] \partial_0^2 + 61 \partial^8}{2^{12} \cdot 3 \cdot 5 \cdot \pi^4}$ \\
\hline
$3$ & $\frac{ \left( -96 a a'^2 H^2 ((3 a -2 a')\partial_0 +4 a  H (a'+a )) 
\partial^2 \partial_0 +6 a a'H \left(8 (a -17 a')\partial_0 + a'H (4 a'-165 a ) 
+ 16 a ^2 H\right)\partial^4 \right)H^2 }{2^{12} \cdot 3 \cdot \pi^4}$ \\
& $\frac{H^2 \left(-192 a ^2 a'^3  H^3 (2 \partial_0+a  H)\partial_0^2 + 
\left(11 a ^2-52 a  a'-179 a'^2\right)\partial^6\right)}{2^{12} \cdot 3 \cdot \pi^4}$ \\
\hline
$5$ & $\frac{\left(+96 a ^2 H^4 a'^2 (a'-a )\partial_0^2 +2 (26 a'+3 a )\partial^6 
\right)H }{2^{11} \cdot 3 \cdot \pi^4}$ \\
& $\frac{\left(96 a ^2 a'^2  H^3 (\partial_0-a  H)+H \left(4 \partial_0 
\left(-2 a ^2+29 a  a'+19 a'^2\right)+a  H a' (125 a -3 a')\right) \partial^2 
\right)H \partial^2}{2^{11} \cdot 3 \cdot \pi^4}$ \\
\hline
$7$ & $\frac{5 H \left(-2 \partial_0 (19 a'+10 a )+H a' (61 a'-52 a )+7 a^2 H\right)
\partial^4 +480 a ^2 a'^2 H^5 (a'-2 a )\partial_0 }{2^{12} \cdot 3 \cdot 5 \cdot \pi^4}$ \\
& $\frac{+960 a ^2 a'^2H^4 \partial^2 -23 \partial^6}{2^{12} \cdot 3 \cdot 5 \cdot \pi^4}$ \\
\hline
$9$ & $\frac{ \left(-2 aa' H \left(70 (a'-a )\partial_0 + 24 a'^2H -265 aa'H + 24 a^2 
H\right) \partial^2 +\left(31 a ^2-224 a  a'+31 a'^2\right)\partial^4 \right)
H^2\partial^2 }{2^{11} \cdot 3 \cdot \pi^4}$ \\
$$ & $+\frac{ 48 a ^2 a'^2  H^4 \left(8 aa'  H^2 -3 \partial_0^2\right)\partial^2 
+ 576 a ^3 H^6 a'^3 \partial_0^2 }{2^{11} \cdot 3 \cdot \pi^4}$ \\
\hline
$11$ & $\tfrac{H \left(48 a a' H^2 \left(a' \left(4 a ^2 H^2-2  (\partial_0+2 
a  H) \partial_0\right)+a  (\partial_0+2 a  H) \partial_0 \right) \partial^2 + 
96 a^2 {a'}^2 H^3 (-H a'+\partial_0+4 a  H)\partial_0^2\right)}{2^{11} \cdot 3 \cdot \pi^4}$ \\
& $+\frac{ \left(+2H\left(\left(5 a ^2+64 a  a'-23 a'^2\right)\partial_0 +aa' H  
(93 a -7 a')\right)+(5 a -13 a')\partial^2\right)H\partial^4}{2^{11} \cdot 3 \cdot \pi^4}$ \\
\hline
$12$ & $-\tfrac{240 a  a' H^2(a' H (\partial_0 + 5 a  H)-(\partial_0+a  H)
\partial_0 ) \partial^2+240 a ^2  a'^2H^4 (2 H a'+\partial_0-2 a H) \partial_0 
+ 61 \partial^6}{2^{11} \cdot 3 \cdot 5 \cdot \pi^4}$ \\
& $-\tfrac{+70 \partial^4 H \left(-a' (\partial_0+5 a  H)+4 H a'^2+a  
(\partial_0+4 a  H)\right)}{2^{11} \cdot 3 \cdot 5 \cdot \pi^4}$ \\
\hline
$13$ & $\frac{a a' H^4  \left(12 a a'H  (3 (a -a')\partial_0 +2 a a' H)\partial^2 
+ 48 a ^2  a'^2H^2 \partial_0^2-\left(a ^2-39 a  a'+a'^2\right)\partial^4 
\right)}{2^8 \cdot \pi^4}$ \\
\hline
$14$ & $\frac{a a'  H^3 \left(-6 aH \left(-9 a' \partial_0+4 a'^2H +4 a  
\partial_0\right)  \partial^2 -48 a ^2 a' H^2\partial_0^2-(41 a'+70 a )
\partial^4 \right)}{2^8 \cdot 3 \cdot \pi^4}$ \\
\hline
$16$ & $\tfrac{-24 a  a' \left(-a' (\partial_0+7 a  H)+2 H a'^2 + a  
(3 \partial_0+4 a  H)\right)H^3 \partial^2 }{2^{11} \cdot 3 \cdot \pi^4}$ \\ 
& $+\tfrac{\left(-48 a ^2 H^2 a'^2 (-4 H a'+\partial_0+4 a  H)\partial_0 + 
\left(75 a ^2+52 a  a'-5 a'^2\right)\partial^4 \right)H^2 }{2^{11} \cdot 3 \cdot \pi^4}$ \\
\hline
$18$ & $-\tfrac{[(16 a^2 + 3 a a' + 16 {a'}^2) H^2 \partial^4 + 24 a a' (a-a') H^3 
\partial_0 \partial^2]}{2^9 \cdot 3 \cdot \pi^4}$ \\
$$ & $-\tfrac{(a a')^2 H^4 [28 \partial^2 + 52 \partial_0^2 - 48 (a-a') H
\partial_0]}{2^9 \cdot 3 \cdot \pi^4}$ \\
\hline
$19$ & $\tfrac{[(19 a + 6 a') H \partial^4 - 24 a a' H^2 (\partial_0 - a' H) 
\partial^2 + 24 (a a')^2 H^4 \partial_0]}{2^{10} \cdot 3 \cdot \pi^4}$ \\
\hline
$21$ & $\tfrac{7 \partial^4}{2^{10} \cdot \pi^4}$ \\
\hline
\end{tabular}
\caption{\footnotesize Primitive double log contributions to $-i [\mbox{}^{\mu\nu}
\Sigma^{\rho\sigma}](x;x')$.}
\label{Primitive3ptFin}
\end{table}

\begin{table}[H]
\setlength{\tabcolsep}{8pt}
\def\arraystretch{1.5}
\centering
\begin{tabular}{|@{\hskip 1mm }c@{\hskip 1mm }||c|}
\hline
$i$ & Coefficient\ of\ $\kappa^2 [\mbox{}^{\mu\nu} D_i^{\rho\sigma}] \ln(\tfrac14 H^2 
\Delta x^2)$ \\
\hline\hline
$$ & $ \frac{10H^2 \left(257 aa'H (a -a')\partial_0 - 2 \left(4 a ^2-181 a  a'+
4 a'^2\right) \partial_0^2+2 aa' \left(54 a ^2-125 a  a'+54 a'^2\right) H^2\right)
\partial^4}{2^{12} \cdot 3 \cdot 5 \cdot \pi^4}$ \\
$1$ & $ \frac{+20 aa'H^3\left(144 aa'H^2 (a -a')\partial_0 + H\left(108 a ^2+7 a a' + 
108 a'^2\right)\partial_0^2 +61  (a -a')\partial_0^3 +144 a ^2 a'^2 H^3\right)
\partial^2}{2^{12} \cdot 3 \cdot 5 \cdot \pi^4}$ \\ 
$$ & $\frac{-5H \left(30(a -a')\partial_0+ a'H(3 a'-142 a )+3 a ^2 H\right)\partial^6 
+ 40 a ^2a'^2H^4(48 H a' (a  H-\partial_0)+\partial_0 (29 \partial_0+48 a  H))
\partial_0^2 - 46 \partial^8}{2^{12} \cdot 3 \cdot 5 \cdot \pi^4} $ \\
\hline
$$ & $-\frac{10 H^2 \left(122 aa'H (a'-a )\partial_0 + \left(-5 a ^2+72 a  a'-5 a'^2
\right)\partial_0^2 + a a'H^2 \left(36 a ^2+41 a  a'+36 a'^2\right)\right) \partial^4}{
2^{11} \cdot 3 \cdot 5 \cdot \pi^4}$ \\
$2$ & $-\frac{+ 20 aa'H^3\left(144 aH^2a' (a -a')\partial_0 - 3 H\left(4 a ^2-11 
a a' + 4 a'^2\right)\partial_0^2+4 (a -a')\partial_0^3 + 168 a ^2 a'^2H^3\right)
\partial^2 }{2^{11} \cdot 3 \cdot 5 \cdot \pi^4} $ \\ 
$$ & $-\frac{+5 H \left( 8(a -a') \partial_0 +a'H (57 a'+50 a )+57 a ^2 H\right)
\partial^6+40 a ^2 a'^2H^4\left(24 H (a -a')\partial_0+72 aa'H^2 + 11 
\partial_0^2\right)\partial_0^2 +61 \partial^8}{2^{11} \cdot 3 \cdot 5 \cdot \pi^4}$ \\
\hline
$3$ & $ \frac{H^2 \left(8 aa'^2H^2\left(12 a' \left(a ^2 H^2-4 \partial_0^2\right)
+ a(79 \partial_0+156 a  H)\partial_0\right)\partial^2+2 aa' \left(21 H \partial_0 
(18 a'+a) + 4 H^2 a' (24 a'+101 a )+18 \partial_0^2\right) \partial^4 \right)}{
2^{11} \cdot 3 \cdot \pi^4} $ \\ 
$$ & $ \frac{H^2 \left(+192 a ^2 a'^3 H^3(\partial_0+a  H)\partial_0^2 +\left(-11 
a^2+199 a  a'+179 a'^2\right)\partial^6 \right)}{2^{11} \cdot 3 \cdot \pi^4} $ \\
\hline
$5$ & $\frac{2 aa'H^3\left(H \left(-60 a ^2+139 a  a'-108 a'^2\right)\partial_0 + 
(7 a'-31 a )\partial_0^2+48 a ^2 a'H^2\right)\partial^2-4 a ^2a'^2 H^4(48 H 
(a -2 a')+11 \partial_0)\partial_0^2}{2^{10} \cdot 3 \cdot \pi ^4}$ \\ 
$$ & $\frac{\left(-H^2 
\left(4 \left(-2 a ^2+62 a  a'+19 a'^2\right) \partial_0 + 15 aa' H(4 a'+9 a )\right)
-2H(26 a' + 3 a)\partial^2\right)\partial^4 }{2^{10} \cdot 3 \cdot \pi^4}$ \\
\hline
$7$ & $-\frac{10 a a'H^2\left(8 H(3 a -2 a')\partial_0+H^2 \left(72 a ^2-91 a a'
+ 108 a'^2\right)-2 \partial_0^2\right) \partial^2+5 H \left(-2(19 a'+10 a )
\partial_0 + a'H(61 a'+41 a )+7 a ^2 H\right)\partial^4}{2^{11} \cdot 3 \cdot 5 \cdot \pi^4}$ \\ 
$$ & $-\frac{+20 a ^2 a'^2 H^4(48 H (a -2 a')+29 \partial_0)\partial_0 -23 
\partial^6}{2^{11} \cdot 3 \cdot 5 \cdot \pi^4} $ \\
\hline
$9$ & $\frac{H^2 \left(2 a a' \left(117 H (a'-a )\partial_0 +H^2 \left(36 a ^2-103 
a  a'+36 a'^2\right)-72 \partial_0^2\right)\partial^4 + 4 a ^2a'^2H^2 \left(204 H 
(a'-a )\partial_0-168 aa'H^2 +91 \partial_0^2\right) \partial^2\right)}{
2^{10} \cdot 3 \cdot \pi^4}$ \\ 
$$ & $\frac{H^2 \left(-576 a ^3a'^3 H^4  \partial_0^2+\left(-31 a ^2+12 a  a' -31 {a'}^2
\right)\partial^6 \right)}{2^{10} \cdot 3 \cdot \pi^4}$ \\
\hline
$11$ & $\frac{H \left(-8 aa'H^2 \left(3 H \left(2 a ^2+3 a  a'-7 a'^2\right)\partial_0 
-(11 a'+2 a )\partial_0^2 +30 a ^2  a'H^2\right)\partial^2+2H \left(2 a a' (8 \partial_0
-51 a  H)+a'^2 (23 \partial_0+64 a  H)-5 a ^2 \partial_0\right)\partial^4 \right)}{
2^{10} \cdot 3 \cdot \pi^4}$ \\ 
$$ & $\frac{H \left(+8 a ^2 H^3 a'^2 \partial_0^2 (-36 H a'+11 \partial_0+24 a  H)
+\partial^6 (13 a'-5 a )\right)}{2^{10} \cdot 3 \cdot \pi^4} $ \\
\hline
$12$ & $\frac{20 a a'H^2\left(11 H (a'-a )\partial_0 +42 H^2 \left(a ^2-a  a'+a'^2\right)
-5 \partial_0^2\right) \partial^2 +10 H \left(7 (a -a')\partial_0 +4 H \left(7 a ^2+4 a a'
+7 a'^2\right)\right) \partial^4}{2^{10} \cdot 3 \cdot 5 \cdot \pi^4}$ \\ 
$ $ & $\frac{+240 a ^2 a'^2 H^4(-6 H a'+\partial_0+6 a  H) \partial_0+61 
\partial^6}{2^{10} \cdot 3 \cdot 5 \cdot \pi^4} $ \\
\hline
$13$ & $\frac{aa'H^2\left(-168 aa'H^3 (3 (a -a')\partial_0 + 2 aa' H ) \partial^2 +
H \left(9  (a'-a )\partial_0+2 a' H(42 a'-155 a )+84 a ^2 H\right)\partial^4 \right)}{
2^9 \cdot 3 \cdot \pi^4}$ \\ 
$ $ & $\frac{aa'H^2\left(-96 a ^2a'^2 H^4\partial_0^2-18 \partial^6\right)}{
2^9 \cdot 3 \cdot\pi^4} $ \\
\hline
$14$ & $\frac{a a'H^2\left(6 a H^2 (3 (4 a -7 a')\partial_0 +4  a'H (5 a'-3 a ))
\partial^2+ (146 H a'+237 a  H+36 \partial_0)\partial^4 -96 a ^2 a' H^3 \partial_0^2
\right)}{2^9 \cdot 3 \cdot \pi^4} $ \\
\hline
$16$ & $\frac{2aa'H^3\left(4 a' \partial_0+26 a  \partial_0-59 a a' H +12 a'^2H +24 a^2 H
\right) \partial^2}{2^{10} \cdot 3 \cdot \pi ^4} $ \\ 
$ $ & $ \frac{H^2 \left(-4 a ^2 H^2 a'^2 (24 H (a -a')+11 \partial_0)\partial_0 +
\left(-75 a ^2-112 a  a'+5 a'^2\right)\partial^4 \right)}{2^{10} \cdot 3 \cdot \pi^4} $ \\
\hline
$18$ & $\tfrac{16 (2 a^2 - a a' + 2 {a'}^2) H^2 \partial^4 + 31 a a' (a - a') H^3 
\partial_0 \partial^2]}{2^8 \cdot 3 \cdot \pi^4}$ \\ 
$ $ & $\tfrac{-a a' H^4 [(42 a^2 - 73 a a' +  42 {a'}^2) \partial^2 + 52 a a' 
\partial_0^2 - 48 a a' (a - a') H \partial_0]}{2^8 \cdot 3 \cdot \pi^4} $ \\
\hline
$19$ & $-\tfrac{[19 a + 6 a') H \partial^4 + 6 a a' H^2 (\partial_0 + a H) \partial^2
+ 24 (a a')^2 H^4 \partial_0]}{2^9 \cdot 3 \cdot \pi^4}$ \\
\hline
$21$ & $-\tfrac{7 \partial^4}{
2^9 \cdot \pi^4}$ \\
\hline
\end{tabular}
\caption{\footnotesize Primitive single log contributions to $-i [\mbox{}^{\mu\nu}
\Sigma^{\rho\sigma}](x;x')$.}
\label{Primitive3ptFinSingleLog}
\end{table} 

\begin{enumerate*}
\item{They take the form of $\kappa^2$ times two graviton fields, joined by 
a tensor differential operator of dimension four.}
\item{The dimensionful factors consist of derivatives and factors of $H$.}
\item{The terms with four derivatives are the same on de Sitter background
as those for flat space background in the same gauge.}
\item{Because our gauge respects homogeneity, isotropy and dilatation 
invariance, counterterms must possess the same symmetries.}
\end{enumerate*}
Principle \#4 is especially interesting. Homoegeneity and isotropy imply
that $h_{00}$, $h_{0i}$ and $h_{ij}$ are distinct tensors. Dilitation 
invariance implies that each counterterm must possess an overall factor of 
$a^D$, with an inverse scale factor for every derivative. 

The Ward Identity imposes severe restrictions on the entire self-energy
$-i [\mbox{}^{\mu\nu} \Sigma^{\rho\sigma}](x;x')$, both primitive and renormalized,
although not necessarily on specific counterterms. We define the Ward Operator,
\begin{equation}
\mathcal{W}^{\mu}_{~\alpha\beta} \equiv \delta^{\mu}_{~(\alpha} \partial_{\beta)} 
+ a H \delta^{\mu}_{~0} \eta_{\mu\nu} \; . \label{Warddef}
\end{equation}
A key distinction exists between matter contributions, which are annihilated
by acting the Ward operator on {\it either} coordinate \cite{Tan:2021ibs},
\begin{equation}
\mathcal{W}^{\mu}_{~\alpha\beta} \!\times\! -i [\mbox{}^{\alpha\beta}
\Sigma^{\rho\sigma}_{\rm mat}](x;x') = 0 \;\; , \;\; 
{\mathcal{W}'}^{\rho}_{~\gamma\delta} \!\times\! -i [\mbox{}^{\mu\nu} 
\Sigma^{\gamma\delta}_{\rm mat}](x;x') = 0 \; . \label{matter}
\end{equation}
On the other hand, gravitational contributions to the graviton self-energy
are only annihilated when the Ward operator acts on {\it both} coordinates
\cite{Tan:2021ibs},
\begin{equation}
\mathcal{W}^{\mu}_{~\alpha\beta} \!\times\! {\mathcal{W}'}^{\rho}_{~\gamma\delta}
\!\times\! -i [\mbox{}^{\alpha\beta} \Sigma^{\gamma\delta}_{\rm grav}](x;x') = 0
\; . \label{gravity}
\end{equation}
This means that matter contributions are considerably more constrained than
gravitational ones. In terms of structure functions, there should be four
structure functions for matter contributions, but ten are allowed for 
gravitational contributions.

\subsection{Invariant Counterterms}

There are two invariant counterterms at 1-loop,
\begin{equation}
\Delta \mathcal{L}_{\rm inv} = c_1 R^2 \sqrt{-g} + c_2 
C_{\alpha\beta\gamma\delta} C^{\alpha\beta\gamma\delta} \sqrt{-g} \; .
\label{DLinv}
\end{equation}
The term proportional to $c_1$ is known as the ``Eddington'' counterterm;
the term proportional to $c_2$ is called the ``Weyl'' counterterm. The 
$D$-dependent coefficients $c_1$ and $c_2$ are background-independent,
but gauge-dependent, so they can be read off from Capper's 1979 flat space 
computation of the graviton self-energy in what is the flat space limit of 
our gauge (\ref{gauge}) \cite{Capper:1979ej},
\begin{eqnarray}
c_1 &\!\!\! = \!\!\!& \tfrac{\mu^{D-4} \Gamma(\frac{D}2-1)}{2^{10} 
\pi^{\frac{D}2}} \times \tfrac{(9 D^4 - 43 D^3 - 4 D^2 + 212 D - 128)}{
(D-1)^2 (D-3) (D-4)} = \tfrac{13}{2^6 \cdot 3^2 \cdot \pi^2} 
\tfrac{\mu^{D-4}}{D-4} + {\rm Finite} \; , \qquad \label{c1} \\
c_2 &\!\!\! = \!\!\!& \tfrac{\mu^{D-4} \Gamma(\frac{D}2-1)}{2^9
\pi^{\frac{D}2}} \times \tfrac{(4 D^4 + 17 D^3 - 97 D^2 - 34 D + 16)}{
(D+1) (D-1) (D-3)^2 (D-4)} = \tfrac{61}{2^6 \cdot 3 \cdot 5 \cdot \pi^2} 
\tfrac{\mu^{D-4}}{D-4} + {\rm Finite} \; . \qquad \label{c2}
\end{eqnarray}
Here $\mu$ is the standard mass scale of dimensional regularization.

It is useful to decompose the Eddington counterterm into three parts,
\begin{equation}
R^2 = (R \!-\! D \Lambda)^2 + 2 D \Lambda [R \!-\! (D\!-\!2) \Lambda]
+ D (D\!-\!4) \Lambda^2 \; . \label{Rsquared}
\end{equation}
We label the associated sub-Lagrangians $\Delta \mathcal{L}_{1a}$, $\Delta
\mathcal{L}_{1b}$ and $\Delta \mathcal{L}_{1c}$, respectively. 
Evaluating their second variations at de Sitter gives the vertices,
\begin{eqnarray}
\lefteqn{ \tfrac{i \delta^2 \Delta S_{1a}}{\delta h_{\mu\nu}(x)
\delta h_{\rho\sigma}(x')} \Bigr\vert_{\rm dS} \!\!\!= 2 c_1 \kappa^2 
\mathcal{F}^{\mu\nu} \Bigl\{ a^{D-4} {\mathcal{F}'}^{\rho\sigma} 
i \delta^D(x \!-\! x') \Bigr\} , \qquad } \label{DL1a} \\
\lefteqn{ \tfrac{i \delta^2 \Delta S_{1b}}{\delta h_{\mu\nu}(x)
\delta h_{\rho\sigma}(x')} \Bigr\vert_{\rm dS} \!\!\!= {\scriptstyle
D (D - 1)} c_1 \kappa^2 H^2 \Bigl\{ [\eta^{\mu (\rho} \eta^{\sigma) \nu} 
\!-\! \eta^{\mu\nu} \eta^{\rho\sigma}] \mathcal{D} i \delta^D(x \!-\! x') } 
\nonumber \\
& & \hspace{2cm} + [2 {\partial'}^{(\mu} \eta^{\nu) (\rho} 
\partial^{\sigma)} \!+\! \eta^{\mu\nu} \partial^{\rho} \partial^{\sigma}
\!+\! {\partial'}^{\mu} {\partial'}^{\nu} \eta^{\mu\nu} ]
[a^{D-2} i \delta^D(x \!-\! x')] \Bigr\} , \qquad \label{DL1b} \\
\lefteqn{ \tfrac{i \delta^2 \Delta S_{1c}}{\delta h_{\mu\nu}(x)
\delta h_{\rho\sigma}(x')} \Bigr\vert_{\rm dS} \!\!\!= \tfrac12 {\scriptstyle D 
(D - 1)^2 (D - 4)} c_1 \kappa^2 H^4 [-\eta^{\mu (\rho} \eta^{\sigma) \nu}
\!+\! \tfrac12 \eta^{\mu\nu} \eta^{\rho\sigma} ] a^D 
i \delta^D(x \!-\! x') . \qquad } \label{DL1c}
\end{eqnarray}
Here and henceforth $\mathcal{D} \equiv \partial^{\alpha} a^{D-2} 
\partial_{\alpha}$, and the second order differential operator 
$\mathcal{F}^{\mu\nu}$ is \cite{Leonard:2014zua},
\begin{equation}
\mathcal{F}^{\mu\nu} \equiv \partial^{\mu\nu} \!-\! \eta^{\mu\nu} \mathcal{D}_1
+ 2 (D\!-\! 1) a H \delta^{(\mu}_{~~0} \partial^{\nu)} + (D\!-\!1) (D\!-\!2)
a^2 H^2 \delta^{\mu}_{~0} \delta^{\nu}_{~0} \; , \label{Fdef}
\end{equation}
where we define,
\begin{equation}
\mathcal{D}_1 \equiv \partial^2 \!+\! (D\!-\! 1) a H \partial_0 \!+\! (D\!-\!1)
a^2 H^2 \; . \label{D1def}
\end{equation} 

The vertex induced by the Weyl counterterm is,
\begin{equation}
\tfrac{i \delta^2 \Delta S_{2}}{\delta h_{\mu\nu}(x) \delta h_{\rho\sigma}(x')} 
\Bigr\vert_{\rm dS} \!\!\!= 2 c_2 \kappa^2 \mathcal{C}^{\alpha\beta\gamma\delta \mu\nu}
\Bigl\{ a^{D-4} {\mathcal{C}'}_{\alpha\beta\gamma\delta}^{~~~~~\rho\sigma} 
i \delta^D(x \!-\! x') \Bigr\} . \label{DL2}
\end{equation}
Here the second order differential operator 
$\mathcal{C}_{\alpha\beta\gamma\delta}^{~~~~~\mu\nu}$ operates on the graviton field
to give the linearized, conformally transformed Weyl tensor \cite{Park:2011ww,
Leonard:2014zua},
\begin{equation}
\widetilde{C}_{\alpha\beta\gamma\delta} = \mathcal{C}_{\alpha\beta\gamma\delta
}^{~~~~~\mu\nu} \!\times\! \kappa h_{\mu\nu} + O(\kappa^2 h^2) \; . 
\label{mathCdef}
\end{equation}
Some tedious but straightforward tensor algebra uncovers its explicit form,
\begin{equation}
\mathcal{C}_{\alpha\beta\gamma\delta}^{~~~~~\mu\nu} = 
\mathcal{D}_{\alpha\beta\gamma\delta}^{~~~~~\mu\nu} - 
\tfrac{[\eta_{\alpha\gamma} \mathcal{D}_{\beta\delta}^{~~\mu\nu} -
\eta_{\gamma\beta} \mathcal{D}_{\delta\alpha}^{~~\mu\nu} +
\eta_{\beta\delta} \mathcal{D}_{\alpha\gamma}^{~~\mu\nu} -
\eta_{\delta\alpha} \mathcal{D}_{\gamma\beta}^{~~\mu\nu}]}{D-2}
+ \tfrac{[\eta_{\alpha\gamma} \eta_{\beta\delta} - \eta_{\alpha\delta}
\eta_{\beta\gamma}] \mathcal{D}^{\mu\nu}}{(D-1) (D-2)} \; , \label{mathCdef2}
\end{equation}
where we define,
\begin{eqnarray}
\mathcal{D}_{\alpha\beta\gamma\delta}^{~~~~~\mu\nu} &\!\!\! \equiv \!\!\!&
-\tfrac12 \Bigl\{ \delta^{(\mu}_{~~\alpha} \delta^{\nu)}_{~~\gamma} 
\partial_{\beta} \partial_{\delta} \!-\! 
\delta^{(\mu}_{~~\gamma} \delta^{\nu)}_{~~\beta} 
\partial_{\delta} \partial_{\alpha} \!+\!
\delta^{(\mu}_{~~\beta} \delta^{\nu)}_{~~\delta} 
\partial_{\alpha} \partial_{\gamma} \!-\!
\delta^{(\mu}_{~~\delta} \delta^{\nu)}_{~~\alpha} 
\partial_{\gamma} \partial_{\beta} \Bigr\} , \qquad \label{MD6} \\
\mathcal{D}_{\beta\delta}^{~~\mu\nu} &\!\!\! \equiv \!\!\!& \eta^{\alpha\gamma}
\mathcal{D}_{\alpha\beta\gamma\delta}^{~~~~\mu\nu} = -\tfrac12 \Bigl\{
\eta^{\mu\nu} \partial_{\beta} \partial_{\delta} \!-\! 2 \partial^{(\mu}
\delta^{\nu)}_{~~(\beta} \partial_{\delta)} \!+\! \delta^{(\mu}_{~~\beta}
\delta^{\nu)}_{~~\delta} \partial^2 \Bigr\} \; , \qquad \label{MD4} \\
\mathcal{D}^{\mu\nu} &\!\!\! \equiv \!\!\!& \eta^{\beta\delta} 
\mathcal{D}_{\beta\delta}^{~~\mu\nu} = \partial^{\mu} \partial^{\nu} \!-\!
\eta^{\mu\nu} \partial^2 \; . \label{MD2}
\end{eqnarray}

The 2-point vertex (\ref{DL1c}) is finite, so we can take its $D=4$ limit,
\begin{equation}
\tfrac{i \delta^2 \Delta S_{1c}}{\delta h_{\mu\nu}(x)
\delta h_{\rho\sigma}(x')} \Bigr\vert_{\rm dS} \longrightarrow 
\tfrac{13 \kappa^2 H^4 a^4}{2^5 \cdot \pi^2} \Bigl\{\tfrac12 \eta^{\mu\nu}
\eta^{\rho\sigma} \!-\! \eta^{\mu (\rho} \eta^{\sigma) \nu} \Bigr\} 
i \delta^4(x \!-\! x') \; . \label{DL1ccont}
\end{equation}
Its contribution to the final result is given in Table~\ref{InvFin}.
Examination of the divergent 2-point vertices (\ref{DL1a}), (\ref{DL1b}) and
(\ref{DL2}) reveals two key qualitative differences with respect to the 
primitive divergences of Tables~\ref{Primitive4pt} and \ref{Primitive3ptDiv}:

\begin{table}[H]
\setlength{\tabcolsep}{8pt}
\def\arraystretch{1.5}
\centering
\begin{tabular}{|@{\hskip 1mm }c@{\hskip 1mm }||c|c|}
\hline
$i$ & $\mathcal{L}_{1b}$ & $\mathcal{L}_{1c}$ \\
\hline\hline
$1$ & $-a a' \ln(\tfrac{4 \mu^2 a a'}{H^2}) H^2 [\partial^2 \!+\! 2 a a' H^2]$ 
& $\tfrac12$ \\
\hline
$2$ & $a a' \ln(\tfrac{4 \mu^2 a a'}{H^2}) H^2 [\partial^2 \!+\! 2 a a' H^2]$ 
& $-1$ \\
\hline
$3$ & $12 a^4 \ln(\tfrac{2 \mu a}{H}) H^4 \!+\! 5 a^4 H^4$ & $0$ \\
\hline
$5$ & $-8 a^3 \ln(\tfrac{2 \mu a}{H}) H^3 \!-\! 2 a^3 H^3 $ & $0$ \\
\hline
$7$ & $2 a^2 \ln(\tfrac{2 \mu a}{H}) H^2$ & $0$ \\
\hline
$9$ & $-2 (a a')^2 \ln(\tfrac{4 \mu^2 a a'}{H^2}) H^4 \!-\! 4 (a a')^2 H^4$ & $0$ \\
\hline
$11$ & $4 a^3 \ln(\tfrac{2 \mu a}{H}) H^3$ & $0$ \\
\hline
$12$ & $-2 a a' \ln(\tfrac{4 \mu^2 a a'}{H^2}) H^2$ & $0$ \\
\hline
\end{tabular}
\caption{\footnotesize Coefficients of $[\mbox{}^{\mu\nu} D_i^{\rho\sigma}]
i \delta^4(x \!-\! x')$ from the finite part of $\mathcal{L}_{1b}$ and from
$\mathcal{L}_{1c}$. The contributions from $\mathcal{L}_{1b}$ and 
$\mathcal{L}_{1c}$ are multiplied by $\tfrac{13 \kappa^2}{2^5 \cdot 3 \cdot 
\pi^2}$ and $\tfrac{13 \kappa^2}{2^5 \cdot \pi^2}$, respectively.}
\label{InvFin}
\end{table}

\begin{enumerate*}
\item{Whereas the dimensional constant of the primitive divergences 
$\mathcal{C}_4$ and $\mathcal{C}_3$ is $(H/2)^{D-4}$, it is $\mu^{D-4}$ in
the counterterms; and}
\item{Whereas the primitive divergences involve only integer powers of the
scale factor, the counterterms each contain a relative factor of $a^{D-4}$.}
\end{enumerate*}
This suggests that we extract the finite factors of $\ln(2 \mu a/H)$,
which constitute a very part important of the final result. For (\ref{DL1a}) 
this process gives,
\begin{eqnarray}
\lefteqn{ \tfrac{i \delta^2 \Delta S_{1a}}{\delta h_{\mu\nu}(x)
\delta h_{\rho\sigma}(x')} \Bigr\vert_{\rm dS} \longrightarrow 
2 (\tfrac{H}{2\mu})^{D-4} c_1 \kappa^2 \mathcal{F}^{\mu\nu} \!\times\! 
{\mathcal{F}'}^{\rho\sigma} i \delta^D(x \!-\! x') } \nonumber \\
& & \hspace{4.5cm} + \tfrac{13 \kappa^2}{2^6 \cdot 3^2 \cdot \pi^2}
\mathcal{F}^{\mu\nu} \!\times\! {\mathcal{F}'}^{\rho\sigma} \Bigl\{
\ln(\tfrac{4 \mu^2 a a'}{H^2}) i \delta^4(x \!-\! x') \Bigr\} . \qquad
\label{DL1aReorg}
\end{eqnarray}
Extracting the scale factors from the divergent part of (\ref{DL2}) allows
us to perform the contractions on $\alpha$, $\beta$, $\gamma$ and $\delta$
\cite{Park:2011ww,Leonard:2014zua},
\begin{eqnarray}
\lefteqn{ \tfrac{i \delta^2 \Delta S_{2}}{\delta h_{\mu\nu}(x)
\delta h_{\rho\sigma}(x')} \Bigr\vert_{\rm dS} \longrightarrow 
2 (\tfrac{D - 3}{D-2}) (\tfrac{H}{2 \mu})^{D-4} c_2 \Bigl\{
\mathcal{D}^{\mu (\rho} \mathcal{D}^{\sigma) \nu} - 
\tfrac{\mathcal{D}^{\mu\nu} \mathcal{D}^{\rho\sigma}}{D-1} \Bigr\} 
i \delta^D(x \!-\! x') } \nonumber \\
& & \hspace{3cm} + \tfrac{61 \kappa^2}{2^6 \cdot 3 \cdot 5 \cdot \pi^2}
\mathcal{C}^{\alpha\beta\gamma\delta \mu\nu} \!\times\!
{\mathcal{C}'}_{\alpha\beta\gamma\delta}^{~~~~~\rho\sigma} \Bigl\{
\ln(\tfrac{4 \mu^2 a a'}{H^2}) i \delta^4(x \!-\! x') \Bigr\} . \qquad
\label{DL2Reorg}
\end{eqnarray}
Making a similar reduction of the Einstein counterterm (\ref{DL1b}) is
complicated by the fact that it, even for $D=4$, it possesses scale factors
that must be moved to the left of any derivatives to put it in standard form.
The finite part is given is the left column of Table~\ref{InvFin}, and the
middle column of Table~\ref{InvCterms} gives its divergent part. The left
and right columns of Table~\ref{InvCterms} give the divergent 
contributions from $\mathcal{L}_{1a}$ and $\mathcal{L}_{2}$, respectively.
Rather than reducing the finite parts to standard form, it is preferable
to leave them in the forms given on the second lines of expressions 
(\ref{DL1aReorg}) and (\ref{DL2Reorg}).

\subsection{Noninvariant Counterterms}

Although the fact that our gauge (\ref{gauge}) breaks de Sitter invariance 
means that noninvariant counterterms are needed, we have still tried to 
build them from geometrical meaningful quantities such as curvatures and
connections. Table~{\ref{Counterterms} gives the complete list, including
the two invariant ones. In each case the coefficient consists of a divergent
part plus a finite remainder chosen to enforce the Ward identity (\ref{gravity}).
As with the invariant counterterms, each of the 17 noninvariant will give
rise to factors of $\ln(2 \mu a/H)$. 

\begin{table}[H]
\setlength{\tabcolsep}{8pt}
\def\arraystretch{1.5}
\centering
\begin{tabular}{|@{\hskip 1mm }c@{\hskip 1mm }||c|c|c|}
\hline
$i$ & $\mathcal{F}^{\mu\nu} \!\times\! {\mathcal{F}'}^{\rho\sigma}
i\delta^D(\Delta x)$ & $H^2 \mathcal{L}^{\mu\nu\rho\sigma} i\delta^D(\Delta x)$ 
& $\mathcal{C}^{\alpha\beta\gamma\delta\mu\nu} \!\times\! {\mathcal{C}'}_{
\alpha\beta\gamma\delta}^{~~~~~ \rho\sigma} i\delta^D(\Delta x)$ \\
\hline\hline
$1$ & $\mathcal{D}_1 \mathcal{D}_1'$ & $-a a' H^2 [\partial^2 \!+\! 
\tfrac14 a a' D (D\!-\!2) H^2]$ & $-\tfrac{(D-3)}{(D-1)(D-2)} \partial^4$ \\
\hline
$2$ & $0$ & $a a' H^2 [\partial^2 \!+\! \tfrac14 a a' D (D\!-\!2) H^2]$ 
& $\tfrac{(D-3)}{(D-1)(D-2)} \partial^4$ \\
\hline
$3$ & $-(D\!-\!2) (D\!-\!1) {a'}^2 H^2 \mathcal{D}_1$ & $\tfrac14 D (D\!-\!2) (a a')^2 H^4$ 
& $0$ \\
\hline
$5$ & $2 (D\!-\!1) a' H \mathcal{D}_1$ & $-(D\!-\!2) a^2 a' H^3$ & $0$ \\
\hline
$7$ & $-\mathcal{D}_1$ & $a a' H^2$ & $\tfrac{(D-3)}{(D-1)(D-2)} \partial^2$ \\
\hline
$9$ & $0$ & $\tfrac12 (D\!-\!2)^2 (a a')^2 H^4$ & $0$ \\
\hline
$11$ & $0$ & $(D\!-\!2) a^2 a' H^3$ & $0$ \\
\hline
$12$ & $0$ & $-2 a a' H^2$ & $-2 (\tfrac{D-3}{D-2}) \partial^2$ \\
\hline
$13$ & $(D\!-\!2)^2 (D\!-\!1)^2 (a a')^2 H^4$ & $0$ & $0$ \\
\hline
$14$ & $-2 (D\!-\!2) (D\!-\!1)^2 a^2 a' H^3$ & $0$ & $0$ \\
\hline
$16$ & $(D\!-\!2) (D\!-\!1) a^2 H^2$ & $0$ & $0$ \\
\hline
$18$ & $-4 (D\!-\!1)^2 a a' H^2$ & $0$ & $0$ \\
\hline
$19$ & $2 (D\!-\!1) a H$ & $0$ & $0$ \\
\hline
$21$ & $1$ & $0$ & $(\tfrac{D-2}{D-1})$ \\
\hline
\end{tabular}
\caption{\footnotesize Coefficients of $[\mbox{}^{\mu\nu} D_i^{\rho\sigma}]
i \delta^D(x \!-\! x')$ in the counterterms. The leftmost is multiplied by 
$2 (\tfrac{H}{2\mu})^{D-4} c_1 \kappa^2$; the middle is multiplied by $D (D-1) 
(\tfrac{H}{2 \mu})^{D-4} c_1 \kappa^2$; and the rightmost is multiplied by $2 
(\tfrac{H}{2 \mu})^{D-4} c_2 \kappa^2$.}
\label{InvCterms}
\end{table}

\subsection{The Fully Renormalized Result}

The fully renormalized result takes the same form (\ref{genform}) as the
primitive result. In fact, the nonlocal contributions are exactly the
same as the primitive nonlocal contributions given in Tables~\ref{Primitive3ptFin}
and \ref{Primitive3ptFinSingleLog}.
The local contributions consist of the second lines of expressions 
(\ref{DL1aReorg}) and (\ref{DL2Reorg}), as well as Table~\ref{InvFin} and
\ref{RenormalizedLocal}.  

\begin{table}[H]
\setlength{\tabcolsep}{8pt}
\def\arraystretch{1.2}
\centering
\begin{tabular}{|@{\hskip 1mm }c@{\hskip 1mm }||c|c|}
\hline
$i$ & Counterterm & Expansion\ of\ $c_i$ \\
\hline\hline
$1$ & $c_1 \!\times\! R^2 \sqrt{-g}$ & $\tfrac{13}{2^4 \cdot 3^2} \!\times\! \mathcal{C}_{3} +
\tfrac{211}{2^8 \cdot 3^3 \cdot \pi^2}$ \\
\hline
$2$ & $c_2 \!\times\! C^{\mu\nu\rho\sigma} C_{\mu\nu\rho\sigma} \sqrt{-g}$ & 
$\tfrac{61}{2^4 \cdot 3 \cdot 5} \!\times\! \mathcal{C}_3 + \tfrac{2113}{2^8 \cdot 3^2 
\cdot 5^2 \cdot \pi^2}$ \\
\hline
$3$ & $c_3 \!\times\! a^{-3} H \kappa h^{\mu\nu}_{~~ ,\mu} R_{\nu 0} \sqrt{-g}$ & 
$-\tfrac{2}{3} \!\times\! \mathcal{C}_3 + \tfrac{11}{2^4 \cdot 3^2 \cdot \pi^2}$ \\
\hline
$4$ & $c_4 \!\times\! a^{-1} H \kappa h_{,0} R \sqrt{-g}$ & $\mathcal{C}_3 + \tfrac{13}{2^5 
\cdot 3 \cdot \pi^2}$ \\
\hline
$5$ & $c_5 \!\times\! a^{-1} H \kappa h^{\mu}_{0 ,\mu} R \sqrt{-g}$ & $-\tfrac43 \!\times\!
\mathcal{C}_3 - \tfrac{85}{2^5 \cdot 3^3 \cdot \pi^2}$ \\
\hline
$6$ & $c_6 \!\times\! H^2 \kappa h_{00} R \sqrt{-g}$ & $\tfrac{23}{4} \!\times\! \mathcal{C}_3
- \tfrac12 \!\times\! \mathcal{C}_4 + \tfrac{337}{2^6 \cdot 3 \cdot \pi^2}$ \\
\hline
$7$ & $c_7 \!\times\! a^{-2} H^2 \kappa h^{\mu\nu} R_{\mu\nu} \sqrt{-g}$ & $-\tfrac73 
\!\times\! \mathcal{C}_3 - 2 \!\times\! \mathcal{C}_4 + \tfrac{155}{2^4 \cdot
3^2 \cdot \pi^2}$ \\
\hline
$8$ & $c_8 \!\times\! a^{-2} H^2 \kappa h R_{00} \sqrt{-g}$ & $-\tfrac13 \!\times\! 
\mathcal{C}_3 - \mathcal{C}_4 + \tfrac{203}{2^5 \cdot 3^2 \cdot \pi^2}$ \\
\hline
$9$ & $c_{9} \!\times\! a^{-4} H^2 \kappa h^{\mu\nu} R_{0\mu 0\nu} \sqrt{-g}$ & $-\tfrac{89}{96}
\!\times\! \mathcal{C}_3 + \mathcal{C}_4 - \tfrac{3259}{2^9 \cdot 3^2 \cdot \pi^2}$ \\
\hline
$10$ & $c_{10} \!\times\! H^2 \kappa h R \sqrt{-g}$ & $\tfrac12 \!\times\! \mathcal{C}_3
- \tfrac{155}{2^9 \cdot \pi^2}$ \\
\hline
$11$ & $c_{11} \!\times\! a^{-2} H^2 \kappa h^{\mu\nu} R_{\mu\nu} \sqrt{-g}$ & $-\tfrac13
\!\times\! \mathcal{C}_3 + \tfrac{173}{2^5 \cdot 3^2 \cdot \pi^2}$ \\
\hline
$12$ & $c_{12} \!\times\! a^{-1} H^3 \kappa h_{00} \eta^{\alpha\beta} \Gamma^0_{~\alpha\beta}
\sqrt{-g}$ & $-\tfrac{13}{3} \!\times\! \mathcal{C}_3 + 4 \!\times\! \mathcal{C}_4
- \tfrac{37}{2^3 \cdot 3^2 \cdot \pi^2}$ \\
\hline
$13$ & $c_{13} \!\times\! a^{-1} H^3 \kappa h \eta^{\alpha\beta} \Gamma^0_{~\alpha\beta} 
\sqrt{-g}$ & $-\tfrac{59}{6} \!\times\! \mathcal{C}_3 + \mathcal{C}_4 + \tfrac{553}{2^8 \cdot
3^2 \cdot \pi^2}$ \\
\hline
$14$ & $c_{14} \!\times\! a^{-1} H^3 \kappa h^{\alpha\beta} \Gamma^0_{~\alpha\beta} \sqrt{-g}$ 
& $\tfrac{17}{6} \!\times\! \mathcal{C}_3 - \mathcal{C}_4 - \tfrac{29}{2 \cdot 3
\cdot \pi^2}$ \\
\hline
$15$ & $c_{15} \!\times\! a^{-1} H^3 \kappa h_{00} \Gamma^0_{~00} \sqrt{-g}$ & 
$\tfrac{14}{3} \!\times\! \mathcal{C}_3 - 4 \!\times\! \mathcal{C}_4 
-\tfrac{59}{2^3 \cdot 3^2 \cdot \pi^2}$ \\
\hline
$16$ & $c_{16} \!\times\! a^{-1} H^3 \kappa h \Gamma^{\alpha}_{~\alpha 0} \sqrt{-g}$ 
& $\tfrac{571}{45} \!\times\! \mathcal{C}_3 - \tfrac{82}{15} \!\times \mathcal{C}_4
+ \tfrac{540923}{2^5 \cdot 3^3 \cdot 5^2 \cdot \pi^2}$ \\
\hline
$17$ & $c_{17} \!\times\! a^{-1} H^3 \kappa h^{\mu}_{~\nu} \Gamma^{\nu}_{~\mu 0} \sqrt{-g}$ 
& $\tfrac{14}{9} \!\times\! \mathcal{C}_3 - \tfrac83 \!\times\! \mathcal{C}_4
\tfrac{107}{2 \cdot 3^3 \cdot \pi^2}$ \\
\hline
$18$ & $c_{18} \!\times\! a^{-1} H^3 \kappa h_{\mu 0} (\Gamma^{\mu}_{~00} - \eta^{\mu\nu}
\Gamma^0_{~\nu 0}) \sqrt{-g}$ & $\tfrac{16}{9} \!\times\! \mathcal{C}_3 + \tfrac82
\!\times\! \mathcal{C}_4 + \tfrac{419}{2^4 \cdot 3^3 \cdot \pi^2}$ \\
\hline
$19$ & $c_{19} \!\times\! a^{-1} H^3 \kappa (h_{00} \Gamma^{\alpha}_{~\alpha 0} + h 
\Gamma^0_{~00}) \sqrt{-g}$ & $\tfrac{128951}{2^7 \cdot 3^2} \!\times\! \mathcal{C}_3
\!-\! \tfrac{34}{3} \!\times\! \mathcal{C}_4 \!-\! \tfrac{7^2 \cdot 41 \cdot 1879}{2^{11}
\cdot 3^3 \cdot \pi^2}$ \\
\hline
\end{tabular}
\caption{\footnotesize Counterterms and their coefficients. The divergent constants
are $\mathcal{C}_3 = \tfrac{\mu^{D-4}}{4 \pi^{\frac{D}2}} \tfrac{\Gamma(\frac{D}2-1)}{
(D-3) (D-4)}$ and $\mathcal{C}_4 = \tfrac{\mu^{D-4}}{16 \pi^{\frac{D}2}} 
\tfrac{\Gamma(D-1)}{\Gamma(\frac{D}2)} \pi {\rm cot}(\tfrac{D \pi}{2})$.}
\label{Counterterms}
\end{table}

\section{Solving the Effective Field Equations}

The purpose of this section is to solve the effective field equation
(\ref{linEin}) for 1-loop corrections to the graviton mode function
(\ref{radiation}), and for the Newtonian potential (\ref{Newtonian}),
in our gauge (\ref{gauge}). The section begins by explaining how to 
convert the in-out graviton self-energy computed in the previous 
two sections to the in-in result required to make the effective field
equation real and causal. We also comment the form the effective
field equation must take when working in a fixed gauge, and on how to
solve it perturbatively. The next topic is solving for the 1-loop
correction to the graviton mode function. The section closes with a
discussion of the most significant 1-loop corrections to the
Newtonian potential at late times. 

\begin{table}[H]
\setlength{\tabcolsep}{8pt}
\def\arraystretch{1.5}
\centering
\begin{tabular}{|@{\hskip 1mm }c@{\hskip 1mm }||c|}
\hline
$i$ & Coefficient\ of\ $ \frac{\kappa^2}{\pi^2} [\mbox{}^{\mu\nu} D_i^{\rho\sigma}] \,
i\delta^4(x \!-\! x')$ \\
\hline\hline
$1$ & $-\frac{\partial^4}{96}-\frac{1}{3} a  a' H^2 \ln \left(\frac{4 a a' \mu ^2}{
H^2}\right) \partial_0^2-\frac{23203 a  a' H^2}{3240}\partial_0^2-\frac{368551 a a' 
H^2}{103680}\partial^2 $\\ & $ +\frac{73}{96} a ^2 a'^2 H^4 \ln \left(\frac{4 a a'\mu^2}{H^2}
\right)-\frac{733081 a ^2  a'^2H^4}{41472}$ \\
\hline
$2$ & $-\frac{1}{48} a a' H^2 \ln \left(\frac{4 a a'\mu ^2}{H^2}\right)
\partial_0^2-\frac{79 a  a' H^2 }{2304}\partial_0^2+\frac{1}{24} a  a'H^2 
\ln\left(\frac{4 a a'\mu ^2}{H^2}\right)\partial^2-\frac{143}{192} a a'H^2\partial^2$ \\ 
$$ & $+\frac{1}{8} a ^2  a'^2 H^4\ln \left(\frac{4 a a'\mu ^2}{H^2}\right)+\frac{2311 
a ^2 a'^2H^4 }{4608}$ \\
\hline
$3$ & $-\frac{19}{24} a ^3 H^3  \ln \left(\frac{2 a  \mu }{H}\right)\partial_0 + 
\frac{550961 a ^3 H^3 }{138240}\partial_0-\frac{3}{32} a ^2 H^2 \partial_0^2 - 
\frac{55}{48} a ^2  H^2 \ln \left(\frac{2 a  \mu }{H}\right)\partial^2-\frac{14857 
a^2 H^2}{9216}\partial^2$ \\ 
$$ & $ -\frac{69}{8} a ^4 H^4 \ln \left(\frac{2 a \mu }{H}\right) + 
\frac{268889 a^4 H^4}{5120}$ \\
\hline
$5$ & $\frac{23}{24} a ^2 H^2 \ln \left(\frac{2 a  \mu }{H}\right) \partial_0 +
\frac{43021 a ^2 H^2 }{11520}\partial_0+\frac{1}{3} a H \ln \left(\frac{2 a  \mu }{H}\right)
\partial^2+\frac{49}{24} a ^3 H^3 \ln \left(\frac{2 a  \mu }{H}\right)-\frac{128203 
a^3 H^3}{11520}$ \\
\hline
$7$ & $-\frac{1}{6} a  H \ln \left(\frac{2 a  \mu }{H}\right) \partial_0 + 
\frac{70943 a H }{51840}\partial_0+\frac{406487 a ^2 H^2}{207360}$ \\
\hline
$9$ & $\frac{3}{4} a  a' H^2 \partial_0^2-\frac{11}{24} a a'H^2  
\ln\left(\frac{4 a a' \mu ^2}{H^2}\right) \partial^2-\frac{881 a a' H^2}{3456} 
\partial^2 - \frac{25}{24} a ^2 {a'}^2 H^4\ln \left(\frac{4 a a' \mu ^2 }{H^2}\right) 
- \frac{2249 a ^2 a'^2 H^4}{1728}$ \\
\hline
$11$ & $\frac{7}{12} a ^2 H^2 \ln \left(\frac{2 a  \mu }{H}\right) \partial_0 + 
\frac{2035 a^2 H^2 }{3456}\partial_0-\frac{1}{12} a H \ln \left(\frac{2 a \mu }{H}\right)
\partial^2 + \frac{5}{4} a ^3 H^3 \ln \left(\frac{2 a \mu }{H}\right) + 
\frac{7039 a^3 H^3}{6912}$ \\
\hline
$12$ & $\frac{19}{12} a a' H^2 -\frac{1}{8} a   a'H^2 
\ln\left(\frac{4 a a'\mu ^2}{H^2}\right)$ \\
\hline
$13$ & $\frac{3}{16} a a' H^2 \partial^2+\frac{335}{24} a ^2 a'^2 H^4
\ln\left(\frac{4 a a' \mu ^2}{H^2}\right)-\frac{517999 a^2 a'^2 H^4 }{3456}$ \\
\hline
$14$ & $-\frac{3}{8} a ^2 H^2 \partial_0-\frac{31}{4} a^3 H^3 
\ln\left(\frac{2 a \mu}{H}\right)+\frac{168635 a^3 H^3}{4608}$ \\
\hline
$16$ & $\frac{7}{8} a ^2 H^2 \ln \left(\frac{2 a  \mu }{H}\right)-\frac{48877 
a^2 H^2}{9216}$ \\
\hline
$18$ & $\frac{1241 a a'H^2}{3456}-\frac{37}{24} a a'H^2 \ln 
\left(\frac{4 \mu ^2 a a'}{H^2}\right)$ \\
\hline
$19$ & $\frac{1}{4} a  H \ln \left(\frac{2 a  \mu }{H}\right)+\frac{a H}{16}$ \\
\hline
$21$ & $\frac{1}{96}$ \\
\hline
\end{tabular}
\caption{\footnotesize Renormalized local contributions to 
$-i [\mbox{}^{\mu\nu} \Sigma^{\rho\sigma}](x;x')$.}
\label{RenormalizedLocal}
\end{table} 

\subsection{The Effective Field Equation}

The graviton self-energy we have derived is technically the in-out 
result provided by the usual Feynman rules. That is perfectly valid
for asymptotic scattering problems, but it suffers from two problems
when the intention is to use the effective field equations to 
understand the evolution of quantum corrections at finite times:
\begin{itemize*}
\item{The in-out self-energy contains both real and imaginary parts;
and}
\item{The in-out self-energy fails to vanish when the point ${x'}^{\mu}$
is outside the past light-cone of $x^{\mu}$.}
\end{itemize*}
Both problems can be surmounted using the Schwinger-Keldysh formalism
\cite{Schwinger:1960qe,Mahanthappa:1962ex,Bakshi:1962dv,Bakshi:1963bn,
Keldysh:1964ud}. This is a diagrammatic formalism for giving true
expectation values, which almost as simple as the Feynman rules. The 
effective field equations of the Schwinger-Keldysh formalism are both
real and causal \cite{Chou:1984es,Jordan:1986ug,Calzetta:1986ey}.

There are a series of simple ruled for converting an in-out 1PI 
2-point function to the in-in formalism \cite{Ford:2004wc}:
\begin{enumerate*}
\item{Every endpoint carries a $\pm$ polarity, so that the conformal
coordinate interval $\Delta x^2(x;x')$ of the in-out formalism, which
was defined in expression (\ref{Deltaxsq}), gives rise to in-in four 
intervals,
\begin{eqnarray}
\Delta x^2_{++} &\!\!\! \equiv \!\!\!& \vert \vec{x} \!-\! \vec{x}'
\vert^2 - (\vert \eta \!-\! \eta'\vert \!-\! i \varepsilon)^2 \; ,
\qquad \label{Dx2++} \\
\Delta x^2_{+-} &\!\!\! \equiv \!\!\!& \vert \vec{x} \!-\! \vec{x}'
\vert^2 - (\eta \!-\! \eta' \!+\! i \varepsilon)^2 \; ,
\qquad \label{Dx2+-} \\
\Delta x^2_{-+} &\!\!\! \equiv \!\!\!& \vert \vec{x} \!-\! \vec{x}'
\vert^2 - (\eta \!-\! \eta' \!-\! i \varepsilon)^2 \; ,
\qquad \label{Dx2-+} \\
\Delta x^2_{--} &\!\!\! \equiv \!\!\!& \vert \vec{x} \!-\! \vec{x}'
\vert^2 - (\vert \eta \!-\! \eta'\vert \!+\! i \varepsilon)^2 \; .
\qquad \label{Dx2--} 
\end{eqnarray}}
\item{Similarly, the 1PI 2-point function of the in-out formalism
gives to four in-in 1PI 2-point functions, each involving the 
appropriate interval.}
\item{Only the $++$ and $--$ 1PI 2-point functions contain terms
involving $\delta^4(x - x')$.}
\item{The 1PI 2-point function has a factor of $-1$ for each $-$
polarity.}
\item{The graviton self-energy that goes in the the Schwinger-Keldysh
effective field equation is the sum of the $++$ and $+-$ ones.} 
\end{enumerate*}
From rule 4 we see that there is no change at all to the local 
contributions of Tables~\ref{InvFin} and \ref{RenormalizedLocal},
or to the important logarithms on the second lines of expressions
(\ref{DL1aReorg}) and (\ref{DL2Reorg}). The only change comes to 
the nonlocal contributions of Tables~\ref{Primitive3ptFin} and
\ref{Primitive3ptFinSingleLog}, and the net effect is to replace the 
two logarithms with,
\begin{eqnarray}
\ln(\tfrac14 H^2 \Delta x^2) &\!\!\! \longrightarrow \!\!\!&
2\pi i \theta(\Delta \eta \!-\! \Delta r) \equiv f_3^{\rm SK}(x \!-\! x')
\; , \qquad \label{Log1} \\
\ln^2(\tfrac14 H^2 \Delta x^2) &\!\!\! \longrightarrow \!\!\!&
4\pi i \theta(\Delta \eta \!-\! \Delta r) \ln[\tfrac14 H^2 (\Delta 
\eta^2 \!-\! \Delta r^2)] \equiv f_2^{\rm SK}(x \!-\! x') \; , \qquad 
\label{Log2}
\end{eqnarray}
where we define the temporal and spatial intervals as,
\begin{equation}
\Delta \eta \equiv \eta \!-\! \eta' \qquad , \qquad \Delta r \equiv
\Vert \vec{x} \!-\! \vec{x}' \Vert \; . \label{intervals}
\end{equation}
The reality and causality of the Schwinger-Keldysh effective field
equation are now manifest. We might define $f_1^{SK}(x \!-\! x') \equiv
i \delta^4(x \!-\! x')$.

Two more issues require discussion concerning the effective field equation
(\ref{linEin}). The first issue derives from the fact that, right now, we
only possess the 1-loop correction to the graviton self-energy. This means 
that the effective field equation must be solved perturbatively,
\begin{equation}
h_{\mu\nu} = h_{\mu\nu}^{(0)} + h_{\mu\nu}^{(1)} + h_{\mu\nu}^{(2)} +
\dots \label{pertexp}
\end{equation}
Here $h_{\mu\nu}^{(0)}$ is the ``classical'' field, which obeys,
\begin{equation}
\mathcal{D}^{\mu\nu\rho\sigma} \kappa h_{\rho\sigma}^{(0)} = 8\pi G
T^{\mu\nu}_{\rm lin} \; . \label{0thorder}
\end{equation}
The 1-loop corrections we seek obey,
\begin{equation}
\mathcal{D}^{\mu\nu\rho\sigma} \kappa h_{\rho\sigma}^{(1)}(x) = 
\int \!\! d^4x' \, [\mbox{}^{\mu\nu} \Sigma^{\rho\sigma}](x;x')
\kappa h_{\rho\sigma}^{(0)}(x') \; . \label{1storder}
\end{equation}

\begin{sidewaystable}[p]
\setlength{\tabcolsep}{8pt}
\def\arraystretch{1.3}
\centering
\begin{tabular}{|@{\hskip 1mm }c@{\hskip 1mm }||c|c|c|}
\hline
$i$ & $[\mbox{}^{00} D_i^{\rho\sigma}] h_{\rho\sigma}(x')$ 
& $[\mbox{}^{0k} D_i^{\rho\sigma}] h_{\rho\sigma}(x')$ 
& $[\mbox{}^{k\ell} D_i^{\rho\sigma}] h_{\rho\sigma}(x')$ \\
\hline\hline
$1$ & $-h$ & $0$ & $\delta_{k\ell} h$ \\
\hline
$2$ & $h_{00}$ & $h_{0k}$ & $h_{k\ell}$ \\
\hline
$3$ & $-h_{00}$ & $0$ & $\delta_{k\ell} h_{00}$ \\
\hline
$4$ & $h$ & $0$ & $0$ \\
\hline
$5$ & $\partial_0 \!\times\! h_{00} \!-\! h_{0m,m}$ & $0$ 
& $\delta_{k\ell} [h_{0m,m} \!-\! \partial_0 \!\times\! h_{00}]$ \\
\hline
$6$ & $-\partial_0 \!\times\! h$ & $\tfrac12 \partial_k \!\times\! h$ 
& $0$ \\
\hline
$7$ & $-\partial_0^2 \!\times\! h_{00} \!+\! 2 \partial_0 \!\times\! h_{0m,m} \!-\! h_{mn,mn}$ 
& $0$ & $\delta_{k\ell} [\partial_0^2 \!\times\! h_{00} \!\!-\! 2 \partial_0 \!\times\!
h_{0m,m} \!\!+\! h_{mn,mn}]$ \\
\hline
$8$ & $\partial_0^2 \!\times\! h$ & $-\partial_0 \partial_k \!\times\! h$ 
& $\partial_k \partial_{\ell} h$ \\
\hline
$9$ & $-h_{00}$ & $\tfrac12 h_{0k}$ & $0$ \\
\hline
$10$ & $\partial_0 \!\times\! h_{00} \!-\! h_{0m,m}$ & $\tfrac12 h_{km,m}
\!-\! \tfrac12 \partial_0 \times\! h_{0k}$ & $0$ \\
\hline
$11$ & $\partial_0 \!\times\! h_{00}$ & $-\tfrac12 \partial_k \!\times\! h_{00}
\!-\! \tfrac12 \partial_0 \!\times\! h_{0k}$ 
& $\partial_{(k} \!\times\! h_{\ell) 0}$ \\
\hline
$12$ & $\partial_0 [h_{0m,m} \!-\! \partial_0 \!\times\! h_{00}]$ 
& $\tfrac12 \partial_k [\partial_0 \!\times\! h_{00} \!-\! h_{0m,m}]$ &
$\partial_0 \partial_{(k} \!\times\! h_{\ell) 0}$ \\
$$ & $$ & $+ \tfrac12 \partial_0 [\partial_0 \!\times\! h_{0k} \!-\! h_{km,m}]$
& $+ \partial_{(k} \!\times\! h_{\ell) m,m}$ \\ 
\hline
$13$ & $h_{00}$ & $0$ & $0$ \\
\hline
$14$ & $-\partial_0 \!\times\! h_{00} \!+\! h_{0m,m}$ & $0$ & $0$ \\
\hline
$15$ & $-\partial_0 \!\times\! h_{00}$ & $\tfrac12 \partial_k \!\times\! h_{00}$ & $0$ \\
\hline
$16$ & $\partial_0^2 \!\times\! h_{00} \!-\! 2 \partial_0 \!\times\! h_{0m,m} \!+\! h_{mn,mn}$ 
& $0$ & $0$ \\
\hline
$17$ & $\partial_0^2 \!\times\! h_{00}$ & $-\partial_0 \partial_k h_{00}$ 
& $\partial_k \partial_{\ell} h_{00}$ \\
\hline
$18$ & $\partial_0 [\partial_0 \!\times\! h_{00} \!-\! h_{0m,m}]$ 
& $-\tfrac12 \partial_k [\partial_0 \!\times\! h_{00} \!-\! h_{0m,m}]$ & $0$ \\
\hline
$19$ & $\!\!-\partial_0 [\partial^2 \!\times\! h_{00} \!\!-\! 2 \partial_0 \!\times\! h_{0m,m}
\!\!+\! h_{mn,mn}]\!\!$ & $\tfrac12 \partial_k [\partial_0^2 \!\times\! h_{00} \!-\! 2
\partial_0 h_{0m,m} \!+\! h_{mn,mn}]$ & $0$ \\
\hline
$20$ & $-\partial_0^2 [\partial_0 \!\times\! h_{00} \!-\! h_{0m,m}]$ 
& $\partial_0 \partial_k [\partial_0 \!\times\! h_{00} \!-\! h_{0m,m}]$ & 
$\partial_k \partial_{\ell} [h_{0m,m} \!-\! \partial_0 \!\times\! h_{00}]$ \\
\hline
$21$ & $\partial_0^2 [\partial^2 \!\times\! h_{00} \!\!-\! 2 \partial_0 h_{0m,m} \!\!+\! 
h_{mn,mn}]$ & $\!-\partial_0 \partial_k [\partial_0^2 \!\times\! h_{00} \!\!-\! 2 
\partial_0 \!\times\! h_{0m,m} \!\!+\! h_{mn,mn}]\!$ & $\!\partial_k \partial_{\ell}
[\partial_0^2 \!\times\! h_{00} \!\!-\! 2 \partial_0 \!\times\! h_{0m,m} \!\!+\!
h_{mn,mn}]\!$ \\
\hline
\end{tabular}
\caption{\footnotesize Contraction of $[\mbox{}^{\mu\nu} D_i^{\rho\sigma}]$ into 
$h_{\rho\sigma}(x')$, with some spatial derivatives reflected and partially 
integrated onto $h_{\rho\sigma}(x')$.}
\label{Dtimesh}
\end{sidewaystable}

Most of the graviton self-energy is contained in Tables~\ref{Primitive3ptFin},
\ref{Primitive3ptFinSingleLog}, \ref{InvFin}, and \ref{RenormalizedLocal}. 
These contributions consist of a sum (\ref{genform}) of terms of the form,
\begin{equation}
C_{ij}(a,a',\partial) \!\times\! [\mbox{}^{\mu\nu} D_i^{\rho\sigma}] \!\times\!
f^{\rm SK}_j(x \!-\! x') \; . \label{genterm}
\end{equation}
In Table~\ref{Dtimesh} we give the result of contracting the basis tenors
$[\mbox{}^{\mu\nu} D_i^{\rho\sigma}]$, given in Table~\ref{Tbasis}, into 
$h_{\rho\sigma}(x')$. In some cases we have used the larger form, of an 
integral of (\ref{genterm}), to reflect spatial derivatives in
$[\mbox{}^{\mu\nu} D_i^{\rho\sigma}]$ from $+\partial_{\mu}$ to 
$-\partial'_{\mu}$, and then partially integrating them to act on the 
graviton field $h_{\rho\sigma}(x')$. The remaining contributions to $-i 
[\mbox{}^{\mu\nu} \Sigma^{\rho\sigma}](x;x')$ come from the second lines 
of expressions (\ref{DL1aReorg}) and (\ref{DL2Reorg}). The results can
be expressed in terms of the zeroth order, linearized curvatures
\cite{Miao:2024atw},
\begin{eqnarray}
\lefteqn{ \tfrac{13 \kappa^2}{2^6 \cdot 3^2 \cdot \pi^2} 
\mathcal{F}^{\mu\nu} \!\! \int \!\! d^4x' \, \ln(\tfrac{4 \mu^2 a a'}{H^2}) 
\delta^4(x \!-\! x') \!\times\! {\mathcal{F}'}^{\rho\sigma} \kappa 
h_{\rho\sigma}(x') } \nonumber \\
& & \hspace{6cm} = \tfrac{13 \kappa^2}{2^5 \cdot 3^2 \cdot \pi^2} 
\mathcal{F}^{\mu\nu} \Bigl\{ \ln(\tfrac{2\mu a}{\mu}) R^{(0)}(x)\Bigr\} ,
\qquad \label{Flog} \\
\lefteqn{ \tfrac{61 \kappa^2}{2^6 \cdot 3 \cdot 5 \cdot \pi^2} \, 
\mathcal{C}^{\alpha\beta\gamma\delta\mu\nu} \!\! \int \!\! d^4x' \, 
\ln(\tfrac{4 \mu^2 a a'}{H^2}) \delta^4(x \!-\! x') \!\times\!
{\mathcal{C}'}_{\alpha\beta\gamma\delta}^{~~~~~\rho\sigma} \kappa
h_{\rho\sigma}(x') } \nonumber \\
& & \hspace{6cm} = -\tfrac{61 \kappa^2 \partial_{\rho} \partial_{\sigma}}{
2^4 \cdot 3 \cdot 5 \cdot \pi^2} \, \Bigl\{\! \ln(\tfrac{2 \mu a}{H})
C^{(0) \, \rho\mu\sigma\nu}(x) \!\Bigr\} . \qquad \label{Clog}
\end{eqnarray}

The final issue derives from how the Ward operator (\ref{Warddef}) acts on 
the three terms in the effective field equation. It annihilates both the 
invariant kinetic operator and the linearized stress tensor,
\begin{equation}
\mathcal{W}^{\mu}_{~\alpha\beta} \!\times\! \mathcal{D}_{\rm inv}^{\alpha\beta
\rho\sigma} = 0 = \mathcal{W}^{\mu}_{~\alpha\beta} \!\times\! 
T^{\alpha\beta}_{\rm lin} \; . \label{WardEqn}
\end{equation}
If we use the kinetic operator of the invariant theory, then consistency
of the equation requires that the Ward operator {\it also} annihilate the
self-energy. That is indeed the case for matter contributions to the graviton
self-energy (\ref{matter}), and one employs the invariant kinetic operator in
solving for how these matter corrections affect gravity \cite{Park:2011ww,
Wang:2015eaa,Miao:2024atw,Foraci:2024vng,Foraci:2024cwi}. Because the Ward
operator annihilates all three terms of the effective field equation, there
are four relations among the ten tensor components, leading to only six
independent equations. 

In contrast, graviton loop corrections to $-i [\mbox{}^{\mu\nu}
\Sigma^{\rho\sigma}](x;x')$ are only annihilated when the Ward operator acts
simultaneously on both coordinates (\ref{gravity}). Were we to use the invariant
kinetic operator, equation (\ref{1storder}) would not be consistent because 
acting the Ward operator on the left hand side would give zero, whereas the
right hand side would not vanish. Consistency is restored by employing the 
gauge-fixed kinetic operator, whose $3+1$ decomposition is,
\begin{eqnarray}
\mathcal{D}^{00\rho\sigma} h_{\rho\sigma} &\!\!\! = \!\!\!& \tfrac14
D_A (h_{00} + h_{kk}) - 2 a^4 H^2 h_{00} \; , \qquad \label{D00} \\
\mathcal{D}^{0i\rho\sigma} h_{\rho\sigma} &\!\!\! = \!\!\!&
-\tfrac12 D_B h_{0i} \; , \qquad \label{D0i} \\
\mathcal{D}^{ij\rho\sigma} h_{\rho\sigma} &\!\!\! = \!\!\!&
\tfrac12 D_A \Bigl[ h_{ij} + \tfrac12 \delta_{ij} (h_{00} - h_{kk}) \Bigr] \; .
\qquad \label{Dij}  
\end{eqnarray}
The operators $D_A$ and $D_B$ are,
\begin{equation}
D_A \equiv - a^2[\partial_0^2 + 2 a H \partial_0 - \nabla^2] \qquad , \qquad
D_B \equiv D_A - 2 a^4 H^2 \; . \label{DADBnew}
\end{equation}
They are, respectively, the kinetic operators for a massless, minimally 
coupled scalar ($D_A$) and for a massless, conformally coupled scalar
($D_B$). Note that all ten tensor components of the gauge-fixed effective 
field equation are independent.

\subsection{The Graviton Mode Function}

Gravitational radiation corresponds to $T^{\mu\nu}_{\rm lin} = 0$ and a 
graviton field strength of the form,
\begin{equation}
\kappa h_{\rho\sigma}(x) = \epsilon_{\rho\sigma}(\vec{k}) \!\times\! u(t,k) 
e^{i \vec{k} \cdot \vec{x}} \; , \label{gravwave}
\end{equation}
where the polarization tensor is purely spatial,  
transverse and traceless,
\begin{equation}
\epsilon_{0\mu}(\vec{k}) = 0 \qquad , \qquad k_i \epsilon_{ij}(\vec{k}) = 
0 = \epsilon_{ii}(\vec{k}) \; . \label{epsilon}
\end{equation}
Acting the gauge-fixed kinetic operator on (\ref{gravwave}) gives,
\begin{equation}
\mathcal{D}^{\mu\nu\rho\sigma} \kappa h_{\rho\sigma}(x) = 
\epsilon^{\mu\nu}(\vec{k}) e^{i \vec{k} \cdot \vec{x}} \times -\tfrac{a^2}{2}
[\partial_0^2 \!+\! 2 a H \partial_0 \!+\! k^2] u(t,k) \; . \label{lefthand}
\end{equation}
Of course the mode function is expanded perturbatively,
\begin{equation}
u(t,k) = u^{(0)}(t,k) + u^{(1)}(t,k) + u^{(2)}(t,k) + \dots \label{expand}
\end{equation}
The zeroth order mode function depends on time only through the scale factor
$a(t)$,
\begin{equation}
u^{(0)}(t,k) = \tfrac{H}{\sqrt{2 k^3}} [1 - \tfrac{i k}{a H}] 
\exp[\tfrac{i k}{a H}] \; . \label{uzero}
\end{equation}

We turn now to the right hand side of equation (\ref{1storder}). Examination 
of Table~\ref{Dtimesh} shows that only the $i=2$ term contributes for 
gravitational radiation,
\begin{eqnarray}
\lefteqn{i \!\! \int \!\! d^4x' \sum_{i=1}^{21} \sum_{j=1}^3 C_{ij}(a,a',\partial)
[\mbox{}^{\mu\nu} D_i^{\rho\sigma}] \!\times\! f_j^{\rm SK}(x \!-\! x') 
\!\times\! \kappa h^{(0)}_{\rho\sigma}(x') } \nonumber \\
& & \hspace{0cm} = i \!\! \int \!\! d^4x' \sum_{j=1}^3 C_{2j}(a,a',\partial)
\!\times\! f_j^{\rm SK}(x \!-\! x') \!\times\! \kappa h^{(0)\, \mu\nu}(x')
\; , \qquad \\
& & \hspace{0cm} = \epsilon^{\mu\nu}(\vec{k}) e^{i \vec{k} \cdot \vec{x}} \!
\times\!\!\int \!\! d^4x' \!\sum_{j=1}^{3} \!C_{2j}(a,a',\partial) \!\times\! 
i f_j^{\rm SK}(x \!-\! x') \!\times\! u^{(0)}(t',k) e^{-i \vec{k} \cdot 
\Delta \vec{x}} \! . \qquad \label{radsum} 
\end{eqnarray}
The linearized Ricci scalar (\ref{Flog}) vanishes for gravitational radiation,
and the contribution from the Weyl tensor (\ref{Clog}) is \cite{Miao:2024atw},
\begin{equation}
-\tfrac{61 \kappa^2 \partial_{\rho} \partial_{\sigma}}{
2^4 \cdot 3 \cdot 5 \cdot \pi^2} \, \Bigl\{ \ln(\tfrac{2 \mu a}{H})
C^{(0) \, \rho\mu\sigma\nu}(x) \Bigr\} = \tfrac{61 \kappa^2 
\epsilon^{\mu\nu}(\vec{k}) e^{i \vec{k} \cdot \vec{x}}}{2^5 \cdot 3 \cdot 5 
\cdot \pi^2} \, i k a^2 H^2 \partial_0 u^{(0)}(t,k) \; . \label{Cmode}
\end{equation}
Because both sides of equation (\ref{1storder}) are proportional to 
$\epsilon^{\mu\nu}(\vec{k}) e^{i \vec{k} \cdot \vec{x}}$, we can drop this
factor and write a scalar equation for the 1-loop correction to the mode
function,
\begin{eqnarray}
\lefteqn{ -\tfrac{a^2}{2} [\partial_0^2 \!+\! 2 a H \partial_0 \!+\! k^2] 
u^{(1)}(t,k) = \tfrac{61 \kappa^2}{2^5 \cdot 3 \cdot 5 \cdot \pi^2} \, i k
a^2 H^2 \partial_0 u^{(0)}(t,k) } \nonumber \\
& & \hspace{2cm} + \int \!\! d^4x' \sum_{j=1}^{3} C_{2j}(a,a',\partial) 
\!\times\! i f_j^{\rm SK}(x \!-\! x') \!\times\! u^{(0)}(t',k) e^{-i \vec{k} 
\cdot \Delta \vec{x}} \! . \qquad \label{modeeqn} 
\end{eqnarray}

At this point it is useful to recall the earlier computation of $u^{(1)}(t,k)$
\cite{Tan:2021lza} that led to the double log enhancement (\ref{guessu}).
That analysis employed the $D=4$ result for the graviton self-energy 
\cite{Tsamis:1996qk} that agrees with the nonlocal contributions we have
obtained in Tables~\ref{Primitive3ptFin} and \ref{Primitive3ptFinSingleLog}, 
but which was not capable of producing reliable local results such as those
we reported in Tables~\ref{InvFin} and \ref{RenormalizedLocal}, and the second 
lines of expressions (\ref{DL1aReorg}) and (\ref{DL2Reorg}). Local 
contributions were guessed in deriving the previous result for $u^{(1)}(t,k)$ 
\cite{Tan:2021lza}, and we can now check those guesses.

Note that a double log enhancement of $u^{(1)}(t,k)$ requires that the right 
hand side of (\ref{modeeqn}) grow like $a^4 \ln(a)$ at late times. This 
cannot come from the local term on the first line of (\ref{modeeqn})
because $\partial_0 u^{(0)}(t,k)$ falls off like $1/a$ at late times. Nor
can it come from the nonlocal contributions of Tables~\ref{Primitive3ptFin} 
and \ref{Primitive3ptFinSingleLog}, which are the same as for the earlier 
study, and were shown there to grow no faster than $a^4$ \cite{Tan:2021lza}.
This leaves only the local contributions from the $i=2$ rows of 
Tables~\ref{InvFin} and \ref{RenormalizedLocal}, and it is simple to show
that the result from each one grows no faster than $a^4$ at late times.
This means that the double log enhancement (\ref{guessu}) found in the
earlier calculation is wrong, and was the result of having incorrectly
inferred the local parts of the graviton self-energy.

\begin{table}[H]
\setlength{\tabcolsep}{8pt}
\def\arraystretch{1.5}
\centering
\begin{tabular}{|@{\hskip 1mm }c@{\hskip 1mm }||c|}
\hline
$i$ & Coeffiecient\ of\ $[\mbox{}^{\mu\nu} D_i^{\rho\sigma}] \!\times\!
i \delta^4(x \!-\! x')$ \\
\hline\hline
$1$ & $-\tfrac{52}{27} a a' H^2 \partial^2 - \tfrac{80}{27} a a' H^2 \partial_0^2
-\tfrac{187}{27} (a a')^2 H^4$ \\
\hline
$2$ & $-(a a')^2 H^4$ \\
\hline
$3$ & $\tfrac1{18} a^2 H^2 \partial^2 - \tfrac{29}{18} a^3 H^3 \partial_0
+ \tfrac{107}{6} a^4 H^4$ \\
\hline
$5$ & $\tfrac{16}{9} a^2 H^2 \partial_0 - \tfrac{46}{9} a^3 H^3$ \\
\hline
$7$ & $\tfrac{16}{27} a H \partial_0 + \tfrac{34}{27} a^2 H^2$ \\
\hline
$11$ & $-\tfrac23 a^3 H^3$ \\
\hline
$13$ & $-\tfrac{176}{3} a^4 H^4$ \\
\hline
$14$ & $15 a^3 H^3$ \\
\hline
$16$ & $-\tfrac{49}{18} a^2 H^2$ \\
\hline
\end{tabular}
\caption{\footnotesize Standard form (\ref{genform}) of the extra counterterm 
$\tfrac{i \delta^2 \Delta S_{\rm Extra}}{\delta h_{\mu\nu}(x) 
\delta h_{\rho\sigma}(x')} \vert_{\rm dS}$. Note that the cases of $i=4, 6,
8, 10, 15$ and $17$ follow from reflection (\ref{Crelations}-\ref{pmfactors}).}
\label{Extra}
\end{table}

Of course it is still interesting to see whether or not there is {\it any}
growth in the late time mode function. If the right hand side of (\ref{modeeqn})
went like $a^4$ at late times, that would induce a single logarithm enhancement.
This analysis is very messy because many terms contribute, so we give only the
final result,
\begin{equation}
 -\tfrac{a^2}{2} [\partial_0^2 \!+\! 2 a H \partial_0 \!+\! k^2] 
u^{(1)}(t,k) \longrightarrow -\tfrac{3041 \kappa^2 H^4 a^4}{2^9 \cdot 3 
\cdot \pi^2} \!\times\! \tfrac{H}{\sqrt{2 k^3}} \; . \label{latesource} 
\end{equation}
It follows that our predicted late time behavior for the mode function is
not (\ref{guessu}) but rather,
\begin{equation}
u(t,k) \longrightarrow \tfrac{H}{\sqrt{2 k^3}} \Bigl\{1 +
\tfrac{3041 \kappa^2 H^2}{2^8 \cdot 3^2 \cdot \pi^2} \!\times\!
\ln[a(t)] + \dots \Bigr\} \; . \label{lateuresult}
\end{equation}
   
Relation (\ref{lateuresult}) is not quite the end of the story because it
could be eliminated by what appears to be an allowable, finite renormalization.
The requirements on counterterms are homogeneity, isotropy, dilatation 
invariance, and the Ward Identity (\ref{gravity}). Each of those requirements
is satisfied by a Lagrangian $\Delta \mathcal{L}_{\rm Extra}$,
\begin{eqnarray}
\lefteqn{ \Delta \mathcal{L}_{\rm Extra} = \Bigl\{-\tfrac{H^2}2 h^{\alpha\beta}
h_{\alpha\beta} \!-\! \tfrac{2 H^3}{3 a} h^{\alpha\beta} h_{0\alpha , \beta}
\!-\! \tfrac{2 H^3}{3 a} [h \eta^{\alpha\beta} \!\!-\! 2 h_{00} (\eta^{\alpha\beta}
\!\!-\! 2 \delta^{\alpha}_0 \delta^{\beta}_0) ] \Gamma^0_{~\alpha\beta} }
\nonumber \\
& & \hspace{3.8cm} + \tfrac{2 H^2}{3} h R \!-\! \tfrac{49 H^2}{18} h_{00} R \!+\! 
\tfrac{16 H^2}{3 a^2} h R_{00} \!-\! \tfrac{16 H}{27 a} h_{,0} R \Bigr\} . 
\qquad \label{DLExtra}
\end{eqnarray}
Table~\ref{Extra} gives the standard form (\ref{genform}) of the associated
counterterm. Because the $i=2$ contribution is $-(a a')^2 H^4 \times i 
\delta^4(x - x')$, adding $\tfrac{3041 \kappa^2}{2^9 \cdot 3 \cdot \pi^2}
\times \Delta \mathcal{L}_{\rm Extra}$ would cancel (\ref{latesource}),
eliminating any secular enhancement of the mode function.

\subsection{Response to a Point Mass}

For a stationary point mass $M$ the stress-energy is,
\begin{equation}
8\pi G T^{\mu\nu}_{\rm lin}(x) = 8 \pi G M \!\times\! a(t) \delta^3(\vec{x}) 
\!\times\! \delta^{\mu}_0 \delta^{\nu}_0 \; . \label{pointM}
\end{equation}
Because we are working in a fixed gauge, four scalar potentials must be
used,
\begin{equation}
\kappa h_{00} = -2\Psi \quad , \quad \kappa h_{0i} = -\partial_i \Omega
\quad , \quad \kappa h_{ij} = -2\delta_{ij} \Phi - 2 \partial_i \partial_j
\chi \; . \label{potentials}
\end{equation}
From equations (\ref{D00}-\ref{Dij}) we see,
\begin{eqnarray}
[\mathcal{D}^{00\rho\sigma} + \mathcal{D}^{kk\rho\sigma}] \kappa h_{\rho\sigma}
&\!\!\! = \!\!\!& -2 D_B \Psi \; , \qquad \label{Psieqn} \\
\mathcal{D}^{0i\rho\sigma} \kappa h_{\rho\sigma} &\!\!\! = \!\!\!& \tfrac12 
\partial_i D_B \Omega \; , \qquad \label{Omegaeqn} \\
~[\mathcal{D}^{kk\rho\sigma} - \tfrac{3 \partial_i \partial_j}{\nabla^2} 
\mathcal{D}^{ij\rho\sigma}] \kappa h_{\rho\sigma} &\!\!\! = \!\!\!& 2 \nabla^2
D_A \chi \; , \qquad \label{Chieqn} \\
~[\mathcal{D}^{kk\rho\sigma} + \tfrac{\partial_i \partial_j}{\nabla^2} 
\mathcal{D}^{ij\rho\sigma} ] \kappa h_{\rho\sigma} &\!\!\! = \!\!\!& 2 D_A 
(\Phi - \Psi) \; . \qquad \label{Slipeqn}
\end{eqnarray}

At zeroth order the potentials are,
\begin{equation} 
\Psi^{(0)}(t,r) = \Phi^{(0)}(t,r) = \tfrac{G M}{a(t) r} \qquad , \qquad
\Omega^{(0)} = \chi^{(0)} = 0 \; . \label{0thorderpotentials}
\end{equation}
The procedure for finding 1-loop corrections is to consult the appropriate
equation from (\ref{Psieqn}-\ref{Slipeqn}), and then use Table~\ref{Dtimesh}
and relations (\ref{Flog}-\ref{Clog}) to compute the source. For example,
the Newtonian potential obeys,
\begin{eqnarray} 
\lefteqn{-2 D_B \Psi^{(1)} = \tfrac{13 \kappa^2 (\mathcal{F}^{00} +
\mathcal{F}^{kk})}{2^5 \cdot 3^3 \cdot \pi^2} \Bigl\{\! \ln(\tfrac{2 \mu a}{H})
R^{(0)} \! \Bigr\} \!-\! \tfrac{61 \kappa^2 \partial_{\rho} \partial_{\sigma}}{2^4 
\cdot 3 \cdot 5 \cdot \pi^2} \Bigl\{\! \ln(\tfrac{2 \mu a}{H}) [C^{(0) \rho 0 \sigma 0}
\!\!+\! C^{(0) \rho k \sigma k}] \! \Bigr\} } \nonumber \\
& & \hspace{-0.5cm} + \!\! \int \!\! d^4x' \! \sum_{i=1}^{21} \sum_{j=1}^{3} 
C_{ij}(a,a',\partial) \Bigl\{ [\mbox{}^{00} D_i^{\rho\sigma}] \!+\! 
\Bigl\{ [\mbox{}^{kk} D_i^{\rho\sigma}] \Bigr\} \!\times\! i f_j^{\rm SK}(\Delta x) 
\!\times\! \kappa h^{(0)}_{\rho\sigma}(x') . \qquad \label{1stNewton}
\end{eqnarray}
The next step would be simplifying the $[\mbox{}^{00} D_i^{\rho\sigma}] \times
h_{\rho\sigma}(x')$ and $[\mbox{}^{k\ell} D_i^{\rho\sigma}] \times
h_{\rho\sigma}(x')$ entries of Table~\ref{Dtimesh}) by summing $k$ on $\ell$ 
and using the 0th order solution. For example, the case of $i=1$ would be,
\begin{equation}
[\mbox{}^{00} D_i^{\rho\sigma}] \times \kappa h^{(0)}_{\rho\sigma}(x') +
[\mbox{}^{kk} D_i^{\rho\sigma}] \times \kappa h^{(0)}_{\rho\sigma}(x') = 
-h^{(0)}(x') + 3 h^{(0)}(x') = -\tfrac{8 G M}{a' r'} \; . \label{1stline}
\end{equation}
However, it isn't really necessary to go through this because the results from
the nonlocal contributions of Tables~\ref{Primitive3ptFin} and 
\ref{Primitive3ptFinSingleLog} are identical to those of the previous calculation 
\cite{Tan:2021lza}, and stronger than anything the local contributions can give.
The result for the Newtonian potential is,
\begin{equation}
-2 D_B \Psi^{(1)} \longrightarrow -\tfrac{3 \kappa^2 H^4 a^4 \ln^2(a)}{\pi^2}
\!\times\! \tfrac{GM}{a r} \;\; \Longrightarrow \;\; \Psi^{(1)}(t,r) 
\longrightarrow -\tfrac{\kappa^2 H^2 \ln^3(a)}{2 \pi^2} \!\times\! \tfrac{GM}{ar}
\; . \label{1loopNewtonian}
\end{equation}
The gravitational slip is \cite{Tan:2021lza},
\begin{eqnarray}
\lefteqn{2 D_A [\Phi^{(1)} \!-\! \Psi^{(1)}] \longrightarrow \tfrac{4 \kappa^2 
H^4 a^4 \ln^2(a)}{\pi^2} \!\times\! \tfrac{GM}{ar} } \nonumber \\
& & \hspace{5cm} \Longrightarrow \Phi^{(1)} \!-\! \Psi^{(1)} \longrightarrow 
-\tfrac{\kappa^2 H^2 \ln^2(a)}{\pi^2} \!\times\! \tfrac{GM}{ar} \; . \qquad
\label{1loopSlip}
\end{eqnarray}
None of the other potentials shows secular enhancement.

\section{Conclusions}

We have computed the 1-loop graviton contribution to the graviton
self-energy on de Sitter background. Unlike contributions from matter
loops \cite{Park:2011ww,Wang:2015eaa,Miao:2024atw,Foraci:2024vng,
Foraci:2024cwi}, the graviton loop contribution is gauge dependent, and 
our result was derived using the simplest gauge \cite{Tsamis:1992xa,
Woodard:2004ut}. In section 3 we computed the primitive result in 
position-space, using dimensional regularization. The result consists
of divergent local contributions in Tables~\ref{Primitive4pt} and
\ref{Primitive3ptDiv}, plus finite, nonlocal contributions in
Tables~\ref{Primitive3ptFin} and \ref{Primitive3ptFinSingleLog}. The 
latter agree with the old, $D=4$ computation \cite{Tsamis:1996qk}. 
In section 4 the ultraviolet divergences were absorbed using BPHZ 
counterterms given in Table~\ref{Counterterms}. The final, fully 
renormalized result consists of the nonlocal contributions in 
Tables~\ref{Primitive3ptFin} and \ref{Primitive3ptFinSingleLog}, plus  
the local contributions of Tables~\ref{InvFin} and \ref{RenormalizedLocal},
and the important local logarithms terms on the second lines of
expressions (\ref{DL1aReorg}) and (\ref{DL2Reorg}). This final result
obeys the Ward identity (\ref{gravity}).

The reason for computing $-i [\mbox{}^{\mu\nu} \Sigma^{\rho\sigma}](x;x')$
is so that we can use it in the linearized Einstein equation (\ref{linEin})
to derive 1-loop corrections to the graviton mode function and to the
Newtonian potential. A previous analysis \cite{Tan:2021lza,Tan:2022xpn}
obtained relations (\ref{guessu}-\ref{guessPsi}) by employing an old, $D=4$
result \cite{Tsamis:1996qk} that got the nonlocal part right but was not 
capable of reliably fixing the local part. The previous analysis was based
on having made a plausible guess for the local part. Now that the full
self-energy has been derived we can see that this guess was not correct.
One consequence is that the graviton mode function does not experience
the double log enhancement (\ref{guessu}), but only a single log enhancement
(\ref{lateuresult}). Even that could be nulled by making what seems to be
an allowable, finite renormalization involving (\ref{DLExtra}). On the other
hand, the triple log enhancement of the Newtonian potential (\ref{guessPsi})
was sourced entirely from the nonlocal contributions that the $D=4$ calculation
gets right. Our result (\ref{1loopNewtonian}) for the Newtonian potential is 
therefore unchanged, as is our result (\ref{1loopSlip}) for the gravitational 
slip.

The possible extra counterterm (\ref{DLExtra}) highlights the importance of
properly constraining couterterms. Even the list of 19 counterterms given
in Table~\ref{Counterterms} seems excessive. The graviton self-energy can 
be expressed as a linear combination (\ref{genform}) of the 21 basis tensors 
$[\mbox{}^{\mu\nu} D_i^{\rho\sigma}]$ given in Table~\ref{Tbasis}. Seven of 
these elements are related by reflection invariance (\ref{Reflect}), which 
leaves 14 algebraically independent terms. The Ward identity (\ref{gravity})
implies relations proportional to the four tensors,
\begin{equation}
\eta^{\mu\rho} \quad , \quad \partial^{\mu} \partial^{\rho} \quad , \quad
\partial^{(\mu} \delta^{\rho)}_{~~0} \quad , \quad \delta^{\mu}_0 
\delta^{\rho}_0 \; . \label{Wardtensors}
\end{equation}
So one expects only ten independent coefficients, not 19. Some attention
should be given to simplifying and consolidating the list of 
Table~\ref{Counterterms}, with the potential payoff of deciding whether or
not the 1-loop correction to the graviton mode function has a logarithmic
enhancement. 

A major application of this work is provided by the formulae for products of
differentiated propagators presented in the Appendix. These formulae can be
used to compute the graviton self-energy in the gauge introduced to facilitate
the nonperturbative resummation of secular effects \cite{Miao:2024shs}. This
computation is vital to check the nonperturbative stochastic equations against
explicit perturbative results derived in the same guage. The same formulae 
from the Appendix also facilitate the purging of graviton gauge dependence from 
$-i [\mbox{}^{\mu\nu} \Sigma^{\rho\sigma}](x;x')$ following the technique that 
has been successfully applied on flat space background for the scalar self-mass 
\cite{Miao:2017feh} and for the vacuum polarization \cite{Katuwal:2021thy}, and 
on de Sitter background for the scalar self-mass \cite{Glavan:2024elz,
Glavan:2026pug,Glavan:2026bfw}.

\begin{center}
{\bf Acknowledgements}
\end{center}

This work was partially supported by NSF grant PHY-2207514 and by 
the Institute for Fundamental Theory at the University of Florida.

\section{Appendix: Products of Propagators}

The purpose of this appendix is to expand products of two propagators,
each with up to two derivatives, into a series of divergent and finite 
terms. The same procedure was applied in computing the scalar loop
contribution \cite{Miao:2024atw} for the special case of doubly 
differentiated $A$-type propogators. Our starting point is the 
observation that each of the three propagators $i\Delta_I(x;x')$, for 
$I = A,B,C$ can be expanded in terms of contributions which are less 
and less singular at $x^{\mu} = {x'}^{\mu}$,
\begin{equation}
i\Delta_I(x;x') = i\Delta_{\rm cf}(x;x') + \sum_{n=0}^{\infty} 
I_n(x;x') \; . \label{propexp}
\end{equation}
The most singular contribution is the propagator os a massless,
conformally coupled scalar,
\begin{equation}
i\Delta_{\rm cf}(x;x') = \tfrac1{4 \pi^{\frac{D}2}} 
\tfrac{\Gamma(\frac{D}2 - 1)}{[a a' \Delta x^2]^{\frac{D}2 -1}} \; . 
\label{conformal}
\end{equation}
It turns out that we need only the first two of the functions
$I_n(x;x')$,
\begin{eqnarray}
A_0(x;x') &\!\!\! = \!\!\!& \tfrac{H^2}{8 \pi^{\frac{D}2}} 
\tfrac{\Gamma(\frac{D}2 + 1)}{D-4} \tfrac1{[a a' \Delta x^2]^{\frac{D}2 -2}}
- \tfrac{H^{D-2}}{(4\pi)^{\frac{D}2}} \tfrac{\Gamma(D-1)}{\Gamma(\frac{D}2)}
\, [\pi {\rm cot}(\tfrac{D \pi}{2}) \!-\! \ln(a a')] \; , \qquad 
\label{A0} \\
A_1(x;x') &\!\!\! = \!\!\!& \tfrac{H^4}{64 \pi^{\frac{D}2}} 
\tfrac{\Gamma(\frac{D}2+2)}{D-6} \tfrac1{[a a' \Delta x^2]^{\frac{D}2-3}}
+ \tfrac{H^D}{(4\pi)^{\frac{D}2}} \tfrac{\Gamma(D)}{\Gamma(\frac{D}2+1)}
\!\times\! \tfrac14 a a' \Delta x^2 \; , \qquad \label{A1} \\
B_0(x;x') &\!\!\! = \!\!\!& \tfrac{H^2}{16 \pi^{\frac{D}2}} 
\tfrac{\Gamma(\frac{D}2)}{[a a' \Delta x^2]^{\frac{D}2 -2}} - 
\tfrac{H^{D-2}}{(4\pi)^{\frac{D}2}} \tfrac{\Gamma(D-2)}{\Gamma(\frac{D}2)}
\; , \qquad \label{B0} \\
B_1(x;x') &\!\!\! = \!\!\!& \tfrac{H^4}{128 \pi^{\frac{D}2}} 
\tfrac{\Gamma(\frac{D}2+1)}{[a a' \Delta x^2]^{\frac{D}2-3}}
- \tfrac{H^D}{(4\pi)^{\frac{D}2}} \tfrac{\Gamma(D-1)}{\Gamma(\frac{D}2+1)}
\!\times\! \tfrac14 a a' \Delta x^2 \; , \qquad \label{B1} \\
C_0(x;x') &\!\!\! = \!\!\!& \tfrac{H^2}{32 \pi^{\frac{D}2}} 
\tfrac{(D-6) \Gamma(\frac{D}2 -1)}{[a a' \Delta x^2]^{\frac{D}2 -2}}
+ \tfrac{H^{D-2}}{(4\pi)^{\frac{D}2}} \tfrac{\Gamma(D-3)}{\Gamma(\frac{D}2)}
\; , \qquad \label{C0} \\
C_1(x;x') &\!\!\! = \!\!\!& \tfrac{H^4}{256 \pi^{\frac{D}2}} 
\tfrac{(D-8) \Gamma(\frac{D}2)}{[a a' \Delta x^2]^{\frac{D}2-3}}
+ \tfrac{H^D}{(4\pi)^{\frac{D}2}} \tfrac{\Gamma(D-2)}{\Gamma(\frac{D}2+1)}
\!\times\! \tfrac12 a a' \Delta x^2 \; . \qquad \label{C1}
\end{eqnarray}
Note that only $A_0(x;x')$ has a nonzero limit for $D=4$,
\begin{equation}
\lim_{D\rightarrow 4} A_0(x;x') = -\tfrac{H^2}{8 \pi^2} \, \ln(\tfrac14 H^2
\Delta x^2) \; . \label{A0D4}
\end{equation}

Our goal is to employ the graviton self-energy in the linearized 
Einstein equation (\ref{linEin}). This has two important consequences:
\begin{itemize}
\item{Any derivatives with respect to the coordinate $x^{\mu}$ can be taken 
{\it outside} the integration over ${x'}^{\mu}$; and}
\item{The integration makes sense provided the coincidence limit of the
argument diverges less strong than $1/\Delta x^4$ for $D=4$ dimensions.}
\end{itemize}
We can accordingly adopt the 5-step reduction procedure that was pioneered
in \cite{Onemli:2002hr,Prokopec:2002uw}:
\begin{enumerate}
\item{Act any derivatives on each of the two propagators and multiply them 
together;}
\item{Extract derivatives with respect to $x^{\mu}$ from factors of 
$1/\Delta x^2$ until reaching a form like $1/\Delta x^{2D-6}$ that is
integrable for $D=4$;}
\item{Isolate divergences by adding factors of zero in the form,
\begin{equation}
0 = \tfrac{4 \pi^{\frac{D}2} \, i\delta^D(x-x')}{\Gamma(\frac{D}2 -1)}
- \partial^2 \tfrac1{\Delta x^{D-2}} \; ; \label{zero}
\end{equation}}
\item{Take the limit $D=4$ on any finite, integrable terms; and}
\item{Eliminate any remaining factors of $1/\Delta x^2$ using the identities,
\begin{eqnarray}
\tfrac1{\Delta x^2} &\!\!\! = \!\!\!& \tfrac{\partial^2}{4} \ln(\tfrac14 H^2 
\Delta x^2) \; , \qquad \label{onelog} \\
\tfrac{\ln(\frac14 H^2 \Delta x^2)}{\Delta x^2} &\!\!\! = \!\!\!& 
\tfrac{\partial^2}{8} \Bigl[\ln^2(\tfrac14 H^2 \Delta x^2) \!-\! 2
\ln(\tfrac14 H^2 \Delta x^2) \Bigr] \; . \qquad \label{twologs}
\end{eqnarray}}
\end{enumerate}
Note that Step 1 can produce delta functions when a constant part of one
propagator multiplies the derivative,
\begin{equation}
\partial_{\mu} \partial'_{\rho} \tfrac1{\Delta x^{D-2}} = -\delta^0_{~\mu}
\delta^0_{\rho} \!\times\! \tfrac{4 \pi^{\frac{D}2} \, i\delta^D(x-x')}{
\Gamma(\frac{D}2 -1)} + (D\!-\!2) \Bigl[ \tfrac{\eta_{\mu\rho}}{\Delta x^D}
- \tfrac{D \Delta x_{\mu} \Delta x_{\rho}}{\Delta x^{D+2}} \Bigr] \; .
\label{deltafs}
\end{equation}
However, these delta functions will always be absorbed when the derivatives
are extracted in Step 2. 

The factor of zero (\ref{zero}) called for in Step 3 typically needs to be 
multiplied by a dimensional factor to achieve consistent units. It will 
simplify our analysis by choosing this factor to be $(H/2)^{D-4}$,
\begin{eqnarray}
\lefteqn{\tfrac{\partial^2}{D-4} \Bigl[\tfrac1{\Delta x^{2D-6}}\Bigr] = 
\tfrac{(\frac{H}{2})^{D-4}}{D-4} \!\times\! \tfrac{4 \pi^{\frac{D}2} \, 
i\delta^D(x-x')}{\Gamma(\frac{D}2 -1)} + \tfrac{\partial^2}{D-4} \Bigl[
\tfrac1{\Delta x^{2D-6}} - \tfrac{(\frac{H}{2})^{D-4}}{\Delta x^{D-2}} \Bigr]
\; ,} \\
& & \hspace{1cm} = \tfrac{(\frac{H}{2})^{D-4}}{D-4} \!\times\! 
\tfrac{4 \pi^{\frac{D}2} \, i\delta^D(x-x')}{\Gamma(\frac{D}2 -1)} -
\tfrac{\partial^2}{2} \Bigl[ \tfrac{\ln(\frac14 H^2 \Delta x^2)}{\Delta x^2}
\Bigr] + O(D\!-\!4) \; , \qquad \\
& & \hspace{0.8cm} \longrightarrow \tfrac{(\frac{H}{2})^{D-4}}{D-4} \!\times\! 
\tfrac{4 \pi^{\frac{D}2} \, i\delta^D(x-x')}{\Gamma(\frac{D}2 -1)} -
\tfrac{\partial^4}{16} \Bigl[ \ln^2(\tfrac14 H^2 \Delta x^2) \!-\! 2
\ln(\tfrac14 H^2 \Delta x^2) \Bigr] \; . \qquad \label{dimfactor}
\end{eqnarray}
We express all divergences as functions of $D$ times the constant $\mathcal{K}_3$,
which was defined in (\ref{K3def}) and is repeated here,
\begin{equation}
\mathcal{K}_3 = \tfrac{H^{D-4}}{(4\pi)^{\frac{D}2}} \tfrac{\Gamma(\frac{D}2 -1)}{
2 (D-3) (D-4)} = \tfrac1{32 \pi^2} \tfrac{H^{D-4}}{D-4} + {\rm Finite} \; .
\label{K3def2}
\end{equation}
For example, the product of two conformal propagators is,
\begin{eqnarray}
\lefteqn{ (a a')^{D-2} i\Delta_{\rm cf} \!\times\! i\Delta_{\rm cf}
= \tfrac{\Gamma^2(\frac{D}2 -1)}{16 \pi^D \Delta x^{2D-4}} =
\tfrac{\Gamma^2(\frac{D}2 - 1)}{16 \pi^D} \tfrac{\partial^2}{2 (D-3) (D-4)}
\tfrac1{\Delta x^{2D-6}} \; , } \\
& & \hspace{1.5cm} \longrightarrow 4 \mathcal{K}_3 i\delta^D(x \!-\! x') -
\tfrac{\partial^4}{2^9 \cdot \pi^4} \Bigl[ \ln^2(\tfrac14 H^2 \Delta x^2)
\!-\! 2 \ln(\tfrac14 H^2 \Delta x^2)\Bigr] \; . \qquad \label{propEG}
\end{eqnarray}

All products of the form $(a a')^{D-2} i\Delta_I(x;x') \times 
i\Delta_J(x;x')$ can be read off of Table~\ref{TableA}.
\begin{table}[H]
\setlength{\tabcolsep}{8pt}
\def\arraystretch{1.7}
\centering
\begin{tabular}{|@{\hskip 1mm }c@{\hskip 1mm }||c|c|c|}
\hline
Two\ Factors & $i\delta^D(x \!-\! x')$ & $\ln^2(\tfrac14 H^2 \Delta x^2)$
& $\ln(\frac14 H^2 \Delta x^2)$ \\
\hline\hline
$i\Delta_{\rm cf} \!\times\! i\Delta_{\rm cf}$ & ${\scriptstyle 4 \mathcal{K}_3}$
& $-\tfrac{\partial^4}{512 \pi^4}$ & $\tfrac{\partial^4}{256 \pi^4}$  \\
\hline
$i\Delta_{\rm cf} \!\times\! A_0$ & $0$ & $-\tfrac{H^2 a a' \partial^2}{256 \pi^4}$ 
& $\tfrac{H^2 a a' \partial^2}{128 \pi^4}$ \\
\hline
$A_0 \!\times\! A_0$ & $0$ & $\tfrac{H^4 (a a')^2}{64 \pi^4}$ & $0$ \\
\hline
\end{tabular}
\caption{\footnotesize Contributions from products of the form $(a a')^{D-2} 
i\Delta_I \!\times\! i\Delta_J$.}
\label{TableA}
\end{table} 
\noindent For example, the case of $I=A$ and $J=A$ is,
\begin{eqnarray}
\lefteqn{(a a')^{D-2} i\Delta_A \!\times i\Delta_A = (a a')^{D-2} \Bigl\{
i\Delta_{\rm cf} \!+\! A_0 \!+\! \dots\Bigr\} \Bigl\{i \Delta_{\rm cf} 
\!+\! A_0 \!+\! \dots\Bigr\} \; , } \\
& & \hspace{1cm} = (a a')^{D-2} \Bigl\{ i\Delta_{\rm cf} \!\times\!
i\Delta_{\rm cf} \!+\! i\Delta_{\rm cf} \!\times\! A_0 \!+\! A_0 \!\times\!
i\Delta_{\rm cf} \!+\! A_0 \!\times\! A_0 \!+\! \dots \Bigr\} , \qquad \\
& & \hspace{1cm} = 4 \mathcal{K}_3 i\delta^D(x \!-\! x') - 
\tfrac{[\partial^4 + 4 H^2 a a' \partial^2 - 4 H^4 (a 'a)^2]}{2^9 \cdot \pi^4}
\ln^2(\tfrac14 H^2 \Delta x^2) \nonumber \\
& & \hspace{5cm} + \tfrac{[\partial^4 + 4 H^2 a a' \partial^2]}{
2^8 \cdot \pi^4} \ln(\tfrac14 H^2 \Delta x^2) + O(D\!-\!4) \; . 
\qquad \label{DADA}
\end{eqnarray}
The higher $A_n$'s are not needed because these terms are integrable, even
when multiplied by $i\Delta_{\rm cf}$, and they vanish for $D=4$. For the same
reason, none of the $B_n$'s or $C_n$'s are necessary --- not even $B_0$ or
$C_0$. For example, the case of $I=A$ and $J = B$ is,
\begin{eqnarray}
\lefteqn{(a a')^{D-2} i\Delta_A \!\times i\Delta_B = (a a')^{D-2} \Bigl\{
i\Delta_{\rm cf} \!+\! A_0 \!+\! \dots\Bigr\} \Bigl\{i \Delta_{\rm cf} 
\!+\! \dots\Bigr\} \; , } \\
& & \hspace{0cm} = (a a')^{D-2} \Bigl\{ i\Delta_{\rm cf} \!\times\!
i\Delta_{\rm cf} \!+\! i\Delta_{\rm cf} \!\times\! A_0 \!+\! \dots \Bigr\} , 
\qquad \\
& & \hspace{0cm} \longrightarrow 4 \mathcal{K}_3 i\delta^D(x \!-\! x') - 
\tfrac{[\partial^4 + 2 H^2 a a' \partial^2]}{2^9 \cdot \pi^4} \Bigl[
\ln^2(\tfrac14 H^2 \Delta x^2) \!-\! 2 \ln(\tfrac14 H^2 \Delta x^2) \Bigr] 
\; . \qquad \label{DADB}
\end{eqnarray}

Products of the form $(a a')^{D-2} i\Delta_I \!\times\! i\Delta_J$ can be
read off from Table~\ref{TableB}.
\begin{table}[H]
\setlength{\tabcolsep}{8pt}
\def\arraystretch{1.7}
\centering
\begin{tabular}{|@{\hskip 1mm }c@{\hskip 1mm }||c|c|c|}
\hline
Two\ Factors & $i\delta^D(x \!-\! x')$ & $\ln^2(\tfrac14 H^2 \Delta x^2)$
& $\ln(\frac14 H^2 \Delta x^2)$ \\
\hline\hline
$i\Delta_{\rm cf} \!\times\! i\Delta_{\rm cf}$ & ${\scriptstyle 2 \mathcal{K}_3 H 
[\partial_{\mu} - (D-2) a H \delta^0_{\mu}]}$ & $-\tfrac{(\partial_{\mu} -
2 a H \delta^0_{\mu}) \partial^4}{1024 \pi^4}$ & $\tfrac{(\partial_{\mu}
- 2 a H \delta^0_{\mu}) \partial^4}{256 \pi^4}$  \\
\hline
$i\Delta_{\rm cf} \!\times\! A_0$ & $0$ & $-\tfrac{H^2 a a' (\partial_{\mu}
- a H \delta^0_{\mu}) \partial^2}{256 \pi^4}$ & $-\tfrac{H^2 a a' 
(\partial_{\mu} + a H \delta^0_{\mu}) \partial^2}{256 \pi^4}$ \\
\hline
$A_0 \!\times\! i\Delta_{\rm cf}$ & $0$ & $0$ & $\tfrac{H^2 a a' 
\partial_{\mu} \partial^2}{128 \pi^4}$ \\
\hline
$A_0 \!\times\! A_0$ & $0$ & $\tfrac{H^4 (a a')^2 \partial_{\mu}}{128 \pi^4}$ & $0$ \\
\hline
\end{tabular}
\caption{\footnotesize Contributions from products of the form $(a a')^{D-2} 
\partial_{\mu} i\Delta_I \!\times\! i\Delta_J$.}
\label{TableB}
\end{table}
\noindent Acting $\partial_{\mu}$ on $i\Delta_I(x;x')$ produces either a 
factor of $a H \delta^0_{\mu}$ or $\Delta x_{\mu}/\Delta^2$. After multiplying 
by $(a a')^{D-2} i\Delta_J(x;x')$, the first sort of factor simply multiplies
a reduction involving Table~\ref{TableA}. The factor of $\Delta x_{\mu}/\Delta x^2$
requires that we extract a derivative using the identities,
\begin{eqnarray}
\tfrac{\Gamma^2(\frac{D}2 -1)}{16 \pi^D} \tfrac{\Delta x_{\mu}}{\Delta x^{2D-4}}
&\!\!\! \longrightarrow \!\!\!& -\tfrac{\partial_{\mu} \partial^2}{2^7 \cdot \pi^4}
\ln(\tfrac14 H^2 \Delta x^2) \; , \qquad \label{DxmI} \\
\tfrac{\Gamma^2(\frac{D}2 -1)}{16 \pi^D} \tfrac{\Delta x_{\mu}}{\Delta x^{2D-2}}
&\!\!\! \longrightarrow \!\!\!& -\tfrac{2 \mathcal{K}_3 \partial_{\mu} i 
\delta^D(\Delta x)}{D-2} \nonumber \\
& & \hspace{2cm} + \tfrac{\partial_{\mu} \partial^4}{2 ^{11} \cdot \pi^4} \Bigl[ 
\ln^2(\tfrac14 H^2 \Delta x^2) \!-\! 2 \ln(\tfrac14 H^2 \Delta x^2) \Bigr] , 
\label{DxmII} \qquad \\
\tfrac{\Gamma^2(\frac{D}2 -1)}{16 \pi^D} \tfrac{\Delta x_{\mu}}{\Delta x^{2D}}
&\!\!\! \longrightarrow \!\!\!& -\tfrac{\mathcal{K}_3 \partial_{\mu} i 
\delta^D(\Delta x)}{(D-1) (D-2)^2} \nonumber \\
& & \hspace{2cm} + \tfrac{\partial_{\mu} \partial^4}{2 ^{13} \cdot 3 \cdot \pi^4} 
\Bigl[ \ln^2(\tfrac14 H^2 \Delta x^2) \!-\! 2 \ln(\tfrac14 H^2 \Delta x^2) \Bigr]
. \qquad \label{DxmIII}
\end{eqnarray}

Products of the form $(a a')^{D-2} \partial_{\mu} i\Delta_I \times \partial_{\nu}
i\Delta_J$ are given by Table~\ref{TableC1}.
\begin{table}[H]
\setlength{\tabcolsep}{8pt}
\def\arraystretch{1.5}
\centering
\begin{tabular}{|@{\hskip 1mm }c@{\hskip 1mm }||c|c|c|}
\hline
Two\ Factors & $i\delta^D(x \!-\! x')$ & $\ln^2(\tfrac14 H^2 \Delta x^2)$
& $\ln(\frac14 H^2 \Delta x^2)$ \\
\hline\hline
$$ & $\tfrac{\mathcal{K}_3 [\eta_{\mu\nu} \partial^2 + (D-2) \partial_{\mu} 
\partial_{\nu}]}{D-1}$ & $-\tfrac{(\eta_{\mu\nu} \partial^2 + 2 \partial_{\mu}
\partial_{\nu}) \partial^4}{2^{11} \cdot 3 \cdot \pi^4}$ & $\tfrac{(\eta_{\mu\nu} 
\partial^2 + 2 \partial_{\mu} \partial_{\nu}) \partial^4}{2^{10} \cdot 3 
\cdot \pi^4}$ \\ 
$i\Delta_{\rm cf} \!\times\! i\Delta_{\rm cf}$ & $- {\scriptstyle 2 (D-2) 
\mathcal{K}_3 H a \delta^0_{(\mu} \partial_{\nu)} }$ & $+\tfrac{H a \delta^0_{(\mu}
\partial_{\nu)} \partial^4}{2^9 \cdot \pi^4}$ & $-\tfrac{H a \delta^0_{(\mu}
\partial_{\nu)} \partial^4}{2^8 \cdot \pi^4}$ \\ 
$$ & ${\scriptstyle +(D-2)^2 \mathcal{K}_3 H^2 a^2 \delta^0_{\mu} 
\delta^0_{\nu} }$ & $-\tfrac{H^2 a^2 \delta^0_{\mu} \delta^0_{\nu} \partial^4}{
2^9 \cdot \pi^4}$ & $+\tfrac{H^2 a^2 \delta^0_{\mu} \delta^0_{\nu} \partial^4}{
2^8 \cdot \pi^4}$ \\
\hline
$i\Delta_{\rm cf} \!\times\! A_0$ & $\tfrac{D (D-2) \mathcal{K}_3 H^2
a a' \eta_{\mu\nu}}{4}$ & $-\tfrac{H^2 a a' \eta_{\mu\nu} \partial^4}{2^{10}
\cdot \pi^4}$ & $\tfrac{H^2 a a' (-\eta_{\mu\nu} \partial^2 + 2 \partial_{\mu}
\partial_{\nu}) \partial^2}{2^9 \cdot \pi^4}$ \\
$$ & $$ & $$ & $-\tfrac{H^3 a^2 a' \delta^0_{\mu} \partial_{\nu} \partial^2}{
2^7 \cdot \pi^4}$ \\
\hline
$i\Delta_{\rm cf} \!\times\! B_0$ & $\tfrac{H^2 a a' \eta_{\mu\nu}}{64 \pi^2}$ 
& $0$ & $0$ \\
\hline
$i\Delta_{\rm cf} \!\times\! C_0$ & $-\tfrac{H^2 a a' \eta_{\mu\nu}}{64 \pi^2}$ 
& $0$ & $0$ \\
\hline
$A_0 \!\times\! i\Delta_{\rm cf}$ & $\tfrac{D (D-2) \mathcal{K}_3 H^2 a a'
\eta_{\mu\nu}}{4}$ & $\tfrac{D (D-2) \mathcal{K}_3 H^2 a a' \eta_{\mu\nu}}{4}$ & 
$\tfrac{H^2 a a' (-\eta_{\mu\nu} \partial^2 + 2 \partial_{\mu} \partial_{\nu}) 
\partial^2}{2^9 \cdot \pi^4}$ \\
$$ & $$ & $$ & $-\tfrac{H^3 a^2 a' \delta^0_{\nu} \partial_{\mu} \partial^2}{
2^7 \cdot \pi^4}$ \\
\hline
$B_0 \!\times\! i\Delta_{\rm cf}$ & $\tfrac{H^2 a a' \eta_{\mu\nu}}{64 \pi^2}$ 
& $0$ & $0$\\
\hline
$C_0 \!\times\! i\Delta_{\rm cf}$ & $-\tfrac{H^2 a a' \eta_{\mu\nu}}{64 \pi^2}$ 
& $0$ & $0$ \\
\hline
$A_0 \!\times\! A_0$ & $0$ & $0$ & $\tfrac{H^4 (-\eta_{\mu\nu} 
\partial^2 + 2 \partial_{\mu} \partial_{\nu}) \partial^4}{2^{7} \cdot \pi^4}$ \\
\hline
\end{tabular}
\caption{\footnotesize Contributions from products of the form $(a a')^{D-2} 
\partial_{\mu} i\Delta_I \!\times\! \partial_{\nu} i\Delta_J$.}
\label{TableC1}
\end{table}
\noindent Note that the higher superficial degree of divergence allows $B_0$
and $C_0$ to contribute. Acting the two derivatives can produce the factor 
$\Delta x_{\mu} \Delta x_{\nu}$, among other things. The other factors can be 
reduced using Tables~\ref{TableA} and \ref{TableB}. We extract derivatives 
from terms involving $\Delta x_{\mu} \Delta x_{\nu}$ using the identities,
\begin{eqnarray}
\tfrac{\Gamma^2(\frac{D}2 -1)}{16 \pi^D} \tfrac{\Delta x_{\mu} \Delta x_{\nu}}{
\Delta x^{2D-2}} &\!\!\! \longrightarrow \!\!\!& \tfrac{\mathcal{K}_3 \eta_{\mu\nu}
i \delta^D(\Delta x)}{2 (D-2)} - \tfrac{\eta_{\mu\nu} \partial^4}{2^{11} \cdot 
\pi^4} \ln^2(\tfrac14 H^2 \Delta x^2) \nonumber \\
& & \hspace{4cm} + \tfrac{[\eta_{\mu\nu} \partial^2 + 2 \partial_{\mu} 
\partial_{\nu}] \partial^2}{2^{10} \cdot \pi^4} \ln(\tfrac14 H^2 \Delta x^2) 
\; , \qquad \label{DxmDxnI} \\
\tfrac{\Gamma^2(\frac{D}2 -1)}{16 \pi^D} \tfrac{\Delta x_{\mu} \Delta x_{\nu}}{
\Delta x^{2D}} &\!\!\! \longrightarrow \!\!\!& \tfrac{\mathcal{K}_3 [\eta_{\mu\nu}
\partial^2 + (D-2) \partial_{\mu} \partial_{\nu}] i\delta^D(\Delta x)}{(D-1)
(D-2)^2} \nonumber \\
& & \hspace{0.8cm} - \tfrac{[\eta_{\mu\nu} \partial^2 + 2\partial_{\mu} 
\partial_{\nu}] \partial^4}{2 ^{13} \cdot 3 \cdot \pi^4} \Bigl[ 
\ln^2(\tfrac14 H^2 \Delta x^2) \!-\! 2 \ln(\tfrac14 H^2 \Delta x^2) \Bigr] , 
\label{DxmDxnII} \qquad \\
\tfrac{\Gamma^2(\frac{D}2 -1)}{16 \pi^D} \tfrac{\Delta x_{\mu} \Delta x_{\nu}}{
\Delta x^{2D+2}} &\!\!\! \longrightarrow \!\!\!& \tfrac{\mathcal{K}_3 
[\eta_{\mu\nu} \partial^2 + D \partial_{\mu} \partial_{\nu}] \partial^2 i 
\delta^D(\Delta x)}{2 D^2 (D-1) (D-2)^2} \nonumber \\
& & \hspace{0.9cm} - \tfrac{[\eta_{\mu\nu} \partial^2 + 4 \partial_{\mu} 
\partial_{\nu} ] \partial^6}{2 ^{18} \cdot 3 \cdot \pi^4} \Bigl[ 
\ln^2(\tfrac14 H^2 \Delta x^2) \!-\! 2 \ln(\tfrac14 H^2 \Delta x^2) \Bigr]
. \qquad \label{DxmDxnIII}
\end{eqnarray}
When both derivatives are primed the result follows from Table~\ref{TableC1}
by merely interchanging primed and unprimed coordinates. When only one of 
the two derivatives is primed the result can be read off from Table~\ref{TableC2}.
\begin{table}[H]
\setlength{\tabcolsep}{8pt}
\def\arraystretch{1.5}
\centering
\begin{tabular}{|@{\hskip 1mm }c@{\hskip 1mm }||c|c|c|}
\hline
Two\ Factors & $i\delta^D(x \!-\! x')$ & $\ln^2(\tfrac14 H^2 \Delta x^2)$
& $\ln(\frac14 H^2 \Delta x^2)$ \\
\hline\hline
$$ & $-\tfrac{\mathcal{K}_3 [\eta_{\mu\rho} \partial^2 + (D-2) \partial_{\mu} 
\partial_{\rho}]}{D-1}$ & $\tfrac{(\eta_{\mu\rho} \partial^2 + 2 \partial_{\mu}
\partial_{\rho})}{2^{11} \cdot 3 \cdot \pi^4}$ & $-\tfrac{(\eta_{\mu\rho} 
\partial^2 + 2 \partial_{\mu} \partial_{\rho})}{2^{10} \cdot 3 \cdot \pi^4}$ \\
$i\Delta_{\rm cf} \!\times\! i\Delta_{\rm cf}$ & $+ {\scriptstyle (D-2) 
\mathcal{K}_3 H [a \delta^0_{\mu} \partial_{\rho} - a' \delta^0_{\rho} 
\partial_{\nu}] }$ & $\tfrac{(-a H \delta^0_{\mu} \partial_{\rho} +
a' H \delta^0_{\rho} \partial_{\mu})}{2^{10} \cdot \pi^4}$ & $\tfrac{(a H 
\delta^0_{\mu} \partial_{\rho} - a' H \delta^0_{\rho} \partial_{\mu})}{
2^{9} \cdot \pi^4}$ \\
$$ & ${\scriptstyle + (D-2)^2 \mathcal{K}_3 H^2 a a' \delta^0_{\mu} 
\delta^0_{\rho} }$ & $-\tfrac{H^2 a a' \delta^0_{\mu} \delta^0_{\rho}}{2^9
\cdot \pi^4}$ & $\tfrac{H^2 a a' \delta^0_{\mu} \delta^0_{\rho}}{2^8 \cdot 
\pi^4}$ \\
\hline
$i\Delta_{\rm cf} \!\times\! A_0$ & $-\tfrac{D (D-2) \mathcal{K}_3 H^2
a a' \eta_{\mu\rho}}{4}$ & $\tfrac{H^2 a a' \eta_{\mu\rho} \partial^4}{
2^{10} \cdot \pi^4}$ & $-\tfrac{H^2 a a' (\eta_{\mu\rho} \partial^2 + 
2 \partial_{\mu} \partial_{\rho}) \partial^2}{2^{9} \cdot \pi^4}$ \\
$$ & $$ & $$ & $-\tfrac{H^3 a^2 a' \delta^0_{\mu} \partial_{\rho} \partial^2}{
2^7 \cdot \pi^4}$ \\
\hline
$i\Delta_{\rm cf} \!\times\! B_0$ & $-\tfrac{H^2 a a' \eta_{\mu\rho}}{64 \pi^2}$ 
& $0$ & $0$ \\
\hline
$i\Delta_{\rm cf} \!\times\! C_0$ & $\tfrac{H^2 a a' \eta_{\mu\rho}}{64 \pi^2}$ 
& $0$ & $0$ \\
\hline
$A_0 \!\times\! i\Delta_{\rm cf}$ & $-\tfrac{D (D-2) \mathcal{K}_3 H^2 a a'
\eta_{\mu\rho}}{4}$ & $\tfrac{H^2 a a' \eta_{\mu\rho} \partial^4}{2^{10} \cdot 
\pi^4}$ & $-\tfrac{H^2 a a' (\eta_{\mu\rho} \partial^2 + 2 \partial_{\mu} 
\partial_{\rho}) \partial^2}{2^{9} \cdot \pi^4}$ \\
$$ & $$ & $$ & $-\tfrac{H^3 a {a'}^2 \delta^0_{\rho} \partial_{\mu} \partial^2}{
2^7 \cdot \pi^4}$ \\
\hline
$B_0 \!\times\! i\Delta_{\rm cf}$ & $-\tfrac{H^2 a a' \eta_{\mu\rho}}{64 \pi^2}$ 
& $0$ & $0$ \\
\hline
$C_0 \!\times\! i\Delta_{\rm cf}$ & $\tfrac{H^2 a a' \eta_{\mu\rho}}{64 \pi^2}$ 
& $0$ & $0$ \\
\hline
$A_0 \!\times\! A_0$ & $0$ & $0$ & $\tfrac{H^4 (a a')^2 (-\eta_{\mu\rho}
\partial^2 + 2 \partial_{\mu} \partial_{\rho})}{2^7 \cdot \pi^4}$ \\
\hline
\end{tabular}
\caption{\footnotesize Contributions from products of the form 
$(a a')^{D-2} \partial_{\mu} i\Delta_I \!\times\! \partial'_{\rho} i\Delta_J$.}
\label{TableC2}
\end{table}

The case of two derivatives of the same coordinate acting on the one propagator 
times an undifferentiated propagator can be read off from earlier tables using 
the relation,
\begin{eqnarray}
\lefteqn{(a a')^{D-2} \partial_{\mu} \partial_{\nu} i\Delta_I \!\times\! 
i\Delta_J = (a a')^{D-2} \Bigl\{ \partial_{\nu} [\partial_{\mu} i\Delta_I
\!\times\! i\Delta_J] - \partial_{\mu} i\Delta_I \!\times\! \partial_{\nu}
i\Delta_J \Bigr\} , } \\
& & \hspace{0cm} = \Bigl[\partial_{\nu} - (D\!-\!2) a H \delta^0_{\nu}\Bigr]
\Bigl\{ (a a')^{D-2} \partial_{\mu} i\Delta_I \!\times\! i\Delta_J \Bigr\} 
\nonumber \\
& & \hspace{7.5cm} - (a a')^{D-2} \partial_{\mu} i\Delta_I \!\times\! 
\partial_{\nu} i\Delta_J . \qquad \label{2DID}
\end{eqnarray} 
Table~\ref{TableD} below gives the case of a primed and unprimed derivative 
acting on the same propagator times an undifferentiated propagator. Note that
the same extraction identities (\ref{DxmDxnI}-\ref{DxmDxnIII}) suffice for 
this case.

\begin{sidewaystable}[p]
\setlength{\tabcolsep}{8pt}
\def\arraystretch{1.5}
\centering
\begin{tabular}{|@{\hskip 1mm }c@{\hskip 1mm }||c|c|c|}
\hline
Two\ Factors & $i\delta^D(x \!-\! x')$ & $\ln^2(\tfrac14 H^2 \Delta x^2)$
& $\ln(\tfrac14 H^2 \Delta x^2)$ \\
\hline\hline
$$ & $\tfrac{\mathcal{K}_3 (\eta_{\mu\rho} \partial^2 - D \partial_{\mu} 
\partial_{\rho})}{D-1}$ & $\tfrac{(-\eta_{\mu\rho} \partial^2 + 2 \partial_{\mu}
\partial_{\rho}) \partial^4}{2^{11} \cdot 3 \cdot \pi^4}$ & $\tfrac{(\eta_{\mu\rho} 
\partial^2 - 2 \partial_{\mu} \partial_{\rho}) \partial^4}{2^{10} \cdot 3 
\cdot \pi^4}$ \\
$i\Delta_{\rm cf} \!\times\! i\Delta_{\rm cf}$ & $ +{\scriptstyle (D-2) 
\mathcal{K}_3 H [a \delta^0_{\mu} \partial_{\rho} - a' \delta^0_{\rho} 
\partial_{\mu}]}$ & $+\tfrac{(-a H \delta^0_{\mu} \partial_{\rho} + a' H
\delta^0_{\rho} \partial_{\mu}) \partial^4}{2^{10} \cdot \pi^4}$ & $+\tfrac{(a H 
\delta^0_{\mu} \partial_{\rho} - a' H \delta^0_{\rho} \partial_{\mu}) \partial^4}{
2^{9} \cdot \pi^4}$ \\
$$ & ${\scriptstyle +(D-2)^2 \mathcal{K}_3 H^2 a a' \delta^0_{\mu} 
\delta^0_{\rho} }$ & $-\tfrac{H^2 a a' \delta^0_{\mu} \delta^0_{\rho}
\partial^4}{2^9 \cdot \pi^4}$ & $+\tfrac{H^2 a a' \delta^0_{\mu} 
\delta^0_{\rho} \partial^4}{2^9 \cdot \pi^4}$ \\
\hline
$$ & $$ & $\tfrac{H^2 a a' (-\eta_{\mu\rho} \partial^2 + 4 \partial_{\mu}
\partial_{\rho}) \partial^2}{2^{10} \cdot \pi^4}$ & $\tfrac{H^2 a a' 
(\eta_{\mu\rho} \partial^2 + 2 \partial_{\mu} \partial_{\rho}) \partial^2}{
2^9 \cdot \pi^4}$ \\
$i\Delta_{\rm cf} \!\times\! A_0$ & $\tfrac{D (D-2) \mathcal{K}_3 H^2
a a' \eta_{\mu\rho}}{4}$ & $+\tfrac{H^3 (a^2 a' \delta^0_{\mu} \partial_{\rho}
- a {a'}^2 \delta^0_{\rho} \partial_{\mu} ) \partial^2}{2^8 \cdot \pi^4}$ & 
$-\tfrac{H^3 (a^2 a' \delta^0_{\mu} \partial_{\rho} - a {a'}^2 \delta^0_{\rho} 
\partial_{\mu} ) \partial^2}{2^6 \cdot \pi^4}$ \\
$$ & $$ & $-\tfrac{H^4 (a a')^2 \delta^0_{\mu} \delta^0_{\rho} \partial^2}{2^8 
\cdot \pi^4}$ & $+ \tfrac{H^4 (a a')^2 \delta^0_{\mu} \delta^0_{\rho} 
\partial^2}{2^7 \cdot \pi^4}$ \\
\hline
$i\Delta_{\rm cf} \!\times\! B_0$ & $\tfrac{H^2 a a' \eta_{\mu\rho}}{64 \pi^2}$ 
& $0$ & $0$ \\
\hline
$i\Delta_{\rm cf} \!\times\! C_0$ & $-\tfrac{H^2 a a' \eta_{\mu\rho}}{64 \pi^2}$ 
& $0$ & $0$ \\
\hline
$A_0 \!\times\! i\Delta_{\rm cf}$ & $\tfrac{D (D-2) \mathcal{K}_3 H^2 a a'
\eta_{\mu\rho}}{4}$ & $-\tfrac{H^2 a a' \eta_{\mu\rho} \partial^4}{2^{10} 
\cdot \pi^4}$ & $\tfrac{H^2 a a' (\eta_{\mu\rho} \partial^2 - 2 \partial_{\mu}
\partial_{\rho}) \partial^2}{2^9 \cdot \pi^4}$ \\
\hline
$B_0 \!\times\! i\Delta_{\rm cf}$ & $\tfrac{H^2 a a' \eta_{\mu\rho}}{64 \pi^2}$ 
& $0$ & $0$ \\
\hline
$C_0 \!\times\! i\Delta_{\rm cf}$ & $-\tfrac{H^2 a a' \eta_{\mu\rho}}{64 \pi^2}$ 
& $0$ & $0$ \\
\hline
$A_0 \!\times\! A_0$ & $0$ & $-\tfrac{H^4 (a a')^2 \partial_{\mu} 
\partial_{\rho}}{2^7 \cdot \pi^4}$ & $\tfrac{H^4 (a a')^2 (\eta_{\mu\nu}
\partial^2 - 2 \partial_{\mu} \partial_{\nu})}{2^7 \cdot \pi^4}$ \\
\hline
\end{tabular}
\caption{\footnotesize Contributions from products of the form 
$(a a')^{D-2} \partial_{\mu} \partial'_{\rho} i\Delta_I \!\times\!
i\Delta_J$.}
\label{TableD}
\end{sidewaystable} 

\begin{sidewaystable}[p]
\setlength{\tabcolsep}{8pt}
\def\arraystretch{1.5}
\centering
\begin{tabular}{|@{\hskip 1mm }c@{\hskip 1mm }||c|}
\hline
Two\ Factors & $i\delta^D(x \!-\! x')$ \\
\hline\hline
$i\Delta_{\rm cf} \!\times\! i\Delta_{\rm cf}$ & $- \tfrac{(D-2) \mathcal{K}_3
\partial_{\mu} \partial_{\nu} \partial_{\rho}}{2 (D-1)} + \tfrac{\mathcal{K}_3
[ \eta_{\mu\rho} \partial_{\nu} - 2 \eta_{\nu (\mu} \partial_{\rho)}] 
\partial^2 }{2 (D-1)} + \tfrac{(D-2) \mathcal{K}_3 H a \delta^0_{\nu} [D
\partial_{\mu} \partial_{\rho} - \eta_{\mu\rho} \partial^2]}{2 (D-1)} +
\tfrac{(D-2) \mathcal{K}_3 H a \delta^0_{\mu} [\eta_{\nu\rho} \partial^2 +
(D-2) \partial_{\nu} \partial_{\rho}]}{2 (D-1)}$ \\
$$ & $- \tfrac{(D-2) \mathcal{K}_3 H a' \delta^0_{\rho} [\eta_{\mu\nu} 
\partial^2 + (D-2) \partial_{\mu} \partial_{\nu}]}{2 (D-1)} + 
\tfrac{(D-2)^2 \mathcal{K}_3 H^2 [2 a a' \delta^0_{\rho} \delta^0_{(\mu} 
\partial_{\nu)} - a^2 \delta^0_{\mu} \delta^0_{\nu} \partial_{\rho}]}{2} 
- \tfrac{(D-2)^3 \mathcal{K}_3 H^3 a^2 a' \delta^0_{\mu} \delta^0_{\nu} \delta^0_{\rho}}{2}$ \\
\hline
$i\Delta_{\rm cf} \!\times\! A_0$ & $+\tfrac{D^2 (D-2)^2 \mathcal{K}_3 H^2 a a' 
[a H \delta^0_{\mu} \eta_{\nu\sigma} \partial_{\rho} - a' H \delta^0_{\rho} 
\eta_{\nu\sigma} \partial_{\mu}]}{16 (D-1)} + \tfrac{D (D-2)^3 \mathcal{K}_3
H^2 a a' [-a H \delta^0_{\mu} \eta_{\rho (\nu} \partial_{\sigma)} + a' H
\delta^0_{\rho} \eta_{\mu (\nu} \partial_{\sigma)} ]}{8 (D-1)} 
+\tfrac{D (D-2)^3 \mathcal{K}_3 H^4
(a a')^2 \delta^0_{\mu} \delta^0_{\rho} \eta_{\nu\sigma}}{16}$ \\
\hline
$i\Delta_{\rm cf} \!\times\! B_0$ & $\tfrac{H^2 a a' [\eta_{\mu\rho}
\partial_{\nu} - 4 \eta_{\nu (\mu} \partial_{\rho)} ]}{192 \pi^2}$ \\
\hline
$i\Delta_{\rm cf} \!\times\! C_0$ & $\tfrac{H^2 a a' [-\eta_{\mu\rho}
\partial_{\nu} + 4 \eta_{\nu (\mu} \partial_{\rho)} ]}{192 \pi^2}$ \\
\hline
$A_0 \!\times\! i\Delta_{\rm cf}$ & $\tfrac{D^2 (D-2) \mathcal{K}_3 H^2 a a'
\eta_{\mu\rho} \partial_{\nu}}{8 (D-1)} - \tfrac{D (D-2)^2 \mathcal{K}_3
H^2 a a' \eta_{\nu (\mu} \partial_{\rho)}}{4 (D-1)} - \tfrac{D (D-2)^2
\mathcal{K}_3 H^3 a^2 a' \eta_{\mu\rho} \delta^0_{\nu}}{8}$ \\
\hline
$B_0 \!\times\! i\Delta_{\rm cf}$ & $\tfrac{H^2 a a' [\eta_{\mu\rho}
\partial_{\nu} - \eta_{\nu (\mu} \partial_{\rho)} ]}{96 \pi^2}
- \tfrac{H^3 a^2 a' \eta_{\mu\rho} \delta^0_{\nu}}{64 \pi^2}$ \\
\hline
$C_0 \!\times\! i\Delta_{\rm cf}$ & $\tfrac{H^2 a a' [-\eta_{\mu\rho}
\partial_{\nu} + \eta_{\nu (\mu} \partial_{\rho)} ]}{96 \pi^2}
+ \tfrac{H^3 a^2 a' \eta_{\mu\rho} \delta^0_{\nu}}{64 \pi^2}$ \\
\hline
$A_0 \!\times\! A_0$ & $0$ \\
\hline
\end{tabular}
\caption{\footnotesize Local contributions from products of the form 
$(a a')^{D-2} \partial_{\mu} \partial'_{\rho} i\Delta_I \!\times\! 
\partial_{\nu} i\Delta_J$.}
\label{TableEDiv}
\end{sidewaystable}

\begin{sidewaystable}[p]
\setlength{\tabcolsep}{8pt}
\def\arraystretch{1.5}
\centering
\begin{tabular}{|@{\hskip 1mm }c@{\hskip 1mm }||c|c|}
\hline
Two\ Factors & $\ln^2(\tfrac14 H^2 \Delta x^2)$ & $\ln(\tfrac14 H^2 \Delta x^2)$ \\
\hline\hline
$$ & $\tfrac{[2 \partial_{\mu} \partial_{\nu} \partial_{\rho} + 2 \eta_{\nu (\mu}
\partial_{\rho)} \partial^2 - \eta_{\mu\rho} \partial_{\nu} \partial^2]
\partial^4}{2^{12} \cdot 3 \cdot \pi^4} + \tfrac{H a' \delta^0_{\rho} [\eta_{\mu\nu} 
\partial^2 + 2 \partial_{\mu} \partial_{\nu}) ] \partial^4}{2^{11} \cdot 3 \cdot 
\pi^4}$ & $\tfrac{[-2 \partial_{\mu} \partial_{\nu} \partial_{\rho} - 2 
\eta_{\nu (\mu} \partial_{\rho)} \partial^2 + \eta_{\mu\rho} \partial_{\nu} 
\partial^2] \partial^4}{2^{11} \cdot 3 \cdot \pi^4} - \tfrac{H a' \delta^0_{\rho} 
[\eta_{\mu\nu} \partial^2 + 2 \partial_{\mu} \partial_{\nu}) ] \partial^4}{2^{10} 
\cdot 3 \cdot \pi^4}$ \\
$i\Delta_{\rm cf} \!\times\! i\Delta_{\rm cf}$ & $+ \tfrac{H [a \delta^0_{\nu} 
(\eta_{\mu\rho} \partial^2 - 4 \partial_{\mu} \partial_{\rho}) - a \delta^0_{\mu} 
(\eta_{\nu\rho} \partial^2 + 2 \partial_{\nu} \partial_{\rho}) ] \partial^4}{2^{11} 
\cdot 3 \cdot \pi^4}$ & $+ \tfrac{H [-a \delta^0_{\nu} (\eta_{\mu\rho} \partial^2 
- 4 \partial_{\mu} \partial_{\rho}) + a \delta^0_{\mu} (\eta_{\nu\rho} \partial^2 
+ 2 \partial_{\nu} \partial_{\rho}) ] \partial^4}{2^{10} \cdot 3 \cdot \pi^4}$ \\
$$ & $+ \tfrac{H^2 [a^2 \delta^0_{\mu} \delta^0_{\nu} \partial_{\rho} - 2 a a'
\delta^0_{\rho} \delta^0_{(\mu} \partial_{\nu)}] \partial^4}{2^{10} \cdot \pi^4}
+ \tfrac{H^3 a^2 a' \delta^0_{\mu} \delta^0_{\nu} \delta^0_{\rho} \partial^4}{
2^9 \cdot \pi^4}$ & $- \tfrac{H^2 [a^2 \delta^0_{\mu} \delta^0_{\nu} \partial_{\rho} 
- 2 a a' \delta^0_{\rho} \delta^0_{(\mu} \partial_{\nu)}] \partial^4}{2^{9} \cdot 
\pi^4} - \tfrac{H^3 a^2 a' \delta^0_{\mu} \delta^0_{\nu} \delta^0_{\rho} \partial^4}{
2^8 \cdot \pi^4}$ \\
\hline
$$ & $\tfrac{H^2 a a' [2 \eta_{\nu (\mu} \partial_{\rho)} - \eta_{\mu\rho}
\partial_{\nu}] \partial^4}{2^{10} \cdot 3 \cdot \pi^4}$ & $-\tfrac{H^2 a a'
\partial_{\mu} \partial_{\nu} \partial_{\rho} \partial^2}{2^7 \cdot 3 
\cdot \pi^4} - \tfrac{H^2 a a' [2 \eta_{\nu (\mu} \partial_{\rho)}
- \eta_{\rho \mu} \partial_{\nu} ] \partial^4}{2^9 \cdot 3 \cdot \pi^4}$ \\
$i\Delta_{\rm cf} \!\times\! A_0$ & $+ \frac{H^3 [-a^2 a' \delta^0_{\mu}
\eta_{\nu \rho} + a {a'}^2 \delta^0_{\rho} \eta_{\mu\nu} ] \partial^4}{
2^{10} \cdot \pi^4}$ & $+ \tfrac{H^3 a^2 a' \delta^0_{\mu} [\eta_{\nu\rho}
\partial^2 - 2 \partial_{\nu} \partial_{\rho}] \partial^2}{2^9 \cdot \pi^4}
- \tfrac{H^3 a^2 a' \delta^0_{\rho} [\eta_{\mu\nu} \partial^2 - 2 
\partial_{\mu} \partial_{\nu}] \partial^2}{2^9 \cdot \pi^4}$ \\
$$ & $$ & $+\tfrac{H^4 (a a')^2 \delta^0_{\mu} \delta^0_{\rho} 
\partial_{\nu} \partial^2}{2^7 \cdot \pi^4}$ \\
\hline
$i\Delta_{\rm cf} \!\times\! B_0$ & $0$ & $0$ \\
\hline
$i\Delta_{\rm cf} \!\times\! C_0$ & $0$ & $0$ \\
\hline
$$ & $\tfrac{H^2 a a' [\eta_{\nu (\mu} \partial_{\rho)} - \eta_{\mu\rho}
\partial_{\nu} ] \partial^4}{2^9 \cdot 3 \cdot \pi^4}$ & $\tfrac{H^2 a a'
[-\partial_{\mu} \partial_{\nu} \partial_{\rho} + \eta_{\rho \mu}
\partial_{\nu} \partial^2 - \eta_{\nu (\mu} \partial_{\rho)} \partial^2] 
\partial^2}{2^8 \cdot 3 \cdot \pi^4}$ \\
$A_0 \!\times\! i\Delta_{\rm cf}$ & $+ \tfrac{H^3 a^2 a' \delta^0_{\nu}
\eta_{\mu\rho} \partial^4}{2^{10} \cdot \pi^4}$ & $-\tfrac{H^3 [a^2 a'
\delta^0_{\nu} \eta_{\mu\rho} + a {a'}^2 \delta^0_{\rho} \eta_{\mu\nu}]
\partial^4}{2^9 \cdot \pi^4}$ \\
$$ & $$ & $+ \tfrac{H^3 a^2 a' [2 \delta^0_{\nu} \partial_{\mu}
\partial_{\rho} + \delta^0_{\mu} \eta_{\nu\rho} \partial^2] \partial^2}{
2^9 \cdot \pi^4}$ \\
\hline
$B_0 \!\times\! i\Delta_{\rm cf}$ & $0$ & $0$ \\
\hline
$C_0 \!\times\! i\Delta_{\rm cf}$ & $0$ & $0$ \\
\hline
$A_0 \!\times\! A_0$ & $0$ & $\tfrac{H^4 (a a')^2 \partial_{\mu} \partial_{\nu}
\partial_{\rho}}{2^7 \cdot \pi^4} + \tfrac{H^4 (a a')^2 [\eta_{\mu\rho}
\partial_{\nu} - 2 \eta_{\nu (\mu} \partial_{\rho)}] \partial^2}{2^8 \cdot
\pi^4}$ \\
\hline
\end{tabular}
\caption{\footnotesize Nonlocal contributions from products of the form 
$(a a')^{D-2} \partial_{\mu} \partial'_{\rho} i\Delta_I \!\times\! 
\partial_{\nu} i\Delta_J$.}
\label{TableEFin}
\end{sidewaystable}

Tables~\ref{TableEDiv} and \ref{TableEFin} give the divergent and finite
parts, respectively, of products of the form $(a a')^{D-2} \partial_{\mu}
\partial'_{\rho} i\Delta_I \!\times\! \partial_{\nu} i\Delta_J$. Acting 
the various derivatives produces up to three factors of $\Delta x$. We
extract derivatives from them using the relations,
\begin{eqnarray}
\tfrac{\Gamma^2(\frac{D}2 -1)}{16 \pi^D} \tfrac{\Delta x_{\mu} \Delta x_{\nu}
\Delta x_{\rho}}{\Delta x^{2D-2}} &\!\!\! \longrightarrow \!\!\!& 
\tfrac{[2 \partial_{\mu} \partial_{\nu} \partial_{\rho} - 3 \eta_{(\mu\nu}
\partial_{\rho)} \partial^2]}{2^9 \cdot \pi^4} \ln(\tfrac14 H^2 \Delta x^2)
\; , \qquad \label{DmDnDrI} \\
\tfrac{\Gamma^2(\frac{D}2 -1)}{16 \pi^D} \tfrac{\Delta x_{\mu} \Delta x_{\nu}
\Delta x_{\rho}}{\Delta x^{2D}} &\!\!\! \longrightarrow \!\!\!& -
\tfrac{3 \mathcal{K}_3 \eta_{(\mu\nu} \partial_{\rho)} i\delta^D(\Delta x)}{
(D-1) (D-2)} + \tfrac{\eta_{(\mu\nu} \partial_{\rho)} \partial^4}{2^{12} 
\cdot \pi^4} \ln^2(\tfrac14 H^2 \Delta x^2) \nonumber \\
& & \hspace{2cm} - \tfrac{[2 \partial_{\mu} \partial_{\nu} \partial_{\rho} 
+ 3 \eta_{(\mu\nu} \partial_{\rho)} \partial^2] \partial^2}{2^{11} \cdot 3 
\cdot \pi^4} \ln(\tfrac14 H^2 \Delta x^2) \; , \qquad \label{DmDnDrII} \\
\tfrac{\Gamma^2(\frac{D}2 -1)}{16 \pi^D} \tfrac{\Delta x_{\mu} \Delta x_{\nu}
\Delta x_{\rho}}{\Delta x^{2D+2}} &\!\!\! \longrightarrow \!\!\!& 
-\tfrac{\mathcal{K}_3 [(D-2) \partial_{\mu} \partial_{\nu} \partial_{\rho} 
+ 3 \eta_{(\mu\nu} \partial_{\rho)} \partial^2] i\delta^D(\Delta x)}{2 D
(D-1) (D-2)^2} \nonumber \\
& & \hspace{-1cm} + \tfrac{[2 \partial_{\mu} \partial_{\nu} \partial_{\rho}
+ 3 \eta_{(\mu\nu} \partial_{\rho)} \partial^2] \partial^4}{2^{16} \cdot 3 
\cdot \pi^4} \Bigl[ \ln^2(\tfrac14 H^2 \Delta x^2) \!-\! 2 \ln(\tfrac14 H^2
\Delta x^2) \Bigr] \; . \qquad \label{DmDnDrIII}
\end{eqnarray} 

Tables~\ref{TableFDiv} and \ref{TableFFin} give the divergent and finite
parts, respectively, of $(a a')^{D-2} \partial_{\mu} \partial'_{\rho} 
i\Delta_I \!\times\! \partial_{\nu} \partial'_{\sigma} i\Delta_J$. There 
are now enough derivatives that $A_1$, $B_1$ and $C_1$ contribute. The four 
derivatives can produce up to four factors of $\Delta x$, from which we can 
extract derivatives using the relations,
\begin{eqnarray}
\lefteqn{\tfrac{\Gamma^2(\frac{D}2 -1)}{16 \pi^D} \tfrac{\Delta x_{\mu} 
\Delta x_{\nu} \Delta x_{\rho} \Delta x_{\sigma}}{\Delta x^{2D}} 
\longrightarrow \tfrac{\mathcal{K}_3 \partial_{\mu} \partial_{\nu} 
\partial_{\rho} \partial_{\sigma} i \delta^D(\Delta x)}{(D-1) (D-2)} - 
\tfrac{\eta_{(\mu\nu} \eta_{\rho\sigma)} \partial^4}{2^{12} \cdot \pi^4} 
\ln^2(\tfrac{H^2 \Delta x^2}{4}) } \nonumber \\
& & \hspace{4cm} + \tfrac{3 \eta_{(\mu\nu} \eta_{\rho\sigma)} \partial^4
+ 12 \eta_{(\mu\nu} \partial_{\rho} \partial_{\sigma)} \partial^2 -
4 \partial_{\mu} \partial_{\nu} \partial_{\rho} \partial_{\sigma}]}{2^{11}
\cdot 3 \cdot \pi^4} \ln(\tfrac{H^2 \Delta x^2}{4}) , \qquad 
\label{DmDnDrDsI} \\
\lefteqn{\tfrac{\Gamma^2(\frac{D}2 -1)}{16 \pi^D} \tfrac{\Delta x_{\mu} 
\Delta x_{\nu} \Delta x_{\rho} \Delta x_{\sigma}}{\Delta x^{2D+2}} 
\longrightarrow \tfrac{\mathcal{K}_3 [3 \eta_{\mu\nu} \eta_{\rho\sigma)} 
\partial^2 + 6 (D-2) \eta_{(\mu\nu} \partial_{\rho} \partial_{\sigma)}] 
i\delta^D(\Delta x)}{2 D (D-1) (D-2)^2} \!+\! \tfrac{\partial_{\mu} 
\partial_{\nu} \partial_{\rho} \partial_{\sigma} \partial^2}{2^{13} 
\cdot 3 \cdot \pi^4} } \nonumber \\
& & \hspace{0cm} \times \ln(\tfrac{H^2 \Delta x^2}{4})
- \tfrac{[3 \eta_{(\mu\nu} \eta_{\rho\sigma)} \partial^2
+ 12 \eta_{(\mu\nu} \partial_{\rho} \partial_{\sigma)}] \partial^4}{2^{16} 
\cdot 3 \cdot \pi^4} \Bigl[\ln^2(\tfrac{H^2 \Delta x^2}{4}) \!-\! 2 
\ln(\tfrac{H^2 \Delta x^2}{4}) \Bigr] , \qquad \label{DmDnDrDsII} \\
\lefteqn{\tfrac{\Gamma^2(\frac{D}2 -1)}{16 \pi^D} \tfrac{\Delta x_{\mu} 
\Delta x_{\nu} \Delta x_{\rho} \Delta x_{\sigma}}{\Delta x^{2D+4}} 
\longrightarrow \tfrac{\mathcal{K}_3 [D (D-2) \partial_{\mu} \partial_{\nu} 
\partial_{\rho} \partial_{\sigma} + 6 D \eta_{(\mu\nu} \partial_{\rho} 
\partial_{\sigma)} \partial^2 + 3 \eta_{(\mu\nu} \eta_{\rho\sigma)} 
\partial^4] i\delta^D(\Delta x)}{4 (D+1) D^2 (D-1) (D-2)^2} } \nonumber \\
& & \hspace{1cm} - \tfrac{[8 \partial_{\mu} \partial_{\nu} \partial_{\rho}
\partial_{\sigma} + 24 \eta_{(\mu\nu} \partial_{\rho)} \partial^2 + 3
\eta_{(\mu\nu} \eta_{\rho\sigma)} \partial^4] \partial^4}{2^{19} \cdot 3 
\cdot 5 \cdot \pi^4} \Bigl[ \ln^2(\tfrac{H^2 \Delta x^2}{4}) \!-\! 2 
\ln(\tfrac{H^2 \Delta x^2}{4}) \Bigr] . \qquad \label{DmDnDrDsIII}
\end{eqnarray} 

\begin{sidewaystable}[p]
\setlength{\tabcolsep}{8pt}
\def\arraystretch{1.5}
\centering
\begin{tabular}{|@{\hskip 1mm }c@{\hskip 1mm }||c|}
\hline
Two\ Factors & $i\delta^D(x \!-\! x')$ \\
\hline\hline
$$ &
${\scriptscriptstyle \tfrac{D (D-2) \mathcal{K}_3 \partial_{\mu} \partial_{\nu} 
\partial_{\rho} \partial_{\sigma}}{8 (D+1) (D-1)}} + \tfrac{6 D \mathcal{K}_3
\eta_{(\mu\nu} \partial_{\rho} \partial_{\sigma)} \partial^2}{4 (D+1) (D-1)}
- \tfrac{\mathcal{K}_3 [\eta_{\mu\rho} \partial_{\nu} \partial_{\sigma} +
\eta_{\nu\sigma} \partial_{\mu} \partial_{\rho}] \partial^2}{2 (D-1)} +
\tfrac{3 \mathcal{K}_3 \eta_{(\mu\nu} \eta_{\rho\sigma)} \partial^4}{4
(D+1) (D-1)} +\tfrac{(D-2)^2 \mathcal{K}_3 [a' H \partial_{\mu} \partial_{\nu} 
\partial_{(\rho} \delta^0_{\sigma)} - a' H \partial_{\rho} \partial_{\sigma} 
\partial_{(\mu} \delta^0_{\nu)}]}{2 (D-1)}$ \\
$i\Delta_{\rm cf} \!\times\! i\Delta_{\rm cf}$ & $+ \tfrac{(D-2)^2 \mathcal{K}_3
\{ a^2 H^2 \delta^0_{\mu} \delta^0_{\nu} [\eta_{\rho\sigma} \partial^2 + 
(D-2) \partial_{\rho} \partial_{\sigma}] - a a' H^2 \delta^0_{\mu} \delta^0_{\sigma}
[\eta_{\nu\rho} \partial^2 + (D-2) \partial_{\nu} \partial_{\rho}] - a a' H^2
\delta^0_{\rho} \delta^0_{\nu} [\eta_{\mu\sigma} \partial^2 + (D-2) \partial_{\mu}
\partial_{\sigma}] + {a'}^2 H^2 \delta^0_{\rho} \delta^0_{\sigma} [\eta_{\mu\nu}
\partial^2 + (D-2) \partial_{\mu} \partial_{\nu}] \} }{4 (D-1)}$ \\
$$ & $+ \tfrac{(D-2) \mathcal{K}_3 [a' H \eta_{\mu\nu} \delta^0_{(\rho} 
\partial_{\sigma)} \partial^2 - a H \eta_{\rho\sigma} \delta^0_{(\mu}
\partial_{\nu)} \partial^2]}{2 (D-1)} + \tfrac{(D-2) \mathcal{K}_3 [a' H 
\delta^0_{[\rho} \eta_{\sigma] [\mu} \partial_{\nu]} - a H \delta^0_{[\mu}
\eta_{\nu] [\rho} \partial_{\sigma]}] \partial^2}{D-1}$ \\
$$ & $+\tfrac{(D-2)^2 \mathcal{K}_3 H^2 a a' \{\delta^0_{\mu} \delta^0_{\rho}
[\eta_{\nu\sigma} \partial^2 - D \partial_{\nu} \partial_{\sigma}] + 
\delta^0_{\nu} \delta^0_{\sigma} [\eta_{\mu\rho} \partial^2 - D \partial_{\mu}
\partial_{\rho}]\} }{4 (D-1)} 
+\tfrac{(D-2)^3 \mathcal{K}_3 H^2 a a' [a H \delta^0_{\mu} \delta^0_{\nu}
\delta^0_{(\rho} \partial_{\sigma)} - a' H \delta^0_{\rho} \delta^0_{\sigma}
\delta^0_{(\mu} \partial_{\nu)}]}{2} + \tfrac{(D-2)^4 \mathcal{K}_3 H^4 (a a')^2
\delta^0_{\mu} \delta^0_{\nu} \delta^0_{\rho} \delta^0_{\sigma}}{4}$ \\
\hline
$$ & $\tfrac{D (D-2)^2 \mathcal{K}_3 H^2 a a'[6 \partial_{(\mu} \partial_{\nu}
\eta_{\rho\sigma)} - 2 \partial_{\nu} \partial_{\sigma} \eta_{\mu\rho}
- \frac{2 D}{D-2} \partial_{\mu} \partial_{\rho} \eta_{\nu\sigma} +
\frac{3}{D-2} \eta_{(\mu\nu} \eta_{\rho\sigma)} \partial^2]}{16 (D-1)} $ \\
$i\Delta_{\rm cf} \!\times\! A_0$ & $+\tfrac{D^2 (D-2)^2 \mathcal{K}_3 H^2 a a' 
[a H \delta^0_{\mu} \eta_{\nu\sigma} \partial_{\rho} - a' H \delta^0_{\rho} 
\eta_{\nu\sigma} \partial_{\mu}]}{16 (D-1)} + \tfrac{D (D-2)^3 \mathcal{K}_3
H^2 a a' [-a H \delta^0_{\mu} \eta_{\rho (\nu} \partial_{\sigma)} + a' H
\delta^0_{\rho} \eta_{\mu (\nu} \partial_{\sigma)} ]}{8 (D-1)} 
+\tfrac{D (D-2)^3 \mathcal{K}_3 H^4
(a a')^2 \delta^0_{\mu} \delta^0_{\rho} \eta_{\nu\sigma}}{16}$ \\
$$ & $+\tfrac{H^3 a^2 a' \delta^0_{\nu} [\eta_{\mu\rho} \partial_{\sigma}
- 4 \eta_{\sigma (\mu} \partial_{\rho)}]}{96 \pi^2} - \tfrac{H^3 a {a'}^2
\delta^0_{\sigma} [\eta_{\mu\rho} \partial_{\nu} - 4 \eta_{\nu (\mu} 
\partial_{\rho)}]}{96 \pi^2} + \tfrac{H^4 a a' [a^2 \delta^0_{\mu} 
\delta^0_{\nu} \eta_{\rho\sigma} - a a' \delta^0_{\mu} \delta^0_{\sigma}
\eta_{\nu\rho} - a a' \delta^0_{\rho} \delta^0_{\nu} \eta_{\mu\sigma}
+ {a'}^2 \delta^0_{\rho} \delta^0_{\sigma} \eta_{\mu\nu}]}{32 \pi^2}$ \\
\hline
$i\Delta_{\rm cf} \!\times\! B_0$ & $\tfrac{H^2 a a' [-4\eta_{\mu\rho}
\partial_{\nu} \partial_{\sigma} - 8 \eta_{\nu\sigma} \partial_{\mu}
\partial_{\rho} + 6 \eta_{(\mu\nu} \partial_{\rho} \partial_{\sigma)}
+ 3 \eta_{(\mu\nu} \eta_{\rho\sigma)} \partial^2]}{768 \pi^2}$ \\
$$ & $+ \tfrac{H^2 a^2 a' \delta^0_{\mu} [2 \eta_{\nu\sigma} \partial_{\rho}
- \eta_{\nu\rho} \partial_{\sigma} - \eta_{\rho\sigma} \partial_{\nu}]}{
192 \pi^2} - \tfrac{H^3 a {a'}^3 \delta^0_{\rho} [2 \eta_{\nu\sigma} 
\partial_{\mu} - \eta_{\mu\nu} \partial_{\sigma} - \eta_{\mu\sigma}
\partial_{\nu}]}{192 \pi^2} + \tfrac{H^4 (a a')^2 \delta^0_{\mu}
\delta^0_{\rho} \eta_{\nu\sigma}}{64 \pi^2}$ \\
\hline
$i\Delta_{\rm cf} \!\times\! C_0$ & $\tfrac{H^2 a a' [4\eta_{\mu\rho}
\partial_{\nu} \partial_{\sigma} + 8 \eta_{\nu\sigma} \partial_{\mu}
\partial_{\rho} - 6 \eta_{(\mu\nu} \partial_{\rho} \partial_{\sigma)}
+ 3 \eta_{(\mu\nu} \eta_{\rho\sigma)} \partial^2]}{768 \pi^2}$ \\
$$ & $- \tfrac{H^2 a^2 a' \delta^0_{\mu} [2 \eta_{\nu\sigma} \partial_{\rho}
- \eta_{\nu\rho} \partial_{\sigma} - \eta_{\rho\sigma} \partial_{\nu}]}{
192 \pi^2} + \tfrac{H^3 a {a'}^3 \delta^0_{\rho} [2 \eta_{\nu\sigma} 
\partial_{\mu} - \eta_{\mu\nu} \partial_{\sigma} - \eta_{\mu\sigma}
\partial_{\nu}]}{192 \pi^2} - \tfrac{H^4 (a a')^2 \delta^0_{\mu}
\delta^0_{\rho} \eta_{\nu\sigma}}{64 \pi^2}$ \\
\hline
$i\Delta_{\rm cf} \!\times\! A_1$ & $\tfrac{3 H^4 (a a')^2 \eta_{(\mu\nu}
\eta_{\rho\sigma)}}{64 \pi^2}$ \\
\hline
$i\Delta_{\rm cf} \!\times\! B_1$ & $-\tfrac{H^4 (a a')^2 \eta_{(\mu\nu}
\eta_{\rho\sigma)}}{64 \pi^2}$ \\
\hline
$i\Delta_{\rm cf} \!\times\! C_1$ & $\tfrac{H^4 (a a')^2 \eta_{(\mu\nu}
\eta_{\rho\sigma)}}{64 \pi^2}$ \\
\hline
$A_0 \!\times\! A_0$ & $\frac{3 D^2 (D-2)^3 \mathcal{K}_3 H^4 
\eta_{(\mu\nu} \eta_{\rho\sigma)}}{64 (D-1)}$ \\
\hline
\end{tabular}
\caption{\footnotesize Local contributions from products of the form 
$(a a')^{D-2} \partial_{\mu} \partial'_{\rho} i\Delta_I \!\times\! \partial_{\nu}
\partial'_{\sigma} i\Delta_J$.}
\label{TableFDiv}
\end{sidewaystable}

\begin{sidewaystable}[p]
\setlength{\tabcolsep}{8pt}
\def\arraystretch{1.5}
\centering
\begin{tabular}{|@{\hskip 1mm }c@{\hskip 1mm }||c|c|}
\hline
Two\ Factors & $\ln^2(\tfrac14 H^2 \Delta x^2)$ & $\ln(\tfrac14 H^2 \Delta x^2)$\\
\hline\hline
$$ & $-\tfrac{\eta_{(\mu\nu} \eta_{\rho\sigma)} \partial^8}{2^{13} \cdot 5 
\cdot \pi^4} - \tfrac{\eta_{(\mu\nu} \partial_{\rho} \partial_{\sigma)}
\partial^6}{2^{10} \cdot 5 \cdot \pi^4} + \tfrac{7 [\eta_{\mu\rho}
\partial_{\nu} \partial_{\sigma} + \eta_{\nu\sigma} \partial_{\mu}
\partial_{\rho}] \partial^6}{2^{13} \cdot 5 \cdot \pi^4}$ & $\tfrac{\eta_{(\mu\nu} 
\eta_{\rho\sigma)} \partial^8}{2^{12} \cdot 5 \cdot \pi^4} + \tfrac{\eta_{(\mu\nu} 
\partial_{\rho} \partial_{\sigma)} \partial^6}{2^{9} \cdot 5 \cdot \pi^4} 
- \tfrac{7 [\eta_{\mu\rho} \partial_{\nu} \partial_{\sigma} + \eta_{\nu\sigma} 
\partial_{\mu} \partial_{\rho}] \partial^6}{2^{12} \cdot 5 \cdot \pi^4}$ \\
$$ & $-\tfrac{\partial_{\mu} \partial_{\nu} \partial_{\rho} \partial_{\sigma}
\partial^4}{2^{10} \cdot 3 \cdot 5 \cdot \pi^4} + \tfrac{H a \delta^0_{(\mu} 
\partial_{\nu)} \partial_{\rho} \partial_{\sigma} \partial^4}{2^{10} \cdot 3 
\cdot \pi^4} - \tfrac{H a' \delta^0_{(\rho} \partial_{\sigma)} \partial_{\mu} 
\partial_{\nu} \partial^4}{2^{10} \cdot 3 \cdot \pi^4} $ & $+\tfrac{\partial_{\mu} 
\partial_{\nu} \partial_{\rho} \partial_{\sigma} \partial^4}{2^{9} \cdot 3 
\cdot 5 \cdot \pi^4} - \tfrac{H a \delta^0_{(\mu} \partial_{\nu)} \partial_{\rho}
\partial_{\sigma} \partial^4}{2^{9} \cdot 3 \cdot \pi^4} +\tfrac{H a' 
\delta^0_{(\rho} \partial_{\sigma)} \partial_{\mu} \partial_{\nu} \partial^4}{
2^{9} \cdot 3 \cdot \pi^4}$ \\
$$ & $+ \tfrac{H a \delta^0_{\mu} [2 \eta_{\rho (\nu} \partial_{\sigma)} - 
\eta_{\nu\sigma} \partial_{\rho}] \partial^4}{2^{11} \cdot 3 \cdot \pi^4} + 
\tfrac{H a \delta^0_{\nu} [2 \eta_{\sigma (\mu} \partial_{\rho)}
- \eta_{\mu\rho} \partial_{\sigma}] \partial^4}{2^{11} \cdot 3 \cdot \pi^4}$ 
& $- \tfrac{H a \delta^0_{\mu} [2 
\eta_{\rho (\nu} \partial_{\sigma)} - \eta_{\nu\sigma} \partial_{\rho}] 
\partial^4}{2^{10} \cdot 3 \cdot \pi^4} - \tfrac{H a \delta^0_{\nu} [2 
\eta_{\sigma (\mu} \partial_{\rho)} - \eta_{\mu\rho} \partial_{\sigma}] 
\partial^4}{2^{10} \cdot 3 \cdot \pi^4}$ \\
$$ & $- \tfrac{H a' \delta^0_{\rho} [2 \eta_{\mu (\nu} \partial_{\sigma)} 
- \eta_{\nu\sigma} \partial_{\mu}] \partial^4}{2^{11} \cdot 3 \cdot \pi^4}
- \tfrac{H a' \delta^0_{\sigma} [2 \eta_{\nu (\mu} \partial_{\rho)} - 
\eta_{\mu\rho} \partial_{\nu}] \partial^4}{2^{11} \cdot 3 \cdot \pi^4}$ & 
$+ \tfrac{H a' \delta^0_{\rho} [2 \eta_{\mu (\nu} \partial_{\sigma)}
- \eta_{\nu\sigma} \partial_{\mu}] \partial^4}{2^{10} \cdot 3 \cdot \pi^4} 
+ \tfrac{H a' \delta^0_{\sigma} [2 \eta_{\nu (\mu} \partial_{\rho)} - 
\eta_{\mu\rho} \partial_{\nu}] \partial^4}{2^{10} \cdot 3 \cdot \pi^4}$ \\
$i\Delta_{\rm cf} \!\times\! i\Delta_{\rm cf}$ & $- \tfrac{H^2 a a' 
\delta^0_{\mu} \delta^0_{\rho} [\eta_{\nu\sigma} \partial^2 - 4 \partial_{\nu} 
\partial_{\sigma}] \partial^4}{2^{11} \cdot 3 \cdot \pi^4} - \tfrac{H^2 a a' 
\delta^0_{\nu} \delta^0_{\sigma} [\eta_{\mu\rho} \partial^2 - 4 \partial_{\mu} 
\partial_{\rho}] \partial^4}{2^{11} \cdot 3 \cdot \pi^4}$ & $+ \tfrac{H^2 a a' 
\delta^0_{\mu} \delta^0_{\rho} [\eta_{\nu\sigma} \partial^2 - 4 \partial_{\nu} 
\partial_{\sigma}] \partial^4}{2^{10} \cdot 3 \cdot \pi^4} + \tfrac{H^2 a a' 
\delta^0_{\nu} \delta^0_{\sigma} [\eta_{\mu\rho} \partial^2 - 4 \partial_{\mu} 
\partial_{\rho}] \partial^4}{2^{10} \cdot 3 \cdot \pi^4}$ \\
$$ & $- \tfrac{H^2 a^2 \delta^0_{\mu} \delta^0_{\nu} [\eta_{\rho\sigma} 
\partial^2 + 2 \partial_{\rho} \partial_{\sigma}] \partial^4}{2^{11} \cdot 
3 \cdot \pi^4} + \tfrac{H^2 a a' \delta^0_{\mu} \delta^0_{\sigma} [\eta_{\nu\rho} 
\partial^2 + 2 \partial_{\nu} \partial_{\rho}] \partial^4}{2^{11} \cdot 3 \cdot 
\pi^4}$ & $+ \tfrac{H^2 a^2 \delta^0_{\mu} \delta^0_{\nu} [\eta_{\rho\sigma} 
\partial^2 + 2 \partial_{\rho} \partial_{\sigma}] \partial^4}{2^{10} \cdot 
3 \cdot \pi^4} - \tfrac{H^2 a a' \delta^0_{\mu} \delta^0_{\sigma} [\eta_{\nu\rho} 
\partial^2 + 2 \partial_{\nu} \partial_{\rho}] \partial^4}{2^{10} \cdot 3 \cdot 
\pi^4} $ \\
$$ & $- \tfrac{H^2 {a'}^2 \delta^0_{\rho} \delta^0_{\sigma} [\eta_{\mu\nu} 
\partial^2 + 2 \partial_{\mu} \partial_{\nu}] \partial^4}{2^{11} \cdot 3 \cdot 
\pi^4} + \tfrac{H^2 a a' \delta^0_{\rho} \delta^0_{\nu} [\eta_{\mu\sigma} 
\partial^2 + 2 \partial_{\mu} \partial_{\sigma}] \partial^4}{2^{11} \cdot 3 
\cdot \pi^4}$ & $+\tfrac{H^2 {a'}^2 \delta^0_{\rho} \delta^0_{\sigma} [\eta_{\mu\nu} 
\partial^2 + 2 \partial_{\mu} \partial_{\nu}] \partial^4}{2^{10} \cdot 3 \cdot 
\pi^4} - \tfrac{H^2 a a' \delta^0_{\rho} \delta^0_{\nu} [\eta_{\mu\sigma} 
\partial^2 + 2 \partial_{\mu} \partial_{\sigma}] \partial^4}{2^{10} \cdot 3 
\cdot \pi^4}$ \\
$$ & $-\tfrac{H^3 a^2 a' \delta^0_{\mu} \delta^0_{\nu} \delta^0_{(\rho}
\partial_{\sigma)} \partial^4}{2^9 \cdot \pi^4} + \tfrac{H^3 a {a'}^2 
\delta^0_{\rho} \delta^0_{\sigma} \delta^0_{(\mu} \partial_{\nu)} \partial^4}{
2^9 \cdot \pi^4} - \tfrac{H^4 \delta^0_{\mu} \delta^0_{\nu} \delta^0_{\rho}
\delta^0_{\sigma} \partial^4}{2^9 \cdot \pi^4}$ & $+\tfrac{H^3 a^2 a' 
\delta^0_{\mu} \delta^0_{\nu} \delta^0_{(\rho} \partial_{\sigma)} \partial^4}{
2^8 \cdot \pi^4} - \tfrac{H^3 a {a'}^2 \delta^0_{\rho} \delta^0_{\sigma} 
\delta^0_{(\mu} \partial_{\nu)} \partial^4}{2^8 \cdot \pi^4} + \tfrac{H^4 
\delta^0_{\mu} \delta^0_{\nu} \delta^0_{\rho} \delta^0_{\sigma} \partial^4}{
2^8 \cdot \pi^4}$ \\
\hline
$$ & $\tfrac{H^2 a a' \eta_{(\mu\nu} \eta_{\rho\sigma)} \partial^6}{
2^{12} \cdot \pi^4} + \tfrac{H^2 a a' [\eta_{\mu\rho} \partial_{\nu}
\partial_{\sigma} + 2 \eta_{\nu\sigma} \partial_{\mu} \partial_{\rho}]
\partial^4}{2^{10} \cdot 3 \cdot \pi^4}$ & $-\tfrac{H^2 a a' \eta_{(\mu\nu}
\eta_{\rho\sigma)} \partial^6}{2^{11} \cdot \pi^4} - \tfrac{H^2 a a'
[\eta_{\mu\rho} \partial_{\nu} \partial_{\sigma} + 2 \eta_{\nu\sigma}
\partial_{\mu} \partial_{\rho}] \partial^4}{2^{10} \cdot 3 \cdot \pi^4}$ \\
$i\Delta_{\rm cf} \!\times\! A_0$ & $-\tfrac{H^2 a a' \eta_{(\mu\nu}
\partial_{\rho} \partial_{\sigma)} \partial^4}{2^{10} \cdot \pi^4} +
\tfrac{H^3 a^2 a' \delta^0_{\mu} (\eta_{\nu\sigma} \partial_{\rho}
- \eta_{\rho (\nu} \partial_{\sigma}) \partial^4}{2^9 \cdot 3 \cdot \pi^4}$ 
& $+ \tfrac{H^2 a a' \eta_{(\mu\nu} \partial_{\rho} \partial_{\sigma)}
\partial^4}{2^9 \cdot \pi^4} + \tfrac{H^2 a a' \partial_{\mu} \partial_{\nu} 
\partial_{\rho} \partial_{\sigma} \partial^2}{2^9 \cdot 3 \cdot \pi^4} 
-\tfrac{H^3 a^2 a' \delta^0_{\mu} (2 \eta_{\nu\sigma} \partial_{\rho}
- \eta_{\rho (\nu} \partial_{\sigma}) \partial^4}{2^8 \cdot 3 \cdot \pi^4}
$ \\
$$ & $-\tfrac{H^3 a {a'}^2 \delta^0_{\rho} (\eta_{\nu\sigma} \partial_{\mu}
- \eta_{\mu (\nu} \partial_{\sigma}) \partial^4}{2^9 \cdot 3 \cdot \pi^4}
- \tfrac{H^4 (a a')^2 \delta^0_{\mu} \delta^0_{\rho} \eta_{\nu\sigma}
\partial^4}{2^{10} \cdot \pi^4}$ 
& $+ \tfrac{H^3 a {a'}^2 \delta^0_{\rho} (2\eta_{\nu\sigma} \partial_{\mu}
- \eta_{\mu (\nu} \partial_{\sigma}) \partial^4}{2^8 \cdot 3 \cdot \pi^4}
+\tfrac{H^4 (a a')^2 \delta^0_{\mu} \delta^0_{\rho} [\eta_{\nu\sigma}
\partial^2 - 2 \partial_{\nu} \partial_{\sigma}] \partial^2}{2^9 \cdot \pi^4}$ \\
\hline
$i\Delta_{\rm cf} \!\times\! B_0$ & $0$ & $0$ \\
\hline
$i\Delta_{\rm cf} \!\times\! C_0$ & $0$ & $0$ \\
\hline
$i\Delta_{\rm cf} \!\times\! A_1$ & $0$ & $0$ \\
\hline
$i\Delta_{\rm cf} \!\times\! B_1$ & $0$ & $0$ \\
\hline
$i\Delta_{\rm cf} \!\times\! C_1$ & $0$ & $0$ \\
\hline
$$ & $$ & $\tfrac{H^4 \eta_{(\mu\nu} \eta_{\rho\sigma)} 
\partial^4}{2^{9} \cdot \pi^4} + \tfrac{H^4 \eta_{(\mu\nu} \partial_{\rho}
\partial_{\sigma)} \partial^2}{2^8 \cdot 3 \cdot \pi^4}$ \\
$A_0 \!\times\! A_0$ & $-\tfrac{H^4 \eta_{(\mu\nu} \eta_{\rho\sigma)} 
\partial^4}{2^{10} \cdot \pi^4}$ & $-\tfrac{H^4 [\eta_{\mu\rho}
\partial_{\nu} \partial_{\sigma} + \eta_{\nu\sigma} \partial_{\mu}
\partial_{\rho}] \partial^2}{2^8 \cdot \pi^4}$ \\
$$ & $$ & $-\tfrac{H^4 \partial_{\mu} \partial_{\nu} \partial_{\rho}
\partial_{\sigma}}{2^7 \cdot 3 \cdot \pi^4}$ \\
\hline
\end{tabular}
\caption{\footnotesize Nonlocal contributions from products of the form 
$(a a')^{D-2} \partial_{\mu} \partial'_{\rho} i\Delta_I \!\times\! \partial_{\nu}
\partial'_{\sigma} i\Delta_J$.}
\label{TableFFin}
\end{sidewaystable}

\end{document}